\documentclass[%
reprint,
amsmath,amssymb,
prl,
]{revtex4-2}

\usepackage[unicode=true,colorlinks=true,citecolor=blue,urlcolor=blue]{hyperref}

\usepackage{gensymb}
\usepackage{graphicx}% Include figure files
\usepackage{dcolumn}% Align table columns on decimal point
\usepackage{bm}% bold math
\makeatletter
\renewcommand*\env@matrix[1][\arraystretch]{%
  \edef\arraystretch{#1}%
  \hskip -\arraycolsep
  \let\@ifnextchar\new@ifnextchar
  \array{*\c@MaxMatrixCols c}}
\makeatother

\begin{document}

%\preprint{APS/123-QED}
%footnotes
%\bibliographystyle{apsrev4-1}
\footnotetext[1]{The notion of QSH effect has subtly diverse meanings in literature. Here, we strictly discriminate the QSH states protected by both FTR and $s_z$ spin conservation from the 2D TIs with only FTR symmetry~\cite{wen2017Colloquium} (See more discussions in SM~\cite{suppl}).} %In this paper, we also follow this definition of QSH systems. }
\footnotetext[2]{The $\mathbb{Z}$ invariant of QSH phases, which we designate as QSH-Chern number, is identical with the spin-Chern number when spin $s_z$ is conserved. However, spin-Chern number can still be defined in the spin-coupled case~\cite{sheng2006quantum,prodan2009robustness}, so we avoid using the term spin-Chern number to indicate the index of the QSH phases (see SM for details~\cite{suppl}).}

\title{
Photonic $\mathbb{Z}_2$ topological Anderson insulators
}

\author{Xiaohan Cui}
\author{Ruo-Yang Zhang}
\email{ruoyangzhang@ust.hk}
\author{Zhao-Qing Zhang}%
\author{C. T. Chan}%
 \email{phchan@ust.hk}
\affiliation{%
Department of Physics, The Hong Kong University of Science and Technology, Hong Kong, China
}%

\date{\today}% It is always \today, today,

\begin{abstract}

That disorder can induce nontrivial topology is a surprising discovery in topological physics. 
As a typical example, Chern topological Anderson insulators (TAIs) have been realized in photonic systems, where the topological phases exist without symmetry protection. In this work, by taking TM and TE polarizations as pseudo-spin degrees of freedom, we theoretically propose a scheme to realize disorder-induced symmetry-protected topological (SPT) phase transitions in two-dimensional photonic crystals (PCs) with a combined time-reversal, mirror and duality symmetry $\mathcal{T}_f=\mathcal{T}M_z\mathcal{D}$. 
In particular, we demonstrate that the disorder-induced SPT phase persists even without pseudo-spin conservation, thereby realizing a photonic $\mathbb{Z}_2$ TAI, in contrast to a $\mathbb{Z}$-classified quantum spin Hall (QSH) TAI with decoupled spins.
By formulating a new scattering approach,
we show that the topology of both the QSH and $\mathbb{Z}_2$ TAIs can be manifested by the accumulated spin rotations of the reflected waves from the PCs. 
Using a transmission structure, we also illustrate the trivialization of a disordered QSH phase with an even integer topological index caused by spin coupling.

\end{abstract}

\maketitle

% \newpage
%\tableofcontents
% \section{Introduction}
\textit{Introduction.}---As the photonic counterparts of the electronic topological insulators (TIs)~\cite{topology-review_Kane_2010_Rev.Mod.Phys.,topology-review_Zhang_2011_Rev.Mod.Phys.,quantumspinHall_Mele_2005_Phys.Rev.Lett.,kane2005z,bernevig2006quantum,QSH_Zhang_2006_Science,sheng2006quantum}, topological photonic crystals (PCs) can support gapless edge states that are backscattering-immune against weak disorders~\cite{topologicalphotonics_Raghu_2008_Phys.Rev.Lett.,QSH_Haldane_2008_Phys.Rev.A, topologicalphotonics-theory_Soljacic_2008_Phys.Rev.Lett.,topologicalphotonics-experiment_Soljacic_2009_Nature,topologicalphotonics-review_Soljacic_2014_NaturePhoton,topologicalphotonics-review_Carusotto_2019_Rev.Mod.Phys.}. However, if the disorders are sufficiently strong, the mobility bandgap closes due to the Anderson localization, making the system topologically trivial~\cite{disorder_Chong_2017_Phys.Rev.B,disorder_Fan_2017_Phys.Rev.B,liu2017DisorderInduced,disorder-experiment_Zhang_2020_LightSciAppl,Disorder_Zhang_2020_Opt.Express,disorder-review_Park_2021_NatRevMater,disorder_Zhang_2019_Phys.Rev.B,Disorder_Zhang_2020_LightSciAppl}.
Interestingly, recent studies report a reverse transition: disorders can drive photonic crystals to change from a topologically trivial phase to a nontrivial phase. The systems in these disorder-induced topological phases are called topological Anderson insulators (TAIs)~\cite{TAI_Shen_2009_Phys.Rev.Lett.,TAI_Beenakker_2009_Phys.Rev.Lett.,xing2011Topological,zhang2012Localization,prodan2011Disordered,titum2015disorder,TAI_Schmidt_2016_SciRep,TAI-experiment_Gadway_2018_Science,TAI-experiment_Mitchell_2018_Nat.Phys.,TAI-experiment_Szameit_2018_Nature,TAI-experiment_Zhang_2020_Phys.Rev.Lett.,zhang2019topological,TAI-experiment_Zhang_2021_Phys.Rev.Lett.,wang2022Structural}. 
% On the other hand, 

While the photonic TAI systems are mainly limited to Chern-type~\cite{TAI-experiment_Szameit_2018_Nature,TAI-experiment_Zhang_2020_Phys.Rev.Lett.}, the original discovery of TAI stemmed from investigating disordered quantum spin Hall (QSH) systems with fermionic time reversal (FTR) symmetry~\cite{TAI_Shen_2009_Phys.Rev.Lett.}.
The initial idea of QSH effect is to build a system with two oppositely spin-polarized quantum Hall copies related by FTR symmetry~\cite{quantumspinHall_Mele_2005_Phys.Rev.Lett.,bernevig2006quantum}. However, Kane and Mele pointed out that even without $s_z$ spin conservation, the FTR symmetry can protect nontrivial $\mathbb{Z}_2$ topological phase~\cite{kane2005z}, which is different from the $\mathbb{Z}$ classified QSH phases with conserved spin current~\cite{shiozaki2014topology,wen2017Colloquium,suppl}. This surprising discovery gave rise to the notion of topological insulators and triggered the topological revolution in physics. 
In photonics, although QSH-like effects in ordered PCs had been widely studied in recent years~\cite{hafezi2011Robust,hafezi2013Imaging,PTI-firsttheory_Shvets_2013_NatureMater,PTI-firstexperiment_Chan_2014_NatCommun,PTI-theory_Li_2015_Phys.Rev.Lett.,PTI-experiment_Chan_2015_NatCommun,PTI-experiment_Shvets_2015_Phys.Rev.Lett.,PTI-theory_Ochiai_2015_J.Phys.Soc.Jpn.,PTI-theory_Chen_2016_PNAS,cheng2016Robust,gao2018Topologically,PTI-theory_Silveirinha_2017_Phys.Rev.B,PTI-th_Silveirinha_2019_Nanophotonics,QSH_Khanikaev_2018_Sci.Adv.,PTI-theory_Chen_2019_Crystals,wu2015Scheme,barik2016Twodimensionally,maczewsky2020Fermionic}, almost all these works are based on finding two decoupled pseudo-spin sectors; and the subtle but important difference between the QSH and $\mathbb{Z}_2$ topology had rarely been explored, aside from the discussion of a similar issue in the Floquet system~\cite{maczewsky2020Fermionic}. Moreover, the study of disorder-induced TAI phase transition in QSH and $\mathbb{Z}_2$ systems also remains absent in photonics. The aim of this work is to fill these gaps.

In this letter, we theoretically and computationally design a PC composed of pseudo-FTR symmetric media. By introducing geometric randomness, we observe the topological transitions from a trivial phase to QSH (with $z$-mirror symmetry) and $\mathbb{Z}_2$ (without $z$-mirror symmetry) TAI phases, which are demonstrated by the bulk transmission spectra and the helical edge states. 
To characterize the topology quantitatively, we connect the disordered PCs to a waveguide lead where we detect the pseudo-spin reflection from the PCs ~\cite{ReflectionPhase_Brouwer_2010_Phys.Rev.B,ReflectionPhase_Brouwer_2011_Phys.Rev.B,Z2-calculation_Brouwer_2014_Phys.Rev.B,Z2-calculation_Akhmerov_2012_Phys.Rev.B,xiao2014surface,gao2015determination,ReflectionPhase_Chong_2014_Phys.Rev.B,ReflectionPhase_Chong_2015_Phys.Rev.X,Z2-calculation_Hafezi_2015_Phys.Rev.A,mittal2016measurement,ReflectionPhase_Chong_2016_Phys.Rev.B,ReflectionPhase_Zhang_2020_Phys.Rev.Lett.} and find that the windings of the reflected spins can characterize both the bulk $\mathbb{Z}$ and $\mathbb{Z}_2$ topological indices.
In addition, we exhibit the difference between the QSH and $\mathbb{Z}_2$   TAI phases via boundary transport effects, which was rarely addressed in previous works. 

\begin{figure*}[t!]
\includegraphics[width=0.98\textwidth]{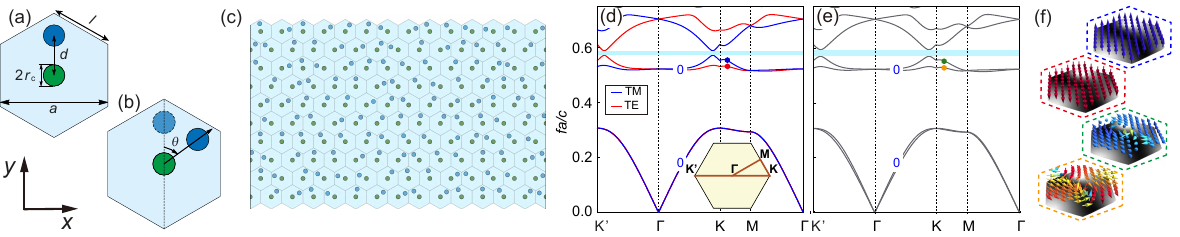}
\caption{\label{fig-1}
(a) Unit cell of an ordered hexagonal PC with a lattice constant $a$. Parameters: $r_c=0.1a$, $d=0.37a$, $l=a/\sqrt{3}$. %The distance between the two cylinders is $d=0.37a$.
(b) One typical cell of the (c) disordered PC  with the off-center cylinder rotated by a random angle $\theta$. 
(d,e) Band structures of two ordered PCs with (d) $M_z$-symmetry ($\kappa =0$) and without (e) $M_z$-symmetry ($\kappa = 0.06$). %are plotted in (d) and (e), respectively. 
(f) The distributions of pseudo-spins $\vec s$ (arrows) and intensity $\langle \psi | \psi \rangle$ (grayscale colormap) of four eigenstates (labeled by the colored dots in (d) and (e)) with conserved and nonconserved pseudo-spins.}
\end{figure*}

%\vspace{-4pt}
\textit{$\mathcal{T}_f $-symmetric photonic crystals.}---Surveying the literature of photonic QSH effects in 2D PCs, the underlying principle of nearly all schemes that are based on the special property of optical materials~\cite{PTI-firsttheory_Shvets_2013_NatureMater,PTI-firstexperiment_Chan_2014_NatCommun,PTI-theory_Li_2015_Phys.Rev.Lett.,PTI-experiment_Chan_2015_NatCommun,PTI-experiment_Shvets_2015_Phys.Rev.Lett.,PTI-theory_Ochiai_2015_J.Phys.Soc.Jpn.,PTI-theory_Chen_2016_PNAS,cheng2016Robust,gao2018Topologically,PTI-theory_Silveirinha_2017_Phys.Rev.B,PTI-th_Silveirinha_2019_Nanophotonics,QSH_Khanikaev_2018_Sci.Adv.,PTI-theory_Chen_2019_Crystals} can be traced to the hidden  
anti-unitary symmetry $\mathcal{T}_f = \mathcal{T} M_z \mathcal{D}$ which combines time reversal $\mathcal{T}$, mirror reflection $M_z: (x,y,z)\rightarrow (x,y,-z)$ and electromagnetic duality transformation $\mathcal{D}: (\mathbf{E}, \mathbf{H} ) \rightarrow (\mathbf{H}, -\mathbf{E})$~\cite{PTI-theory_Silveirinha_2017_Phys.Rev.B}. Since $\mathcal{T}_f^2=-1$,
$\mathcal{T}_f$ serves as a pseudo-FTR operator for photons. 
In the basis of the wave function $\Psi = (\mathbf{E}, \mathbf{H})^\intercal$, the operator takes the matrix representation $\mathcal{T}_f = -i\sigma_y \otimes m_z \mathcal{K}$, where $\sigma_i$ ($i=x,y,z$) denote the Pauli matrices, $m_z = \rm diag(1,1,-1)$, and $\mathcal{K}$ denotes complex-conjugate.
The constitutive tensors of a $\mathcal{T}_f$-invariant photonic medium should respect
$\tensor{\varepsilon}/\varepsilon_0 =\begin{pmatrix}
\tensor{{\varepsilon}} _T & \mathbf g\\
\mathbf g^\dagger &{{\varepsilon _z}}
\end{pmatrix}
$,
$\tensor{\mu}/\mu_0 =
\begin{pmatrix}
\tensor{{\varepsilon}}^*_T & -\mathbf g^*\\
-\mathbf g^\intercal &{{\varepsilon _z}}
\end{pmatrix}
$, where $\varepsilon_0$ and $\mu_0$ are the vacuum permittivity and permeability. Hereinafter, we let 
$\tensor{\varepsilon}_T= %\sigma_0-\alpha\sigma_y$ %
\begin{pmatrix}[0.7]
1 &i \beta \\-i \beta  &1 
\end{pmatrix}$
so that $\tensor{\varepsilon}$ and $\tensor{\mu}$ are gyrotropized in opposite manners~\cite{PTI-theory_Chen_2019_Crystals}, and we use $\mathbf{g}=(\kappa,\kappa)^\intercal$ to control the coupling between TE and TM modes. We stress that the use of gyrotropic media is only for computational convenience. Indeed, our theory is applicable to all $\mathcal{T}_f$-invariant systems, and the photonic $\mathbb{Z}_2$ TAIs can also be realized by reciprocal materials via a $SU(2)$ gauge transformation~\cite{Tf-symmetry_Chan_2019_NatCommun,suppl}.

Figure~\ref{fig-1}(a) shows the unit cell of an ordered $\mathcal{T}_f$-invariant PC. The central gyrotropic cylinder (green) with $\beta=0.7$, $\varepsilon_z=25$, $\kappa =0$ breaks the $\mathcal{T}$ symmetry, while the off-center reciprocal cylinder (blue)  with $\beta=0$, $\varepsilon_z=5$, $\kappa =0$ breaks the spatial inversion  ($\mathcal{P}$) symmetry of the PC. 
If the background medium (light blue) has $\tensor{\varepsilon}/\varepsilon_0 = \tensor{\mu}/\mu_0=1, \kappa=0$, 
the PC has $M_z$ symmetry, and therefore the eigenstates can always be selected as TM ($M_z$-odd) or TE ($M_z$-even) polarized (see Fig.~\ref{fig-1}(d)). In analogy to spin-$1/2$ fermions, we define the pseudo-spin of electromagnetic fields using the two-component spinor $|\psi\rangle =(E_z, \eta_0 H_z)^\intercal$ ($\eta_0=\sqrt{\mu_0/\varepsilon_0}$ is the vacuum impedance) as
\begin{equation}
    \vec s \equiv \langle \psi | \vec \sigma| \psi \rangle / \langle \psi | \psi \rangle,\label{pseudo-spin}
\end{equation}
with $\vec \sigma=\{\sigma_x,\sigma_y,\sigma_z\}$. Then TM and TE modes serves as pseudo-spin up and down states, respectively. The 2D Maxwell's equations with $M_z$ symmetry are invariant under a $U(1)$ pseudo-spin-rotation $U_s=\exp(i\vartheta\sigma_z/2)$ with an arbitrary angle $\vartheta$~\cite{suppl}, which guarantees the  conservation of the $s_z$ spin component (see Fig.~\ref{fig-1}(f)). 
The SPT phases with $\mathcal T_f$ and $M_z$ symmetries are $\mathbb{Z}$ classified~\cite{shiozaki2014topology} by a QSH-Chern number $C_s$~\cite{suppl} (labeled on the bands in Fig.~\ref{fig-1}(d)). 
The sum of $C_s$ below the gap concerned (light blue) is 0, indicating the gap is topologically trivial. This is because the $\mathcal{P}$-breaking effect due to the off-center cylinder beats the $\mathcal{T}$-breaking effect induced by the central gyrotropic cylinder in each spin sector~\cite{suppl}. 

\begin{figure}[b!]
\includegraphics[width=0.48\textwidth]{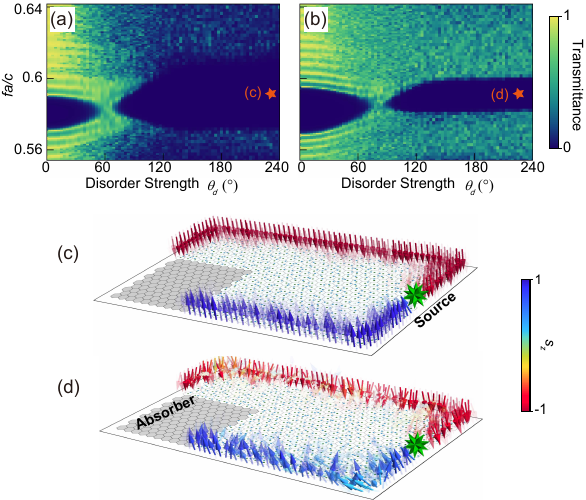}
\caption{\label{fig-2}
Typical bulk transmittance averaged over $N=10$ samples (size: $L_x \times L_y =30a\times 30l$) for (a) spin-decoupled ($\kappa=0$) and (b) coupled cases ($\kappa=0.06$).  
(c,d) Helical edge states in (c) a QSH TAI PC and in (d) a $\mathbb{Z}_2$ TAI PC (corresponding to the orange stars at $(\theta_d,f)=(220\degree,0.59c/a)$ in (a,b)). 
Arrows' color and transparency represent the $s_z$ component of pseudo-spins and the field intensity, respectively. 
}
\end{figure}

Next, we break $M_z$ symmetry of the PC via adding the coupling term $\mathbf{g}=(\kappa,\kappa)^\intercal$ into the background medium. As shown in Fig.~\ref{fig-1}(e), the topologically trivial bandgap remains open at $\kappa=0.06$~\cite{suppl}, but the pseudo-spins of each Bloch mode are no longer uniformly polarized (see Fig.~\ref{fig-1}(f)).
Though the TM-TE coupling makes the QSH phases ill-defined~\cite{fu2006Time}, %we will see that 
$\mathcal{T}_f$ symmetry itself can support a nontrivial SPT phase characterized by the Kane-Mele $\mathbb{Z}_2$ index $\nu$~\cite{kane2005z}, which is connected with the QSH-Chern number by $\nu=C_s\bmod 2$ in the weak coupling limit.

\begin{figure*}[tbh!]
\includegraphics[width=0.98\textwidth]{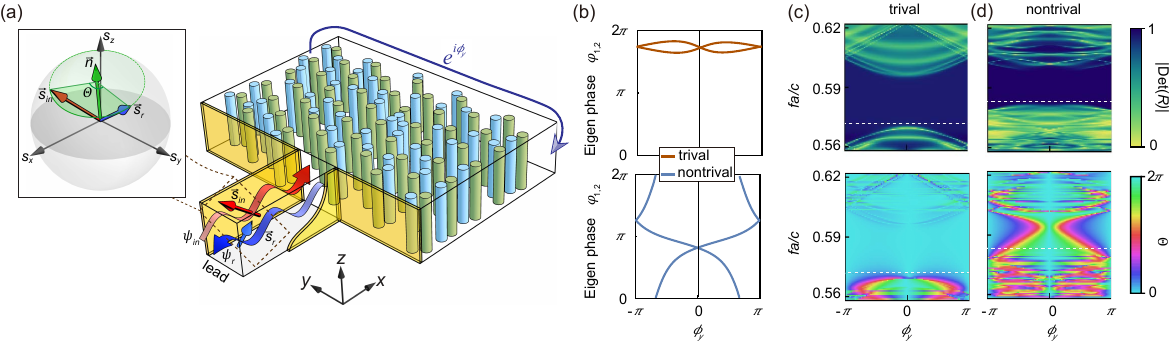}
\caption{\label{fig-4}
(a) Schematic for retrieving the $\mathbb{Z}_2$ index from the spin reflection inside the waveguide lead connected to the disordered PC (size: $18a \times 15l$) with a twisted boundary condition $e^{i\phi_y}$. 
Inset (Bloch sphere): the rotation between the reflected and incident spins. 
(b) Eigen reflection phases in the trivial ($fa/c=0.572$) and nontrivial ($fa/c=0.583$) gaps. The total reflection magnitude $|\mathrm{Det}(R)|$ and the spin rotation angle $\Theta$ in the (c) trivial  ($\theta_d=20\degree$), and (d) nontrivial ($\theta_d=220\degree$) gaps. 
 The white dashed lines mark the frequencies of the eigen reflection phases in (b).
 }
\end{figure*}

% \section{Disorder-induced topological phase transitions}
\textit{Disorder-induced topological phase transitions.}--- 
As shown in Fig. \ref{fig-1}(b) and (c), we introduce disorder into the PCs with $\kappa=0$ by rotating the the off-center cylinder around the center one through a random angle $\theta$ (uniformly distributed in the interval $[-\theta_d/2,\theta_d/2]$) in each unit cell. In this way, we build  disordered PC samples whose top and bottom boundaries are glued together continuously, and plane wave ports are imposed on the left and right boundaries. In the presence of $M_z$ symmetry, one only need to consider the TM modes, since the TE modes correspond bijectively to the TM ones by $\mathcal{T}_f$ symmetry. We simulate the TM bulk transmission between the two ports over $N$ random samples using COMSOL, and calculate the typical (geometric mean) transmittance $\langle T\rangle_\mathrm{typ}=\exp\big[\sum_{n=1}^{N}{\ln t_n}/N\big]$ as a function of the disorder strength $\theta_d$, as shown in Fig. \ref{fig-2}(a), where $t_n$ represents the TM transmittance of the $n$-th sample. When disorder is weak, the transmission gap (dark region) corresponds to the topologically trivial bandgap of the ordered PC in Fig. \ref{fig-1}(d). As we increase $\theta_d$, the mobility gap closes and reopens at around $\theta_d=60$\degree, indicating the PC ensemble transitions into a QSH TAI phase, which can be verified by calculating the local QSH-Chern number~\cite{LocalChernNumber_Kitaev_2006_AnnalsofPhysics,TAI-experiment_Mitchell_2018_Nat.Phys.,suppl}. The QSH TAI is nothing but two copies of Chern TAIs (TE and TM) related by $\mathcal{T}_f$ symmetry.  Intuitively, the emergence of TAI phase in each spin sector is because strengthening the disorders can smooth out the $\mathcal{P}$-violations in different unit cells and makes the $\mathcal{T}$-breaking effect comparatively more dominant~\cite{suppl}.

Turning now to the spin-coupled case ($\kappa=0.06$), we let an obliquely polarized plane wave with $\mathbf{E}\propto(0,1,1)$ incident from the left port of the disorder PCs and collect the total transmittance $\langle T\rangle_\mathrm{typ}=\exp\big[\sum_{n=1}^{N}{\ln(t_n+t'_n)}/N\big]$ of both TE ($t_n$) and TM ($t'_n$) polarizations at the right port. The phase diagram in Fig. \ref{fig-2}(b) also shows the process of mobility gap closing and reopening, as the evidence of the disorder-induced transition from the trivial phase to the nontrivial $\mathbb{Z}_2$ TAI phase.

\begin{figure}[t!]
\includegraphics[width=0.48\textwidth]{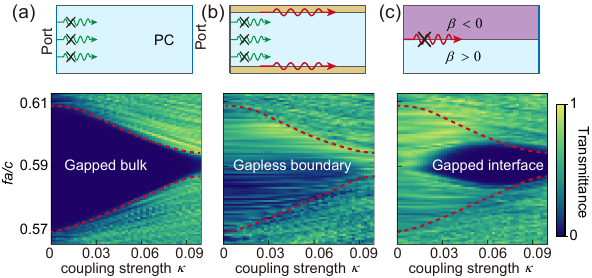}
\caption{\label{fig-5} Typical transmittance spectra (averaged over 15 samples) for the disordered PCs ($\theta_d=150\degree$) with different transverse boundary conditions.
(a) PC with continuously glued top and bottom boundaries. (b) PC sandwiched by TSIC layers (orange).
(c) Two PCs consisting of oppositely gyrotropized center cylinders ($\beta=\pm0.7$ respectively) joined at two domain walls (center interface and top-bottom boundary). The red dashed curves denote the bulk mobility edges. }
\end{figure}

To demonstrate the two gapped phases after disorder-induced transitions indeed have nontrivial topology, we attempt to observe helical edge states  on the boundaries of two disordered PCs in these phases with (Fig.~\ref{fig-2}(c)) and without (Fig.~\ref{fig-2}(d)) $M_z$ symmetry %with $\theta_d=220$ at the frequency $f=5.9c/a$  
(labeled by orange stars in Fig.~\ref{fig-2}(a,b)).
As show in Fig.~\ref{fig-2}(c-d), a point source (green star) with superposed electric and magnetic monopoles is placed on the right boundary of each PC encased by a $\mathcal{T}_f$-symmetric insulation cladding (TSIC) with $\tensor{\varepsilon}/\varepsilon_0=\tensor{\mu}/\mu_0=\mathrm{diag}(1,1,-1)$. 
For the $M_z$-symmetric case (Fig.~\ref{fig-2}(c)),
we observe that a purely spin-up surface wave (blue arrows) is emitted from the source and propagates clockwise along the boundaries, while a purely spin-down wave (red arrows) propagates counterclockwise along the boundaries due to $\mathcal{T}_f$ symmetry. This photonic QSH effect undoubtedly confirms the disorder-induced mobility gap is a TAI. 
In the case of breaking $M_z$ symmetry, Fig.~\ref{fig-2}(d) shows that $\mathcal{T}_f$ symmetry can  sustain the two counter-propagating surfaces waves; however, their spins are no longer uniformly polarized.
Indeed, the whole disorder-induced phase permits bidirectional gapless boundary transportation, demonstrating that $\mathcal{T}_f$-invariant TAIs are compatible with spin coupling, which will be explained in Fig.~\ref{fig-5}(b).

\textit{Scattering approach for classifying disordered topological phases.}---
Inspired by the scattering approaches to retrieving the bulk topology~\cite{ReflectionPhase_Brouwer_2010_Phys.Rev.B,ReflectionPhase_Brouwer_2011_Phys.Rev.B,Z2-calculation_Brouwer_2014_Phys.Rev.B,Z2-calculation_Akhmerov_2012_Phys.Rev.B,xiao2014surface,gao2015determination,ReflectionPhase_Chong_2014_Phys.Rev.B,ReflectionPhase_Chong_2015_Phys.Rev.X,Z2-calculation_Hafezi_2015_Phys.Rev.A,mittal2016measurement,ReflectionPhase_Chong_2016_Phys.Rev.B,ReflectionPhase_Zhang_2020_Phys.Rev.Lett.},
 we propose a new spin-reflection method for calculating the topological invariants. 
As shown in Fig.~\ref{fig-4}(a), we connect a $\mathcal{T}_f$-symmetric dual-mode waveguide lead %(gray) %whose width is less than half wavelength 
to the left end of the PC, and impose a twisted boundary condition with a twist angle $\phi_y$ to the two edges along the $x$ direction  of the PC~\cite{ReflectionPhase_Brouwer_2011_Phys.Rev.B,Z2-calculation_Brouwer_2014_Phys.Rev.B}  (see details and a possible experimental realization in Supplemental materials (SM)~\cite{suppl}).

Considering the scattering process of an incident field $\psi_\mathrm{in}(\phi_y,\omega)$ at frequency $\omega$ impinging upon the PC in the lead, the reflected spinor $\psi_\mathrm{r}(\phi_y,\omega)$ is related to the incident one by a reflection matrix $R$: 
$    \psi_{\rm r}\left( \phi_{y},\omega\right)=R\left(\phi_{y},\omega\right) \psi_{\rm in}\left( \phi_{y},\omega\right)
$. 
Within the mobility gap, $R(\phi_y,\omega)$ is a $U(2)$ matrix for the general PCs without $M_z$ symmetry, and can be expressed as
$%\begin{equation}
R = e^{i q} \exp \left[i \alpha\, \vec n \cdot \vec \sigma\right] %= e^{i q}\left(\cos \alpha \sigma_{0}+i \hat{n} \cdot \vec{\sigma} \sin \alpha\right)
$ %\end{equation}
($\vec n$ is a unit vector), whose eigenvalues  $r_{1,2}=e^{i(q\pm\alpha)}$ represent two eigen reflection coefficients. 
%We can define a reflection phase matrix as $A=-i\ln R$, which is Hermitian and $T_f$-symmetric.
Thanks to $\mathcal{T}_f$ symmetry, the reflection matrix satisfies~\cite{suppl}
\begin{equation}
    \sigma_y R(\phi_y,\omega)^*\sigma_y=R(-\phi_y,\omega)^\dagger, 
\end{equation}
which gives rise to %the spectral symmetry and 
the Kramers' degeneracy of the two eigen reflection phases
$\varphi_{1,2}=\arg(r_{1,2}) = q\pm \alpha$ at $\mathcal{T}_f$-invariant points ($\phi_0=0, \pi$), implying %$q(\phi_y)=q(-\phi_y)$, $\alpha(\phi_y)=\alpha(-\phi_y)$ and 
%$\alpha(\phi_0)=(0\; \text{or}\; \pi) \bmod 2\pi$.
$R(\phi_0) = e^{iq} \sigma_0 \in U(1)$. As a consequence, the loops of the $\mathcal{T}_f$-symmetric reflection matrices over a cycle of $\phi_y$ are topologically classified by the relative homotopy group~\cite{suppl,hatcher2005algebraic}%the topological classification of $R$ is
\begin{equation}\label{eq:relative-homotopy}
    \pi_1(U(2),U(1)) = %\pi_1(SU(2)/\mathbb Z_2) = 
    \pi_1(SO(3)) =\mathbb Z_2.
\end{equation}
Akin to the Wilson loop approach~\cite{yu2011Equivalent}, the two classes of $R(\phi_y)$ can be visually distinguished by the trivial and nontrivial connectivities of the concatenated trajectories of the two eigen reflection phases $\varphi_{1,2}$ in a half-period ($\phi_y\in[0,\pi]$) (see the two examples in Fig.~\ref{fig-4}(b)). 
Choosing a smooth gauge of $\alpha(\phi_y)$, the $\mathbb{Z}_2$ index can also be calculated by the winding number of the relative phase $\Theta=\varphi_2-\varphi_1=-2\alpha$ over a half-period:
\begin{equation}
    \tilde{\nu}=\left[\frac{1}{2\pi}\int_0^\pi d\phi_y \frac{\partial(\varphi_2-\varphi_1)}{\partial\phi_y}\right]\bmod 2= 0\;\text{or}\;1.
\end{equation}

From the perspective of spin, after reflection from the disordered PC, the spin $\vec{s}_\mathrm{r}$ of reflected wave rotates from the initial orientation $\vec{s}_\mathrm{in}$ about the axis $\vec{n}$ by the angle $\Theta=-2\alpha$, as depicted in the inset of Fig.~\ref{fig-4}(a),
\begin{equation}
    \vec s_\mathrm{r} = G(\Theta,\vec n) \vec s_\mathrm{in}=\exp\big[\Theta\, \vec{n}\cdot\vec{\mathfrak{L}}\big]\vec{s}_\mathrm{in},
\label{rotation-SO3}
\end{equation}
where $G(\Theta,\vec n) \in SO(3)$ is the spin reflection matrix % with the rotation angle $\Theta = -2 \alpha$ and rotation axis $\vec n$ after reflection, 
with $\vec{\mathfrak{L}}$ denoting the $\mathfrak{so}(3)$ generators. 
It is intriguing that the spin rotation angle is precisely the phase difference $\Theta=\varphi_2-\varphi_1$ of the two eigen reflection coefficients. Therefore, although % TM-TE coupling renders 
the rotation axis $\vec{n}(\phi_y)$ is generally unfixed caused by TM-TE coupling, the accumulated spin rotation angle $\Delta\Theta$ during a half cycle of $\phi_y$ remains quantized but can only take two stable values $0$ and $2\pi$, which reveals the physical meaning of the homotopy group $\pi_1(SO(3))=\mathbb{Z}_2$ in Eq.~\eqref{eq:relative-homotopy}~\cite{suppl}. 
% By building the correspondence of the edge states and the relative phase~\cite{suppl}, 
We also established the correspondence of the edge states and the relative phase and hence proved that the bulk topology is indeed equivalent to the $\mathbb{Z}_2$ classified accumulated spin rotation of the reflected waves~\cite{suppl}.

Figure~\ref{fig-4}(c-d) plots $|\mathrm{Det}(R)|$ and the relative phase $\Theta$ for spin-coupled PCs ($\kappa=0.06$) inside the trivial and TAI phases, respectively. By $\mathcal{T}_f$ symmetry, $\Theta(\phi_y)$ is symmetric about $\mathcal{T}_f$-invariant points at which it reduces to zero due to Kramers' degeneracy. Inside the mobility gaps ($|\mathrm{Det}(R)|=1$) with weak (Fig.~\ref{fig-4}(c)) and strong (Fig.~\ref{fig-4}(d)) disorders, the winding numbers of $\Theta(\phi_y)$ over a half-period are fixed to 0 and 1, respectively. This result corroborates that the spin-coupled TAI phase in Fig.~\ref{fig-2}(b) possesses a well-defined $\mathbb{Z}_2$ topology.

For the PCs with $M_z$ symmetry, due to the $s_z$ spin conservation, the rotation axis $\vec{n}$ between the reflected and incident spins is fixed along $s_z$. As such, the spin rotation matrix reduces to $G=\exp\left[\Theta\mathfrak{L}_z\right]\in SO(2)$. Thus, after $\phi_y$
% the twisted boundary 
evolves over a half-period, the reflected spin will rotate about the $s_z$ axis an integer number of times ${C}_s\in\pi_1(SO(2))=\mathbb{Z}$, offering a spin rotation route to extracting the QSH-Chern numbers of the spin-decoupled TAI phases (see the numerical results in SM~\cite{suppl}).
Note that our method does not require ensemble averaging, making it more efficient for computing topological indices of disordered systems than other approaches~\cite{Z2-calculation_Hastings_2010_EPL,QSH_Liu_2018_Phys.Rev.Lett.,Z2-calculation_Mong_2019_Phys.Rev.B}.

% \section{Distinction between QSH and $\mathbb{Z}_2$ TAIs}
\textit{Distinction between QSH and $\mathbb{Z}_2$ TAIs.}---According to the above discussions, the  $\mathcal{T}_f$-symmetric PCs with and without $M_z$ symmetry have different topological classifications, {\it i.e.}, $\mathbb{Z}$ versus $\mathbb{Z}_2$. 
Now we examine this difference via edge transport effects.

We consider a strongly disordered PC whose bulk mobility gap shown in Fig.~\ref{fig-5}(a)  is in the  TAI phase with $C_s=1$ at $\kappa=0$. 
Turning on the spin coupling ($\kappa\neq0$), the gap remains nontrivial with the $\mathbb{Z}_2$ index $\nu=1$. 
The persistence of nontrivial topology is manifested by the gapless transmittance spectrum in Fig.~\ref{fig-5}(b), where the mobility gap (region between the two red dashed curves) is filled up by the gapless edge transport. 
For comparison,  we build a domain wall between this PC ($\beta>0$) and its time reversal copy ($\beta<0$) (see Fig.~\ref{fig-5}(c)). 
At $\kappa=0$, the two domains have opposite QSH-Chern numbers $C_{s} =\pm1$. The nontrivial domain wall index $\Delta C_s=2$ protects gapless interface transport at $\kappa=0$. 
However, the emergence of spin coupling makes the two domains fall into the same $\mathbb{Z}_2$ class of $\nu=1$, and immediately reduces the domain wall charge from $\Delta C_s=2$ to $\Delta\nu=0$. Consequently, 
the gapless transmission at $\kappa=0$ opens a mobility gap as $\kappa\neq0$, 
confirming that the domain wall topology is trivialized by the spin coupling.

% \section{Conclusion}
\textit{Conclusion.}---We proposed a scheme for realizing photonic QSH and $\mathbb{Z}_2$ TAIs supporting gapless helical edge transport in 2D disordered  PCs with $\mathcal{T}M_z\mathcal{D}$ symmetry, where geometric disorder induces the transition from the trivial phase to the topological phases. 
Through observing the gapless to gapped transition of interface transport, we also verified that the absence or presence of the spin coupling changes the topological classification of the TAIs. Our system offers a prototypical platform to study photonic SPT phases with strong disorders.
%The TAIs are characterized by spin Chern numbers for decoupled cases and by a $\mathbb{Z}_2$ index for coupled cases. 
Furthermore, we developed a new approach to retrieve the QSH-Chern and $\mathbb{Z}_2$ indices of the disordered PCs from spin reflection, which is not only remarkably efficient for system without periodicity but also endows the topological invariants with the explicit physical meaning through the quantized spin rotation angles of the reflected waves during a half period variation of the PCs' twisted boundaries. This approach is applicable to any systems with (pseudo-)spins and would also be used to characterizing the $\mathbb{Z}_2$ charged 3D Dirac points~\cite{yang2014classification,guo2017three,cheng2020discovering}. 

\begin{acknowledgements}
\textit{Acknowledgements.}---We thank Prof. Lei Zhang and Dr. Tianshu Jiang for the helpful discussions.
This work is supported by the Research Grants Council of Hong Kong (Grant No. 16307420).
% and by the Croucher Foundation (Grant No. CAS20SC01)
\end{acknowledgements}

%\bibliography{Reference}% Produces the bibliography via BibTeX.


\end{document}

\preprint{APS/123-QED}

\title{
Supplemental materials for ``Photonic $\mathbb{Z}_2$ topological Anderson insulators''
%Topological Anderson quantum spin Hall effect in disordered photonic systems
% The calculation of $\mathbb{Z}_2$ index based on reflection matrix in disordered photonic crystals
}

\author{Xiaohan Cui}
\author{Ruo-Yang Zhang}
\email{ruoyangzhang@ust.hk}
\author{Zhao-Qing Zhang}%
\author{C. T. Chan}%
 \email{phchan@ust.hk}
\affiliation{%
Department of Physics, The Hong Kong University of Science and Technology, Hong Kong, China
}%


\maketitle

\tableofcontents

\section{Difference and topological classifications of QSH insulators and TIs}
\textbf{Terminology and classification.} The notion of quantum spin Hall (QSH) effect has subtly diverse meanings in the literature. In the early stage, the QSH insulator is usually noted as a synonym of 2D fermionic time reversal (FTR)-invariant topological insulator (TI). However, since the 2D FTR-invariant TIs do not guarantee the quantization of spin Hall conductance, it is now accepted that the QSH states protected by both FTR and $s_z$-spin-rotation symmetries and the 2D TIs with only FTR symmetry should be regarded as different symmetry-protected topological (SPT) classes~\cite{wen2017Colloquium,wen2020,shiozaki2014topology}. In this paper, we also follow this definition of QSH systems. Incidentally, the QSH states can also be generalized to the FTR-broken systems only with spin rotation symmetry~\cite{yang2011time}, where the spin-up and spin-down states constitute two independent subsystems characterized by two separate Chern numbers, so the topological phases in these generalized QSH insulators follow a $\mathbb{Z}\oplus\mathbb{Z}$ classification~\cite{shiozaki2014topology,wen2017Colloquium}. In the following table, we list the symmetries and topological classifications of the three subtly different kinds of systems that used to be called QSH systems in different articles. And we will only investigate the first two kinds of systems in this work.
\begin{equation*}\arraycolsep=3pt\def\arraystretch{1.5}
    \begin{array}{c|c|c|c}\hline\hline
             & \text{Fermionic time reversal} & U(1)\ \text{spin rotation} & \text{classification}  \\\hline
         \text{QSH insulator} &  \checkmark    &  \checkmark &  \mathbb{Z} \\\hline
         \text{2D FTR-invariant TI}  &  \checkmark    &  \times     &  \mathbb{Z}_2 \\\hline
         \text{Generalized QSH} &  \times & \checkmark  & \mathbb{Z}\oplus\mathbb{Z} \\\hline\hline
    \end{array}
\end{equation*}

\textbf{Difference between QSH-Chern number and spin-Chern number.} 
In this paper, we adopt the term ``QSH-Chern number'' to refer to the $\mathbb{Z}$ topological invariant of the QSH phases protected by $\mathcal{T}_f$ and $M_z$ symmetries, but avoid the widely used term ``spin-Chern number'' due to the following concern.

The spin-Chern number was first introduced as an integer topological invariant for the purpose of generalizing the spin-Hall conductivity to the spin coupled systems~\cite{sheng2006quantum,prodan2009robustness,prodan2011Disordered,sheng2013Spin,lv2021measurement}. Therefore, although the spin-Chern number is identical to the QSH-Chern number when  spin $s_z$ is conserved, it can be defined in a much broader sense beyond the ``Chern number of a uniformly spin-polarized subsystem''. {\color{Blue}Without spin conservation, the quantization of spin-Chern number requires both the energy gap of Hamiltonian and the spectral gap of the spin projection operator $Ps_zP$ to remain open, and if so, the spin-Chern number is given by the Chern number of a set of states that are isolated from others by both the energy gaps and the spin gaps~\cite{prodan2009robustness,prodan2011Disordered}. This elaborate setting implies that the spin-Chern number and the $\mathbb{Z}_2$ index indeed characterize the topology under \textbf{different definitions of bandgaps}  (it has become an important inspiration from the study of non-Hermitian physics that before talking about topology, the definition of gaps should be clarified in the first instance~\cite{kawabata2019Symmetry}).} As a consequence, they lead to entirely different topological classifications, and only the $\mathbb{Z}_2$ index is relevant to the existence of gapless boundary excitations~\cite{prodan2009robustness,prodan2011Disordered}. Hence, to avoid confusion, we introduce a new term, QSH-Chern number, instead of the spin-Chern number, to indicate the $\mathbb{Z}$-classified QSH phases without spin coupling~\cite{shiozaki2014topology}.

\section{Gauge transformation of $\mathcal{T}M_z\mathcal{D}$-symmetric materials}
\subsection{$\mathcal{T}M_z\mathcal{D}$-symmetric optical media}
We consider generic lossless bianisotropic materials whose constitutive relations are
\begin{equation}
    \mathbf{D} = \tensor{\varepsilon}_r \varepsilon_0 \mathbf{E} + \frac{\tensor{\chi}}{c} \mathbf{H},\quad
    \mathbf{B} = \tensor{\mu}_r \mu_0 \mathbf{H} + \frac{\tensor{\chi}^\dagger}{c} \mathbf{E}
\end{equation}
where $c$ is the speed of light in vacuum, $\varepsilon_0$ and $\mu_0$ are the relative permittivity and permeability of the materials, $\tensor{\varepsilon}_r$ and $\tensor{\mu}_r$ are the relative permittivity and permeability of the materials, and $\tensor{\chi}$ represents the magnetoelectric coupling.
When the bianisotropic materials are homogeneous along the $z$ direction, the 2D macroscopic Maxwell's equations at frequency $\omega$ can be expressed as:
\begin{equation}
\arraycolsep=4pt\def\arraystretch{1.3}
    \underbrace{\begin{pmatrix}
0 & i\nabla_T\times\\
-i\nabla_T\times & 0
\end{pmatrix}}_{\displaystyle=\hat{\mathcal{N}}}
\underbrace{\begin{pmatrix}
 \mathbf{E} \\
 \eta_0 \mathbf{H}
\end{pmatrix}}_{\displaystyle=\Psi}
=\frac{\omega}{c} \underbrace{\begin{pmatrix}
\tensor {\varepsilon}_r(\mathbf{r}) & \tensor{\chi}(\mathbf{r})\\
\tensor{\chi}(\mathbf{r})^\dagger & \tensor{\mu}_r(\mathbf{r})
\end{pmatrix}}_{\displaystyle=\hat{\mathcal{M}}(\mathbf{r})}
\begin{pmatrix}
 \mathbf{E} \\ \eta_0 \mathbf{H}
\end{pmatrix},
\label{Maxwell eq}
\end{equation}
where 
% $c$ is the speed of light, 
$ \nabla_T = \partial_x \hat x + \partial_y \hat y$, $\eta_0 = \sqrt{\mu_0 / \epsilon_0}$ is the vacumm impedance.
% $\tensor{\varepsilon}_r$ and $\tensor{\mu}_r$ are the relative permittivity and permeability of the materials, and $\tensor{\chi}$ represents the coupling between the electric and magnetic fields.  
For this six-component wave function $\Psi$, the operators of time reversal, mirror-$z$, and electromagnetic duality, are given by, respectively, $\mathcal{T}=\sigma_z\otimes\tensor{I}_{3\times3}\mathcal{K}$, $M_z=\sigma_z\otimes{m}_z (z\rightarrow-z)$ with $m_z=\mathrm{diag}(1,1,-1)$, and  $\mathcal{D}=-i\sigma_y\otimes\tensor{I}_{3\times3}$. Therefore, the combined symmetry operator reads $\mathcal{T}_f=\mathcal{T}M_z\mathcal{D}=-i\sigma_y\otimes m_z\mathcal{K} (z\rightarrow -z)$. Since the curl tensor $\hat{\mathcal{N}}$ is invariant under the $\mathcal{T}_f$ operation, a photonic system respects $\mathcal{T}_f$ symmetry as long as the constitutive tensor satisfies
\begin{equation}
\begin{split}
    (\sigma_y\otimes m_z) \hat{\mathcal{M}}(m_z\mathbf{r})^*(\sigma_y\otimes m_z)&=\hat{\mathcal{M}}(\mathbf{r}).
\end{split}\label{Tf symmetry}
\end{equation}

For 2D PCs that are homogeneous along the $z$ direction, $\hat{\mathcal{M}}(m_z\mathbf{r})=\hat{\mathcal{M}}(\mathbf{r})$, hence $\mathcal{T}_f$ can be regarded as a 2D local operator, and Eq.~\eqref{Tf symmetry} leads to
\begin{equation}
 % keep the change local
\arraycolsep=4pt\def\arraystretch{1.5}\begin{pmatrix}
     m_z\tensor {\mu}(\mathbf{r})^*m_z & -m_z\tensor{\chi}(\mathbf{r})^\intercal m_z\\
-m_z\tensor{\chi}(\mathbf{r})^* m_z &  m_z\tensor{\varepsilon}_r(\mathbf{r})^* m_z
    \end{pmatrix}
    =\begin{pmatrix}
\tensor {\varepsilon}_r(\mathbf{r}) & \tensor{\chi}(\mathbf{r})\\
\tensor{\chi}(\mathbf{r})^\dagger & \tensor{\mu}_r(\mathbf{r})
\end{pmatrix}\quad\Rightarrow\quad
\left\{\begin{aligned}
m_z\tensor{\mu}_r(\mathbf{r})^*m_z=\tensor{\varepsilon}_r(\mathbf{r}),\\
m_z\tensor{\chi}(\mathbf{r})^\intercal m_z=-\tensor{\chi}(\mathbf{r}),
\end{aligned}\right.
\end{equation}
which restricts the form of a self-$\mathcal{T}_f$-symmetric medium as~\cite{PTI-theory_Silveirinha_2017_Phys.Rev.B}
\begin{equation}\arraycolsep=4pt\def\arraystretch{1.3}
    \tensor{\varepsilon}_r=\begin{pmatrix}
     \tensor{\varepsilon}_T^R+i\beta \tensor{\epsilon}  &\mathbf{g} \\
     \mathbf{g}^\dagger & \varepsilon_z
    \end{pmatrix},\quad
    \tensor{\mu}_r=\begin{pmatrix}
     \tensor{\varepsilon}_T^R-i\beta \tensor{\epsilon} &-\mathbf{g}^* \\
     -\mathbf{g}^\intercal & \varepsilon_z
    \end{pmatrix},\quad
    \tensor{\chi}=\begin{pmatrix}
    h_z\tensor{\epsilon} & \mathbf{t}\\
     \mathbf{t}^\intercal &{0}
    \end{pmatrix} ,
\end{equation}
where $\tensor{\varepsilon}^R_T$, $\varepsilon_z$, $\beta$ take real values, while $\mathbf{g}=\mathbf{g}^R+i\mathbf{g}^I$, $\mathbf{t}=\mathbf{t}^R+i\mathbf{t}^I$, $h_z=h_z^R+ih_z^I$ are generally complex, and $\tensor{\epsilon}=\begin{pmatrix} 0&1\\-1&0\end{pmatrix}$ denotes the 2D Levi-Civita symbol.

\subsection{Correspondence between $\mathcal{T}_f$-symmetric gyrotropic media and bianisotropic media}
\begin{figure}[b!]
\includegraphics[width=0.99\textwidth]{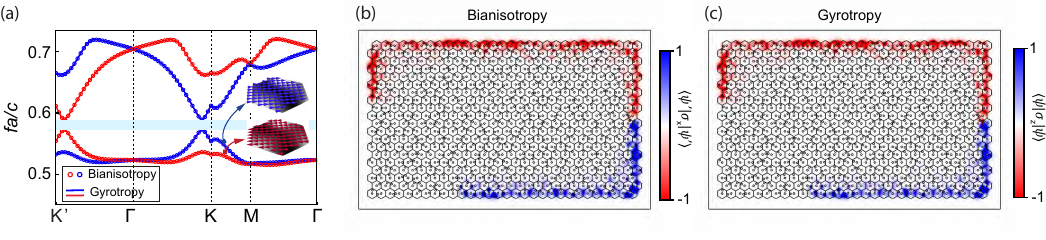}
\vspace{-7pt}
\caption{\label{fig-s0}
 (a) The identical band structures of two $\mathcal{T}_f$-symmetric PCs consisted of bianisotropic materials $\mathcal{M}_b$ (circles) and gyrotropic materials $\mathcal{M}_g$ (lines), respectively, satisfying $\widetilde{U}_t\widetilde{\mathcal{M}}_g(\mathbf{r})\widetilde{U}_t^\dagger=\widetilde{\mathcal{M}}_b(\mathbf{r})$. Insets: distributions of pseudo-spins $\vec s$ (arrows) and intensity $\langle \psi | \psi \rangle$ (grayscale colormap) of two Bloch states of the bianisotropic PC, where the pseudo-spins are all along the $s_x$ axis. (b,c) Distributions of the pseudo-spin densities of (b) $s_x$-projected  $\langle \psi' | \sigma_x | \psi' \rangle = \eta_0 E_z'^* H_z'- \eta_0 H_z'^*E_z'$ and (c) $s_z$-projected $\langle \psi | \sigma_z | \psi \rangle = |E_z|^2 -|\eta_0 H_z|^2 $ of the helical edge states in (b) the bianisotropic PC and in (c) the gyrotropic PC, respectively.
 }
\end{figure}

The $\mathcal{T}_f$-symmetric PCs discussed in the main text are nonreciprocal,
% \begin{equation}
%     \mathcal{M}_{g}=\begin{pmatrix}[c|c]
%      \begin{matrix}\tensor{\varepsilon}^R_T +i\beta \tensor{\epsilon}& \mathbf{0}\\
%      \mathbf{0} & \varepsilon_z \end{matrix} & {\mathlarger{\mathlarger{\mathlarger{0}}}}
%      \\\hline {\mathlarger{\mathlarger{\mathlarger{0}}}} &
%      \begin{matrix} \tensor{\varepsilon}^R_T-i\beta \tensor{\epsilon} & \mathbf{0}\\
%      \mathbf{0} & \varepsilon_z
%      \end{matrix}
%     \end{pmatrix}
% \end{equation}
whose nontrivial topology originates from the gyrotropic terms $\tensor{\varepsilon}_g=-\tensor{\mu}_g=i\beta\hat{z}\times\tensor{I}$ in the permittivity and permeability, which breaks the time reversal symmetry in the pseudo-spin-up (TM) and -down (TE) subspaces in opposite manners. However, the strict restriction between the nonreciprocal gyrotropic terms in $\tensor{\varepsilon}$ and $\tensor{\mu}$ might raise doubt about whether our scheme can be implemented by ordinary materials. Here, we will show that {\color{Blue} this gyrotropic material (labeled by its constitutive tensor $\mathcal{M}_g$) can be mapped to a reciprocal bianisotropic material ($\mathcal{M}_b$) respecting $\mathcal{T}_f$-symmetry by a $SU(2)$ gauge transformation,} 
(see below for the detailed derivation) 
% The constitutive tensor of the reciprocal bianisotropic material is given by
% \begin{equation}
%     %\begin{tikzcd}
%     \color{Blue}
%     \mathcal{M}_{g}=\begin{pmatrix}[c|c]
%      \begin{matrix}\tensor{\varepsilon}^R_T +i\beta \tensor{\epsilon}& {0}\\
%      {0} & \varepsilon_z \end{matrix} & {\mathlarger{\mathlarger{\mathlarger{0}}}}
%      \\\hline {\mathlarger{\mathlarger{\mathlarger{0}}}} &
%      \begin{matrix} \tensor{\varepsilon}^R_T-i\beta \tensor{\epsilon} & {0}\\
%      {0} & \varepsilon_z
%      \end{matrix}
%     \end{pmatrix}
%     \ \stackrel{\displaystyle U_s}{\pspicture[shift=3pt](2,0)\psline[linecolor=Blue,linewidth=1pt,arrowinset=0,arrowscale=1.5]{<->}(2,0)\endpspicture}\ 
%     \mathcal{M}_b=\begin{pmatrix}[cc|cc]
%      \tensor{\varepsilon}^R_T& \mathbf{0} & h_z \tensor{\epsilon}  & \mathbf{0}\\
%      \mathbf{0} & \varepsilon_z & \mathbf{0}& 0
%      \\\hline -h_z^*\tensor{\epsilon} &  \mathbf{t}^* & \tensor{\varepsilon}^R_T-i\beta \tensor{\epsilon} & -\mathbf{g}^*
%      \\
%      \mathbf{t}^\dagger &  0 &-\mathbf{g}^\intercal & \varepsilon_z
%     \end{pmatrix}
%     %\end{tikzcd}
% \end{equation}
% indeed the first proposed optical material that can  
\begin{equation}\label{SU(2) correlation}
\begin{tikzpicture}
[baseline=(current  bounding  box.center),
decision/.style={diamond,draw=blue, thick, fill=blue!20,
text width=4.5em,align=flush center,
inner sep=1pt},
block/.style ={rectangle, draw=Blue, thick, fill=blue!0,
text width=9em,align=center, rounded corners,
minimum height=3em},
line/.style ={draw=Blue, line width=2pt, >=latex'},
cloud/.style ={draw=red, thick, ellipse, fill=red!20,
minimum height=2em}]

\matrix [column sep=23mm,row sep=7mm]
{
% row 3
%&\node [block,text width=13em] (aaa) {Reciprocal bianisotropy\\[2pt] $\tensor{\chi}=\begin{pmatrix}0 & ih_z^I & 0\\-ih_z^I & 0 & 0\\0 & 0 & 0\end{pmatrix}$};&
 \node [block,text width=14em] (gyrotropy){{\bf Gyrotropy}\\[4pt]$\tensor{\varepsilon}_g=-\tensor{\mu}_g=\begin{pmatrix}0 & i\beta & 0\\-i\beta & 0 & 0\\0 & 0 & 0\end{pmatrix}$}; &
  &\node [block,text width=14em] (bianisotropy) {{\bf Reciprocal bianisotropy}\\[4pt]$\tensor{\chi}=\begin{pmatrix}0 & i\beta & 0\\-i\beta & 0 & 0\\0 & 0 & 0\end{pmatrix}$};\\
};
\begin{scope}[every path/.style=line]
\path[triangle 45-triangle 45] (gyrotropy) -- node [pos=0.56,text width=3.5cm,text height=0.5cm,fill=White,text=Blue,above=0.5em ] {\large\bf $U_t=\exp[{- i\frac{\pi}{4} \sigma_y}]$}(bianisotropy);
\end{scope}
\end{tikzpicture}
\end{equation}
{\color{Blue} such that the 2D Maxwell's equations in these two media can be exactly transformed to one another by the $SU(2)$ rotation $U_t=\exp[{- i\frac{\pi}{4} \sigma_y}]$ on the two-component spinors, $\psi=(E_z,\eta_0 H_z)^\intercal\rightarrow \psi'=(E_z',\eta_0 H_z' )=U_t\psi = \frac{1}{\sqrt{2}} (E_z-\eta_0 H_z,E_z+\eta_0 H_z)^\intercal$. In consequence, for two $\mathcal{T}_f$-symmetric PCs comprised of $\mathcal{M}_g$ and $\mathcal{M}_b$, respectively, and correlated by the transformation~\eqref{SU(2) correlation}, the band structures in these two PCs are exactly identical (see Fig.~\ref{fig-s0}(a)) and the solutions of Maxwell's equations in them are one-to-one correspondent via the $U_t$ transformation.} 

Therefore, the $\mathcal{T}_f$-symmetric topological Anderson insulators (TAIs) proposed in the main text can also be realized using the reciprocal bianisotropic materials. As shown in Figs.~\ref{fig-s0}(b) and (c), we numerically simulated the helical transportation of pseudo-spin polarized edge states in the bianisotropic TAI, where the field distribution is exactly the same as that in the gyrotropic TAI except that the pseudo-spins of the edge states are changed from $s_z$ polarization ($\psi\propto (1,0)$ and $(0,1)$) to $s_x$ polarization ($\psi'\propto (1,\pm1)$). Indeed, this type of reciprocal bianisotropic metamaterials is the first proposed material that can realize photonic topological insulators (PTIs)~\cite{PTI-firsttheory_Shvets_2013_NatureMater}, and has 
become a well-developed experimental platform for observing PTI-related effects~\cite{PTI-experiment_Chan_2015_NatCommun,PTI-experiment_Shvets_2015_Phys.Rev.Lett.,cheng2016Robust,gao2018Topologically}.
Hence, our theoretical framework, which is valid for all $\mathcal{T}M_z\mathcal{D}$-symmetric systems, is experimentally achievable. 

More broadly, we can prove that the 2D Maxwell's equations in general $\mathcal{T}M_z\mathcal{D}$-symmetric media are $SU(2)$ gauge covariant. And all proposed $\mathcal{T}M_z\mathcal{D}$-symmetric materials for PTIs can be classified into three classes; the materials in each class can be transformed to each other by gauge transformations, and hence the physical effects in them are essentially equivalent. The comprehensive discussions on this issue will be posted as a separate work~\cite{zhang2022gauge}.\\

\textbf{Derivation of the $SU(2)$ correspondence.}\\
First, we explicitly express the constitutive tensors in the $\mathcal{T}_f$-symmetric gyrotropic medium and bianisotropic medium, respectively,
\begin{equation}
    \mathcal{M}_{g}=\begin{pmatrix}[c|c]
     \begin{matrix}\tensor{\varepsilon}^R_T +i\beta \tensor{\epsilon}& {0}\\
     {0} & \varepsilon_z \end{matrix} & {\mathlarger{\mathlarger{\mathlarger{0}}}}
     \\\hline {\mathlarger{\mathlarger{\mathlarger{0}}}} &
     \begin{matrix} \tensor{\varepsilon}^R_T-i\beta \tensor{\epsilon} & {0}\\
     {0} & \varepsilon_z
     \end{matrix}
    \end{pmatrix},\quad
    \mathcal{M}_b=\begin{pmatrix}[cc|cc]
     \tensor{\varepsilon}^R_T& \mathbf{0} & i\beta \tensor{\epsilon}  & {0}\\
     {0} & \varepsilon_z & {0}& 0
     \\\hline i\beta\tensor{\epsilon} &  {0} & \tensor{\varepsilon}^R_T & {0}
     \\
     {0} &  0 & {0} & \varepsilon_z
    \end{pmatrix}.
\end{equation}
Then, we transform the wave function into a new basis for convenience~\cite{Tf-symmetry_Chan_2019_NatCommun}:
\begin{equation}\label{new basis}
\arraycolsep=4pt\def\arraystretch{1.3}
    \widetilde{\Psi}=\begin{pmatrix}
     \mathbf{E}_T\\ \eta_0\mathbf{H}_T\\\hline {E}_z \\ \eta_0{H}_z
    \end{pmatrix}=U_0\Psi=\begin{pmatrix}[cc|cc]
     1 & 0 & 0 & 0\\
     0 & 0 & 1 & 0\\\hline
     0 & 1 & 0 & 0\\
     0 & 0 & 0 & 1
    \end{pmatrix}\begin{pmatrix}
     \mathbf{E}_T\\ {E}_z\\\hline \eta_0\mathbf{H}_T \\ \eta_0{H}_z
    \end{pmatrix},
\end{equation}
where the new wave function is partitioned into four blocks, and all the following deviations in this section are in this 4-dimensional block form. Performing this transformation to the curl tensor and the constitutive tensor, we have
\begin{equation}
    \widetilde{\mathcal{N}}=U_0\mathcal{N}U_0
    =U_0\begin{pmatrix}[cc|cc]
     0 & 0 & 0 & i\tensor{\epsilon}\cdot\nabla_T\\
     0 & 0 & (i\tensor{\epsilon}\cdot\nabla_T)^\intercal & 0\\\hline
     0 & -i\tensor{\epsilon}\cdot\nabla_T & 0 & 0\\
     -(i\tensor{\epsilon}\cdot\nabla_T)^\intercal & 0 & 0 & 0
    \end{pmatrix}U_0
    =\begingroup 
\arraycolsep=4pt\def\arraystretch{2}
\begin{pmatrix}[c|c]
     \mathbf{0}  &- \sigma_y (\tensor{\epsilon}\cdot\nabla_T)\\\hline
     -\sigma_y(\tensor{\epsilon}\cdot \nabla_T)^\intercal & 0 \\
    \end{pmatrix}\endgroup,
\end{equation}
\begin{equation}
    \widetilde{\mathcal{M}}_g=U_0\mathcal{M}_gU_0=
    \begingroup % keep the change local
\arraycolsep=4pt\def\arraystretch{2}
\begin{pmatrix}[c|c]
     \sigma_0\tensor{\varepsilon}^R_T  +{\color{red}\sigma_z}(i\beta\tensor{\epsilon})&  \mathbf{0}\\\hline
    \mathbf{0} & \sigma_0\varepsilon_z \\
    \end{pmatrix}
    \endgroup,\quad
        \widetilde{\mathcal{M}}_b=U_0\mathcal{M}_bU_0=
    \begingroup % keep the change local
\arraycolsep=4pt\def\arraystretch{2}
\begin{pmatrix}[c|c]
     \sigma_0\tensor{\varepsilon}^R_T  +{\color{red}\sigma_x}(i\beta\tensor{\epsilon})&  \mathbf{0}\\\hline
    \mathbf{0} & \sigma_0\varepsilon_z \\
    \end{pmatrix}.\label{constitutive of Mb}
    \endgroup
\end{equation}
The 2D Maxwell's equations in the two kinds of materials can be expressed in the new basis as $\widetilde{\mathcal{N}}\widetilde{\Psi}=\frac{\omega}{c}\widetilde{\mathcal{M}}_g\widetilde{\Psi}$ and $\widetilde{\mathcal{N}}\widetilde{\Psi}=\frac{\omega}{c}\widetilde{\mathcal{M}}_b\widetilde{\Psi}$, respectively.
% In addition, the $\mathcal{T}_f=\mathcal{T}M_z\mathcal{D}$ operator takes the matrix form in the new basis
% \begin{equation}\arraycolsep=4pt\def\arraystretch{1.3}
%     \widetilde{\mathcal{T}}_f=U_0\mathcal{T}_f U_0=\begin{pmatrix}[c|c]
%          -i\sigma_y & 0 \\\hline
%          0 & i\sigma_y
%     \end{pmatrix}\mathcal{K}.
% \end{equation}

For a global $2\times2$ $SU(2)$ gauge transform $U$ acting on the two-component spinor function $\psi=(E_z,\eta_0 H_z)^\intercal$, we may introduce the corresponding $4\times4$-block $SU(2)$ transform for the full wave function $\widetilde{\Psi}$~\cite{Tf-symmetry_Chan_2019_NatCommun}
\begin{equation}
\arraycolsep=4pt\def\arraystretch{1.3}
    \widetilde{U}=\begin{pmatrix}[c|c]
        U^* & 0 \\\hline
        0 & U
    \end{pmatrix}=\begin{pmatrix}[c|c]
        \sigma_yU\sigma_y & 0 \\\hline
        0 & U
    \end{pmatrix}.\label{SU(2) transform}
\end{equation}
It is easy to check that the curl tensor is invariant under any such $SU(2)$ transformation
\begin{equation}\arraycolsep=4pt\def\arraystretch{1.3}
    \widetilde{U}\widetilde{\mathcal{N}}\widetilde{U}^\dagger=
    \begin{pmatrix}[c|c]
        0 & -(\sigma_y U\sigma_y)\sigma_y U^\dagger(\tensor{\epsilon}\cdot\nabla_T)\\\hline
        -U\sigma_y(\sigma_yU^\dagger\sigma_y)(\tensor{\epsilon}\cdot\nabla_T)^\intercal & 0
    \end{pmatrix}=\widetilde{\mathcal{N}}.
\end{equation}
More importantly, since
\begin{equation}\arraycolsep=4pt\def\arraystretch{1.5}
    \widetilde{U}\widetilde{\mathcal{M}}_g\widetilde{U}^\dagger=
    \begin{pmatrix}[c|c]
     \sigma_0\tensor{\varepsilon}^R_T  +{\color{red}(U^*{\sigma_z}U^\intercal)}(i\beta\tensor{\epsilon})&  \mathbf{0}\\\hline
    \mathbf{0} & \sigma_0\varepsilon_z \\
    \end{pmatrix},
\end{equation}
taking $U=U_t=\exp[{-i\frac{\pi}{4} \sigma_y}]$, we have $U_t^*\sigma_z U_t^\intercal=\sigma_x$ and hence $\widetilde{U}_t\widetilde{\mathcal{M}}_g(\mathbf{r})\widetilde{U}_t^\dagger=\widetilde{\mathcal{M}}_b(\mathbf{r})$.
Thus, we have proved that the 2D Maxwell's equations in the two materials can be transformed to each other via the $SU(2)$ transformation $\widetilde{U}_t$. For an arbitrary solution $\widetilde{\Psi}_g$ of the 2D EM fields in material $\widetilde{\mathcal{M}}_g(\mathbf{r})$, there is a corrersponding wave solution in $\widetilde{\mathcal{M}}_b(\mathbf{r})$ given by $\widetilde{\Psi}_b(\mathbf{r})=\widetilde{U}_t\widetilde{\Psi}_g(\mathbf{r})$.

\subsection{$M_z$ symmetry as a $U(1)$ spin rotation symmetry}
Under the new basis $\widetilde{\Psi}$ (Eq.~\eqref{new basis}), the mirror-$z$ operator can be written as the following $4\times4$-block form
\begin{equation}
    \widetilde{M}_z=U_0M_zU_0=\begin{pmatrix}[c|c]
        \sigma_z & 0 \\\hline
        0 & -\sigma_z
    \end{pmatrix}.
\end{equation}
%where $\tensor{I}_T$ denotes 2D unit tensor. 
Comparing with Eq.~\eqref{SU(2) transform}, we find that the $\widetilde{M}_z$ operator performs as the generator of the following one-parameter 4D $SU(2)$ transformation:
\begin{equation}\arraycolsep=4pt\def\arraystretch{1.3}
    \widetilde{U}_s(\vartheta)=\exp\left(i\frac{\vartheta}{2}\widetilde{M}_z\right)=
    \begin{pmatrix}[c|c]
        U_s(\vartheta)^* & 0 \\\hline
        0 & U_s(\vartheta)
    \end{pmatrix}=\begin{pmatrix}[c|c]
        \exp\left(i\frac{\vartheta}{2}{\sigma_z}\right) & 0 \\\hline
        0 & \exp\left(-i\frac{\vartheta}{2}{\sigma_z}\right)
    \end{pmatrix}
\end{equation}
with $U_s(\vartheta)=\exp(-i\frac{\vartheta}{2}{\sigma_z})$ denoting the corresponding $SU(2)$ transform for the two-component spinor $\psi$. Since the constitutive tensor $\widetilde{\mathcal{M}}_g$ of the gyrotropic material is invariant about these rotations, 
\begin{equation}
    \widetilde{U}_s(\vartheta)\widetilde{\mathcal{M}}_g\widetilde{U}_s(\vartheta)^\dagger=\widetilde{\mathcal{M}}_g,
\end{equation}
$\widetilde{U}_s(\vartheta)$ is also a symmetry of the 2D Maxwell's equations in $\widetilde{\mathcal{M}}_g$. Since the one-parameter group of $\widetilde{U}_s(\vartheta)$ is isomorphic to $U(1)$, it can be referred to as a $U(1)$ spin rotation symmetry of the Maxwell's equations.  Physically, the $U(1)$ spin rotation symmetry ensures the conservation of the $s_z$ spin component for waves propagating in such systems. 

For the bianisotropic media $\widetilde{\mathcal{M}}_b$ in Eq.~\eqref{constitutive of Mb}, the counterparts of  $\widetilde{M}_z$ can be obtained using $\widetilde{U}_t$
\begin{equation}
    \widetilde{U}_t\widetilde{M}_z\widetilde{U}_t^\dagger=
    \begin{pmatrix}[c|c]
        \sigma_x & 0 \\\hline
        0 & -\sigma_x
    \end{pmatrix}
    =-\widetilde{U}_0 (M_z\mathcal{D})\widetilde{U}_0.
\end{equation}
Therefore, $M_z\mathcal{D}$ serves as the generator of the $U(1)$ spin rotation symmetry  about $s_x$ axis, and accordingly the $s_x$ spin current will be conserved.

\subsection{Analogy between photonic wave equation and electronic Schr\"{o}dinger equation}
In this section, we offer an intuitive understanding of the mechanism of the photonic  QSH and $\mathbb{Z}_2$ topological insulator by mapping the wave equation for the effective 2-component  spinor $\psi=(E_z,\eta_0 H_z)^\intercal$ to the Schr\"{o}dinger equation in a spin-dependent magnetic field with spin-orbit coupling.

First, We recast the Maxwell's equations $\widetilde{\mathcal{N}}\widetilde{\Psi}=\frac{\omega}{c}\widetilde{\mathcal{M}}_g\widetilde{\Psi}$ in the $\mathcal{T}M_z\mathcal{D}$ material:
\begin{gather}
    -\sigma_y\qty(\tensor{\epsilon}\cdot\nabla_T)\begin{pmatrix} E_z\\\eta_0H_z\end{pmatrix}
    =k_0\qty(\sigma_0\tensor{\varepsilon}^R_T+\sigma_z(i\beta\tensor{\epsilon}))\cdot\begin{pmatrix} \mathbf{E}_T\\\eta_0\mathbf{H}_T\end{pmatrix}+k_0\sigma_z\mathbf{g}\begin{pmatrix} E_z\\\eta_0H_z\end{pmatrix},\\
    -\sigma_y\qty(\tensor{\epsilon}\cdot\nabla_T)^\intercal\cdot\begin{pmatrix} \mathbf{E}_T\\\eta_0\mathbf{H}_T\end{pmatrix}
    =k_0\sigma_0\varepsilon_z\begin{pmatrix} E_z\\\eta_0H_z\end{pmatrix}+k_0\sigma_z\mathbf{g}^\intercal\cdot\begin{pmatrix} \mathbf{E}_T\\\eta_0\mathbf{H}_T\end{pmatrix},
\end{gather}
where $k_0=\omega/c$, $\tensor{\varepsilon}^R_T=\tensor{I}_{2\times2}$ and $\mathbf{g}=(\kappa,\kappa)^\intercal\in\mathbb{R}$ (for simplicity, $\mathbf{g}$ is assumed to be independent of the coordinates in the following derivation). Eliminating $(\mathbf{E}_T,\eta_0\mathbf{H}_T)^\intercal$ in the above two equations, we obtain
\begin{equation}
\begin{gathered}
    \qty[\sigma_y\qty(\tensor{\epsilon}\cdot\nabla_T)^\intercal+k_0\sigma_z\mathbf{g}]\cdot\qty[\sigma_0\tensor{\varepsilon}^R_T+\sigma_z(i\beta\tensor{\epsilon})]^{-1}\cdot\qty[\sigma_y\qty(\tensor{\epsilon}\cdot\nabla_T)+k_0\sigma_z\mathbf{g}]\psi=k_0^2\sigma_0\varepsilon_z\psi, \\
    \Rightarrow\quad\qty[\nabla_T^\intercal+ik_0\sigma_x\qty(\mathbf{g}^\intercal\cdot\tensor{\epsilon})]\cdot\qty(\sigma_y\tensor{\epsilon})\cdot\frac{1}{\tilde{\varepsilon}_T}\qty[\sigma_0\tensor{I}_{2\times2}-\sigma_z\qty(i\beta\tensor{\epsilon})]\cdot\qty(\sigma_y\tensor{\epsilon})\cdot\qty[\nabla_T+ik_0\sigma_x\qty(\tensor{\epsilon}^\intercal\cdot\mathbf{g})]\psi=k_0^2\varepsilon_z\psi,\\
    \Rightarrow\quad\qty[\qty(\nabla_T+i\sigma_x\mathbf{q})\cdot\frac{1}{\tilde{\varepsilon}_T}\qty[\sigma_0\tensor{I}_{2\times2}+\sigma_z\qty(i\beta\tensor{\epsilon})]\cdot\qty(\nabla_T+i\sigma_x\mathbf{q})+k_0^2\varepsilon_z]\psi=0,\\
    % \Rightarrow\quad \frac{1}{\tilde{\varepsilon}_T}\Big[\nabla_T^2+\nabla_T\ln(\tilde{\varepsilon}_T)+i\sigma_z\mathcal{A}\cdot\nabla_T+i\sigma_y\beta(\mathbf{g}\cdot\nabla_T)+i\sigma_x\mathbf{q}\cdot\qty(\nabla_T+i\sigma_z\mathcal{A}) +i\sigma_x\qty(\nabla_T\ln\tilde{\varepsilon}_T)\cdot\mathbf{q}-|\mathbf{q}|^2+k_0^2\varepsilon_z \Big]\psi=0
    \end{gathered}
\end{equation}
where $\tilde{\varepsilon}_T=\det\tensor{\varepsilon}_T=(1-\beta^2)$, and $\mathbf{q}=k_0\hat{\mathbf{z}}\times\mathbf{g}$. Using a scaled spinor function $\tilde{\psi}=\psi/\sqrt{\tilde{\varepsilon}_T}$, the wave equation can be rewritten as
\begin{equation}
    \begin{gathered}
    \qty[\qty(\nabla_T+i\sigma_x\mathbf{q})\cdot\qty[\sigma_0\tensor{I}_{2\times2}+\sigma_z\qty(i\beta\tensor{\epsilon})]\cdot\qty(\sigma_0\nabla_T+i\sigma_x\mathbf{q})+k_0^2\tilde{\varepsilon}_T\varepsilon_z-\frac{1}{4}|\nabla_T\ln\tilde{\varepsilon}_T|^2+\frac{1}{2}\nabla_T^2\ln\tilde{\varepsilon}_T]\tilde{\psi}=0,\\
    \Rightarrow\quad \Big[\nabla_T^2+i\sigma_z(\nabla{\beta}\times\hat{\mathbf{z}})\cdot\nabla_T+i\sigma_x\mathbf{q}\cdot\nabla_T +k_0^2\tilde{\varepsilon}_T\varepsilon_z-\frac{1}{4}|\nabla_T\ln\tilde{\varepsilon}_T|^2+\frac{1}{2}\nabla_T^2\ln\tilde{\varepsilon}_T +\mathcal{O}(\beta^2,\mathbf{q}^2,\beta\mathbf{q})\Big]\tilde\psi=0,
    \end{gathered}
\end{equation}
where we assume $\beta$ and $\kappa$ are small enough and only take the first-order approximation for simplicity. Then the wave equation for $\tilde{\psi}$ can be expressed as a Schr\"{o}dinger-like form
\begin{equation}
    \Big[\Big(\hat{\mathbf{p}}+{\color{red}\underbrace{\sigma_z\frac{1}{2}(\nabla{\beta}\times\hat{\mathbf{z}})}_{\substack{\text{spin-dependent}\\\text{vector potential}}}} \Big)^2
    +{\color{Blue}\underbrace{\sigma_x\mathbf{q}\cdot\hat{\mathbf{p}}}_{\substack{\text{SOC induced by}\\ M_z\ \text{breaking}}}}+\underbrace{\qty(\frac{1}{4}|\nabla_T\ln\tilde{\varepsilon}_T|^2-\frac{1}{2}\nabla_T^2\ln\tilde{\varepsilon}_T-k_0^2\tilde{\varepsilon}_T\varepsilon_z)}_{\displaystyle V(\mathbf{r},\omega)} +\mathcal{O}(\beta^2,\mathbf{q}^2,\beta\mathbf{q})\Big]\tilde\psi=0,
\end{equation}
with the canonical momentum operator $\hat{\mathbf{p}}=-i\nabla$. Here, $\color{red}\mathcal{A}=\sigma_z\mathcal{A}=\sigma_z\frac{1}{2}(\nabla \beta\times\hat{\mathbf{z}})$ is an effective spin-dependent vector potential induced by the gyrotropic term $\beta$ in the permittivity and permeability tensors, 
which gives rise to an effective spin-dependent magnetic field $\mathcal{B}_{\uparrow/\downarrow}=\pm\nabla\times\mathcal{A}$ and breaks the time reversal symmetry with opposite manners in the pseudo-spin up and down sectors, hence leading to the photonic QSH effect. On the other hand, the real off-diagonal term
$\mathbf{g}=(\kappa,\kappa)^\intercal$ in $\tensor{\varepsilon}$ and $\tensor{\mu}$ introduces the spin-orbit coupling (SOC) like term, $\color{Blue}H_{\mathrm{SOC}}=\sigma_x\mathbf{q}\cdot\hat{\mathbf{p}}$. This SOC-like term breaks the $z$-mirror symmetry (i.e., $\sigma_z H_\mathrm{SOC} \sigma_z\neq H_\mathrm{SOC}$) and hence mixes the spin-up and down sectors. Nonetheless, the $\mathcal{T}_f=\mathcal{T}M_z\mathcal{D}$ symmetry (which takes the form $\tilde{\mathcal{T}}_f=\hat{\tau}\mathcal{K}=-i\sigma_y\mathcal{K}$ for $\tilde{\psi}$) is preserved, $\hat{\tau} H_\mathrm{SOC}^*\hat{\tau}^\dagger=H_\mathrm{SOC}$, so the $\mathcal{T}_f$-symmetry protected $\mathbb{Z}_2$ topological phases remain well-defined.

It is worth noting that the 2-by-2 wave equation cannot be expressed as a linear eigenvalue problem since $q(\omega)$ and $V(r,\omega)$ both implicitly depend on the frequency. As a result, the analogy between the photonic wave equations and the Schr\"{o}dinger equation can only be understood in an intuitive manner, the rigorous correspondence between the electronic and photonic QSH and $\mathbb{Z}_2$ topological insulators is rooted in the symmetry.

\section{Topology of ordered $\mathcal{T}_f$-symmetric photonic crystals}

\subsection{The Wilson loop of ordered photonic crystals}

\begin{figure}[h!]
\includegraphics[width=0.7\textwidth]{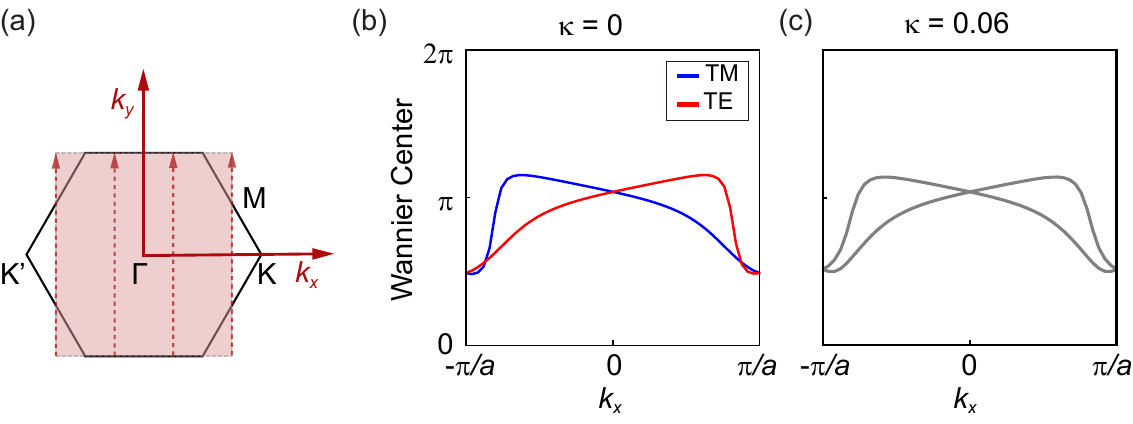}
\caption{\label{fig-s1}
 (a) The deformed BZ (pink shadow) for calculating the hybird Wannier centers along the dashed lines of the hexagonal lattice. 
The Wannier centers of the two bands below the concerned gap for the (b) spin-decoupled and (c) spin-coupled PCs in Fig. 1(a) of the main text. 
 }
\end{figure}

For an ordered photonic crystal, the QSH-Chern number and $\mathbb{Z}_2$ index of Bloch bands are associated with the evolution of the hybrid Wannier centers along closed loops in the 2D Brillouin zone (BZ)~\cite{yu2011Equivalent}. 
As shown in Fig.~\ref{fig-s1}(a), we discretize the each loop along $k_x=\mathrm{const.}$ (dashed lines) into $N$ points %with each point denoted as 
at $\mathbf{k}_j=(k_x, k_{yj}=-b/2 +j b/(N-1))$ with $b=4\pi/(\sqrt{3}a)$. For a group of bands that are isolated from other bands or can be decoupled from others by symmetry, the hybrid Wannier centers of these bands at $k_x$ are characterized by the arguments of the eigenvalues of the Wilson loop matrix whose element is~\cite{Wilsonloop_Garcia-Etxarri_2020_Adv.QuantumTechnol.}
\begin{equation}\label{Wmn}
W_{mn}\left(k_{x}\right)=\lim _{N \rightarrow \infty} \prod_{j=0}^{N} M_{mn}^{(j, j+1)}(k_x),
\end{equation}
where $M_{mn}^{(j, j+1)}(k_x) = \langle \mathcal L_m(\mathbf{k}_j)|\mathcal R_n(\mathbf{k}_{j+1}) \rangle $ represents the inner product of the left and right eigenstates on bands $m$ and $n$, respectively. Here, the effective Hamiltonian of the Maxwell's equation~\eqref{Maxwell eq} is $\mathcal{H}=\hat{\mathcal{M}}^{-1}\cdot\hat{\mathcal{N}}$, which is a quasi-Hermitian operator with real eigenvalues since $\hat{\mathcal{M}}$ is positive-definite. As a result, the right and left eigenstates refering to $\mathcal{H}$ are given by the periodic parts of the Bloch electromagnetic states:  $|\mathcal{R}_m(\mathbf{k})\rangle =\Psi_m(\mathbf{k})e^{-i\mathbf{k}\cdot\mathbf{r}}=(u\mathbf{E},u\mathbf{H})$ and $|\mathcal{L}_m(\mathbf{k}) \rangle = \hat{\mathcal{M}}\cdot\Psi_m(\mathbf{k})e^{-i\mathbf{k}\cdot\mathbf{r}}=(u\mathbf{D},u\mathbf{B})$  satisfying the biorthonormality $ \langle \mathcal L_m(\mathbf{k})|\mathcal R_n(\mathbf{k}) \rangle =\int_{\mathrm{unit\ cell}} dxdy\,\Psi_m^\dagger(\mathbf{k})\cdot\hat{\mathcal{M}}\cdot\Psi_n(\mathbf{k})= \delta_{mn}$. And the periodic gauge, \textit{i.e.}, $\Psi_m(\mathbf{k}_0)=\Psi_m(\mathbf{k}_N)$, are imposed to ﬁx the gauge choice.

\begin{figure}[b!]
\includegraphics[width=0.7\textwidth]{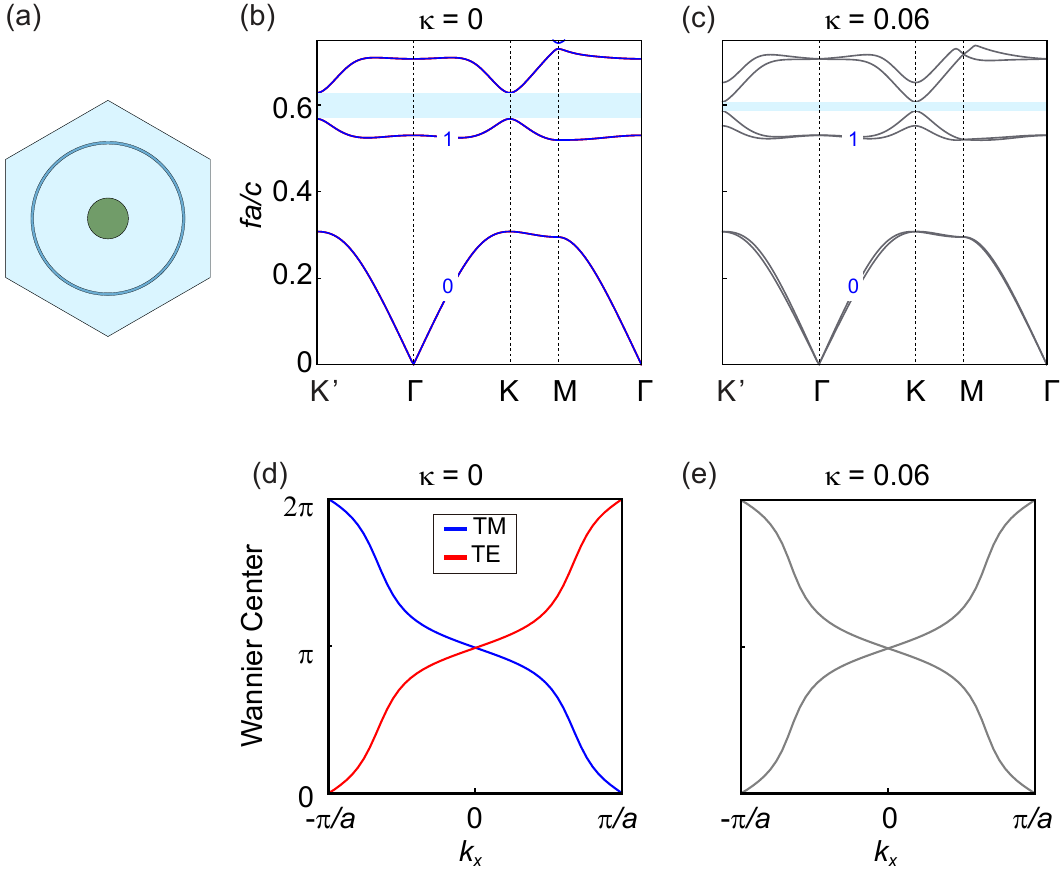}
\caption{\label{fig-s2}
(a)The unit cell of the PC composed of a gyrotropic cylinder (green) and a dielectric annulus (blue) with area $\pi r_c^2$. (b) The band structures of the PCs (b) without and (c) with spin coupling. The Wannier centers of the two valence bands below the second gap (light blue) of the PCs (d) without and (e) with spin coupling.
}
\end{figure}

% \subsection{The topologically trivial ordered PCs }

%For an ordered photonic crystal, the topological invaraint can be calculated directly from the Bloch states. 
In a 2D $\mathcal{T}_f$-invariant system with $M_z$ symmetry, the QSH-Chern number is defined as
\begin{equation}
    C_s=(C_+-C_-)/2 = C_+,
\end{equation}
where $C_\pm$ represent the Chern number for the TM and TE modes, respectively. 
Indeed, since TM and TE modes are two copies of the Chern insulator related by the $\mathcal{T}_f$ symmetry, we only need to study the topology of one type of modes.
% The TM and TE modes can be studied separately because they are fully decoupled. 
We consider the TM modes (blue) plotted in Fig. 1(d) of the main text.
The gyrotropic center cylinder breaks the $\mathcal{T}$ symmetry and the corner cylinder breaks the 2D inversion symmetry $\mathcal{P}$. %For a $\mathcal{T}$-symmetric ($\alpha=0$) system, all gaps have trivial Chern number. As we increase $\alpha$ to $0.7$, 
Since the first gap remains open when $\beta$ changes from 0 to $0.7$, it is a topologically trivial gap.
Therefore, the topology of the concerned gap (light blue) is determined by the second TM band whose Wilson loop matrix in Eq.~\eqref{Wmn} can be reduced to a scalar $W(k_x)$. The Bloch states $|\mathcal{R}\rangle$, $|\mathcal{L}\rangle$ are obtained from the COMSOL, and in the calculation we discretize each loop along $k_y$ into $N=21$ points. 
The phases of $W(k_x)$ which denote the Wannier centers are plotted in Fig.\ref{fig-s1}(b). As $k_x$ varies from $-\pi/a$ to $\pi/a$, the winding number of the Wannier centers is $0$, demonstrating that the gap is trivial.
We also plot the Wannier centers of TE modes, which are symmetric to those of TM modes about the $\mathcal{T}_f$-invariant point $k_x=0$ or $k_x=\pi/a$, as a consequence of $\mathcal{T}_f$ symmetry.

In a 2D $\mathcal{T}_f$-invariant system without $M_z$ symmetry ($\kappa=0.06$), the TM and TE modes are coupled and the QSH-Chern numbers are ill-defined. As shown in Fig. 1(e) of the main text, the two valence bands below the second gap are degenerate at the time-reversal invariant (TRI) points. We calculate the Wilson loop matrix of these two bands and plot the phases of its two eigenvalues in Fig. \ref{fig-s1}(c). The flow of the Wannier centers indicates the second gap of this coupled case is also topologically trivial characterized by the zero $\mathbb{Z}_2$ index. 
In summary, for both spin-coupled and spin-decoupled cases, the PCs are in topologically trivial phases in the absence of disorder ($\theta_d=0\degree$). Therefore, in the Fig. 2(a-b) of the main text, the gaps of PCs with weak disorder strength are topologically trivial. 
As the disorder strength increases, the gap closes and reopens, and the PCs are driven into the topologically nontrivial phase, becoming TAIs.

To understand the mechanism of realizing the photonic TAIs, we replace the corner cylinder with an annulus of equal area to effectively mimic the average property of the disordered PC with maximum random strength $\theta_d=360\degree$, whose unit is shown in Fig. \ref{fig-s2}(a). 
This PC is $\mathcal{P}$-symmetric and $\mathcal{T}$-breaking. 
In Fig. \ref{fig-s2}(b,c), we plot the band structures in the spin-decoupled and coupled ($\kappa=0.06$) cases, respectively.
% and find that all bands are doubly degenerate because of the $\mathcal P\mathcal{T}\mathcal{D}$ symmetry. 
% The TM mode is just a $\mathcal P\mathcal{T}\mathcal{D}$ copy of the TE mode. 
In both cases, the second gaps (light blue) are topologically nontrivial as demonstrated by the evolutions of the Wannier centers in Fig. \ref{fig-s2}(d,e). 
%When the TM and TE modes are coupled in PCs with $\kappa=0.06$, the Wannier centers are plotted in Fig. \ref{fig-s2}(e), demonstrating the second gap is still topologically nontrivial. 
Comparing with the ordered PCs in the trivial phase, we find that the TAI phase transition comes from the competition between the $\mathcal{P}$-breaking and $\mathcal{T}$-breaking.
In disordered PCs, randomly rotating the corner cylinder can smooth out the $\mathcal{P}$-breaking effect.
As the disorder strength increases to a critical value, the $\mathcal{P}$-breaking effect is overridden by the $\mathcal{T}$-breaking effect, and hence the topological phase transition occurs.

\subsection{Topological edge (interface) states of the $\mathbb{Z}$ and $\mathbb{Z}_2$ PC ribbons}

\begin{figure}[t!]
\includegraphics[width=0.8\textwidth]{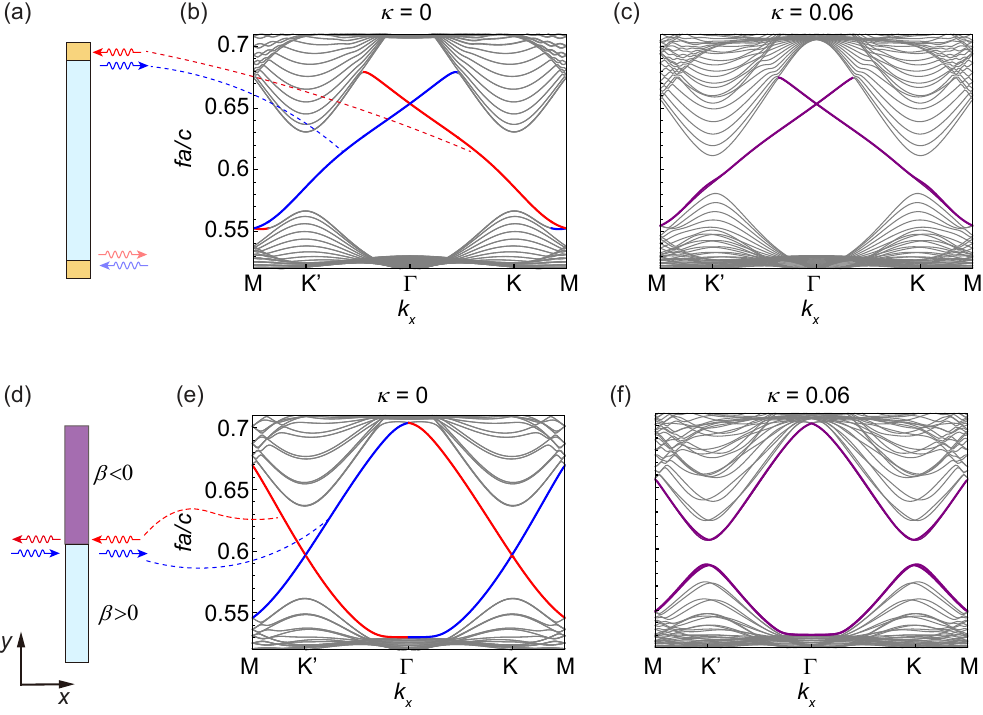}
\caption{\label{fig-projection}
The projected band structure and topological edge (interface) states of PC ribbons with the unit cells in Fig.~\ref{fig-s2}(a).
(a) The schematic supercell (30 layers) of a PC ribbon with $\beta=0.7$ whose top and bottom boundaries are covered by TSIC layers (orange) with $\tensor{\varepsilon}/\varepsilon_0=\tensor{\mu}/\mu_0=\mathrm{diag}(1,1,-1)$.
(b,c) Band structures of PC ribbons in (a) with (b)  and without (c) $M_z$ symmetry, where 
only the bands of edge states localized on the top edge are shown. 
(d) The schematic supercell of PC ribbons with two domains ($\beta = \pm 0.7$) with 15 layers of each. (e) Band structures of the PC ribbons in (d) with (e)  and without (f) $M_z$ symmetry. 
% For the spin-decoupled case, the top (bottom) domain has QSH-Chern number $1$ ($-1$), and the domain wall with charge $\Delta C_s = 2$ can support two pairs of gapless helical interface states.
% (f) For the spin-coupled case, the two domains both enter into the topological phase with $\tilde{\nu}=1$, and the domain wall index becomes trivial $\Delta \nu = 0$, implying the interface states are no longer gapless.
The other parameters of the PCs are the same with the PC in Fig. \ref{fig-s2}. }
\end{figure}

Although the spin-decoupled and spin-coupled PCs in Fig. \ref{fig-s2} are both topologically nontrivial, they belong to different topological classifications.
To demonstrate that, we study the topological edge (interface) states of PC ribbons in two different systems in Fig.~\ref{fig-projection}, which are indeed the ordered counterparts of the systems shown by Figs.~4(b) and (c) of the main text.

The first system is shown in Fig. \ref{fig-projection}(a), where we cover the top and bottom boundaries of a nontrivial PC ribbon by $\mathcal{T}_f$-symmetric insulation cladding (TSIC) layers. In the spin-decoupled case (Fig.~\ref{fig-projection}(b)), the spin-up and spin-down edge states localized at the top boundary are denoted by blue and red lines, respectively, which propagate in opposite directions and bridge the gap between the upper and lower bulk bands, demonstrating the QSH effect. 
% (blue) and the spin-down (red) edge states localized at the top boundary have opposite propagating directions, demonstrating the QSH effects. 
Note that there is also a pair of edge states localized at the bottom boundary, which is not shown in the figure.
When the $M_z$ symmetry of the PC ribbon is broken (Fig.~\ref{fig-projection}(c)), the pseudo-spins of the Bloch eigenmodes are not directed up or down uniformly, implying that the QSH effect is not well-defined. %And the $\mathbb{Z}$-classified QSH-Chern phases reduce to the $\mathbb{Z}_2$ classified two phases. 
Nevertheless, the gapless edge states persist with intact band crossings at TRI points ($\Gamma$ and $\mathrm M$), thanks to the $\mathcal{T}_f$ symmetry.

In the second system shown in Fig. \ref{fig-projection}(d), we construct a domain wall composed of two domains with $\beta = \pm 0.7$. When $\kappa = 0$, the two domains have opposite QSH-Chern numbers, $C_s =\pm 1$. Therefore, the domain wall with topological charge $\Delta C_s = 2$ supports two pairs of gapless interface states as shown in Fig. \ref{fig-projection}(e).
When introducing spin coupling ($\kappa \neq 0$) into the PC ribbon, the two domains drop into the same $\mathbb Z_2$ nontrivial phase ($\nu =1$), implying the domain wall is relegated to being trivial ($\Delta \nu =0$) and the bands of interface states open gaps near $K'$ and $K$, as displayed in Fig.~\ref{fig-projection}(f). From the perspective of band degeneracy protected by symmetry, the crossings of the two interface bands in Fig.~\ref{fig-projection}(e) are guaranteed by the $M_z$ symmetry (or equivalently the $U_1$ spin-rotation symmetry), which hence cannot survive when $M_z$ is broken. 
% Comparing to the gapless edge states in Fig. \ref{fig-projection}(c), 
%we note that the $\mathcal{T}_f$-symmetry can only protect band degeneracies at the $\mathcal{T}_f$-invariant points . 

\section{Theoretical estimation of the topological phase transition threshold}
%In the main text, we choose the rotation angle as the random disorders and provided an intuitive explanation for the topological transition of TAI based on the competition mechanism between the $\mathcal{P}$-breaking effect and $\mathcal{T}$-breaking effect. Furthermore, 
In this section, we will build a theoretical model
to give an estimation of the topological phase transition threshold of the disordered PCs.
Inspired by the standard self-energy approach for disordered systems (see \textit{e.g.}~\cite{sheng2006introduction}), here we formulate a general theoretical framework for averaging the disordered PCs and demonstrate that the topological phase transition is governed by the first order averaging effect in our system.

Staring with the Maxwell's equation Eq. \eqref{Maxwell eq}, 
for the initial unperturbed PC (Fig.~1(a) in the main text),  the constitutive tensor is denoted by $\hat{\mathcal{M}} _0 (\mathbf r)$.%, the Bloch states $\Psi_{n, \boldsymbol{k}_{0}}^{0}=\boldsymbol{u}_{n, \boldsymbol{k}_{0}}^{0} e^{i \boldsymbol{k}_{0} \cdot \mathbf{r}}$ at $\mathbf{k}_0$ (we will choose $\mathbf{k}_0$ at $K$ point) satisfies $\hat{\mathcal{N}} \Psi_{n, \boldsymbol{k}_{0}}^{0}=\omega_{n, \boldsymbol{k}_{0}}^{0} \hat{\mathcal{M}}_{0}(\mathbf{r}) \Psi_{n, \boldsymbol{k}_{0}}^{0}$. 
When we rotate the off-center cylinder around the center one by an angle $\theta$, the constitutive tensor of the perturbed PC (Fig. 1(b) in the main text) is given by
$\hat{\mathcal{M}}_{\theta}(\mathbf{r})=\hat{\mathcal{M}}_{0}(\mathbf{r})+\delta \hat{\mathcal{M}}_{\theta}(\mathbf{r})$.
Consider the retarded electromagnetic Green’s functions of the unperturbed PC and the perturbed one, $\hat G_0 (\mathbf{r},\mathbf{r}')$ and $\hat G_\theta (\mathbf{r},\mathbf{r}')$, respectively, which are $6\times 6$ $2^\mathrm{nd}$-order tensors and are determined by the following equations
\begin{equation}
\begin{gathered}
\left(\hat{\mathcal{N}}-\omega \hat{\mathcal{M}}_{0}(\mathbf{r})\right) \hat{G}_{0}\left(\mathbf{r}, \mathbf{r}^{\prime}\right)=\hat{I} \delta\left(\mathbf{r}-\mathbf{r}^{\prime}\right), \\
\left(\hat{\mathcal{N}}-\omega\left(\hat{\mathcal{M}}_{0}(\mathbf{r})+\delta \hat{\mathcal{M}}_{\theta}(\mathbf{r})\right)\right) \hat{G}_{\theta}\left(\mathbf{r}, \mathbf{r}^{\prime}\right)=\hat{I} \delta\left(\mathbf{r}-\mathbf{r}^{\prime}\right).
\end{gathered}
\end{equation}
Since $\left(\hat{\mathcal{N}}-\omega \hat{\mathcal{M}}_{0}(\mathbf{r})\right) \hat{G}_{\theta}\left(\mathbf{r}, \mathbf{r}^{\prime}\right)=\hat{I} \delta\left(\mathbf{r}-\mathbf{r}^{\prime}\right)+\omega \delta \hat{\mathcal{M}}_{\theta}(\mathbf{r}) \hat{G}_{\theta}\left(\mathbf{r}, \mathbf{r}^{\prime}\right)$, 
the two Green's functions can be related by
\begin{equation}\label{eq-Gtheta}
\begin{aligned}
 \hat{G}_{\theta}\left(\mathbf{r}, \mathbf{r}^\prime\right)
 &=\int d^{2}{r}_{1} \hat{G}_{0}\left(\mathbf{r}, \mathbf{r}_{1}\right)\left(\hat{I} \delta\left(\mathbf{r}_{1}-\mathbf{r}^{\prime}\right)+\omega \delta \hat{\mathcal{M}}_{\theta}\left(\mathbf{r}_{1}\right) \hat{G}_{\theta}\left(\mathbf{r}_{1}, \mathbf{r}^{\prime}\right)\right) \\
&=\hat{G}_{0}\left(\mathbf{r}, \mathbf{r}^{\prime}\right)+\omega \int d^{2} r_{1} \hat{G}_{0}\left(\mathbf{r}, \mathbf{r}_{1}\right) \delta \hat{\mathcal{M}}_{\theta}\left(\mathbf{r}_{1}\right) \hat{G}_{\theta}\left(\mathbf{r}_{1}, \mathbf{r}^{\prime}\right). 
\end{aligned}
\end{equation}
Iteration of $\hat G_\theta (\mathbf{r},\mathbf{r}')$ back to the above relation yields
\begin{equation}\label{eq-Gtheta_iteration}
    \hat{G}_{\theta}=\hat{G}_{0}+\omega \hat{G}_{0} \delta \hat{\mathcal{M}}_{\theta} \hat{G}_{0}+\omega^{2} \hat{G}_{0} \delta \hat{\mathcal{M}}_{\theta} \hat{G}_{0} \delta \hat{\mathcal{M}}_{\theta} \hat{G}_{0}+\cdots=\sum_{n=0}^{\infty}\left(\omega \hat{G}_{0} \delta \hat{\mathcal{M}}_{\theta}\right)^{n} \hat{G}_{0},
\end{equation}
where $\left(\omega \hat{G}_{0} \delta \hat{\mathcal{M}}_{\theta}\right)^{n} \hat{G}_{0}$ is short for $\prod_{i=1}^{n}\left(\int d^{2} r_{i}\left(\omega \hat{G}_{0}\left(\mathbf{r}_{i-1}, \mathbf{r}_{i}\right) \delta \hat{\mathcal{M}}_{\theta}\left(\mathbf{r}_{i}\right)\right)\right) \hat{G}_{0}\left(\mathbf{r}_{n}, \mathbf{r}^{\prime}\right)$ (note that $\mathbf{r}_{0}=\mathbf r$).

For any given external excitation, the ensemble average of the observed electromagnetic fields in the disordered PC samples is determined by the averaged Green’s function $\langle \hat{G}_\theta \rangle = \frac{1}{\theta_{d}} \int_{-\theta_{d} / 2}^{\theta_{d} / 2} d \theta \hat{G}_{\theta}$. By analogy with the definition of self-energy for Schrodinger equation with random impurities~\cite{sheng2006introduction}, we may introduce an effective spatially \textbf{nonlocal} ``self-modification” $\delta \hat{\mathpzc{M} }( \mathbf{r},\mathbf{r}')$ of the constitutive tensor from the configuration average of the perturbed Green’s functions:
\begin{equation} \label{eq-perturbed_Green}
\left(\hat{\mathcal{N}}-\omega \hat{\mathcal{M}}_{0}(\mathbf{r})\right)\left\langle\hat{G}_{\theta}\left(\mathbf{r}, \mathbf{r}^{\prime}\right)\right\rangle-\omega \int d^{2} r_{1}  {\color{red}\delta\hat{\mathpzc{M}}\left(\mathbf{r}, \mathbf{r}_1\right)}\left\langle\hat{G}_{\theta}\left(\mathbf{r}_{1}, \mathbf{r}^{\prime}\right)\right\rangle=\hat{I} \delta\left(\mathbf{r}-\mathbf{r}^{\prime}\right).
\end{equation}
And akin to Eq. \eqref{eq-Gtheta}, we have
\begin{equation}
    \left\langle\hat{G}_{\theta}\left(\mathbf{r}, \mathbf{r}^{\prime}\right)\right\rangle=\hat{G}_{0}\left(\mathbf{r}, \mathbf{r}^{\prime}\right)+\omega \int d^{2} r_{1} d^{2} r_{2} \hat{G}_{0}\left(\mathbf{r}, \mathbf{r}_{1}\right) {\color{red}\delta \hat{\mathpzc{M}}\left(\mathbf{r}_{1}, \mathbf{r}_{2}\right)}\left\langle\hat{G}_{\theta}\left(\mathbf{r}_{2}, \mathbf{r}^{\prime}\right)\right\rangle,
\end{equation}
which is parallel to the well-known Dyson equation for self-energy. Then, substituting Eq. \eqref{eq-Gtheta_iteration} into Eq. \eqref{eq-perturbed_Green}, we obtain 
\begin{align*}
&\left(\hat{\mathcal{N}}-\omega \int d^{2} r_{1}\left(\hat{\mathcal{M}}_{0} \delta\left(\mathbf{r}-\mathbf{r}_{1}\right)+\delta \hat{\mathpzc{M}}\left(\mathbf{r}, \mathbf{r}_{1}\right)\right)\right) \hat{G}_{0}\left(\hat{I}+\omega\left\langle\delta \hat{\mathcal{M}}_{\theta}\right\rangle \hat{G}_{0}+\omega^{2}\left\langle\delta \hat{\mathcal{M}}_{\theta} \hat{G}_{0} \delta \hat{\mathcal{M}}_{\theta}\right\rangle \hat{G}_{0}+\cdots\right) =\hat{I} \delta\left(\mathbf{r}-\mathbf{r}^{\prime}\right), \\
\Rightarrow \quad & \omega \int d^{2} r_{1}\left(\left\langle\delta \hat{\mathcal{M}}_{\theta}(\mathbf{r})\right\rangle \delta\left(\mathbf{r}-\mathbf{r}_{1}\right)-\delta \hat{\mathpzc{M}}\left(\mathbf{r}, \mathbf{r}_{1}\right)\right) \hat{G}_{0}\left(\mathbf{r}, \mathbf{r}^{\prime}\right) 
+\omega^{2} \int d^{2} r_{1}\left\langle\delta \hat{\mathcal{M}}_{\theta}(\mathbf{r}) \hat{G}_{0}\left(\mathbf{r}, \mathbf{r}_{1}\right) \delta \hat{\mathcal{M}}_{\theta}\left(\mathbf{r}_{1}\right)\right\rangle \hat{G}_{0}\left(\mathbf{r}_{1}, \mathbf{r}^{\prime}\right) \\
&-\omega^{2} \int d^{2} r_{1}d^2r'' \delta\hat{\mathpzc{M}}(\mathbf{r},\mathbf{r}'') \hat{G}_{0}\left(\mathbf{r}'', \mathbf{r}_{1}\right)\left\langle\delta \hat{\mathcal{M}}_{\theta}\left(\mathbf{r}_{1}\right)\right\rangle \hat{G}_{0}\left(\mathbf{r}_{1}, \mathbf{r}^{\prime}\right)+\mathcal{O}\left(\delta M^{3}\right) = 0,\\
\Rightarrow \quad &\int d^{2} r_{1}\left[\omega\left(\left\langle\delta \hat{\mathcal{M}}_{\theta}(\mathbf{r})\right\rangle \delta\left(\mathbf{r}-\mathbf{r}_{1}\right)-\delta \hat{\mathpzc{M}}\left(\mathbf{r}, \mathbf{r}_{\mathbf{1}}\right)\right)+\omega^{2}\left\langle\delta \hat{\mathcal{M}}_{\theta}(\mathbf{r}) \hat{G}_{0}\left(\mathbf{r}, \mathbf{r}_{1}\right) \delta \hat{\mathcal{M}}_{\theta}\left(\mathbf{r}_{1}\right)\right\rangle\right. \\
&\left.-\omega^{2}\int d^2r''\delta\hat{\mathpzc{M}}(\mathbf{r},\mathbf{r}'') \hat{G}_{0}\left(\mathbf{r}'', \mathbf{r}_{1}\right)\left\langle\delta \hat{\mathcal{M}}_{\theta}\left(\mathbf{r}_{1}\right)\right\rangle+\mathcal{O}\left(\delta M^{3}\right)\right] \hat{G}_{0}\left(\mathbf{r}_{1}, \mathbf{r}^{\prime}\right) = 0.
\end{align*}
Therefore, the $2^\mathrm{nd}$-order approximation of the self-modification of the constitutive tensor is given by
\begin{equation}
\begin{split}
&\qquad\delta \hat{\mathpzc{M}}\left(\mathbf{r}, \mathbf{r}_{\mathbf{1}}\right) \\
&=\left\langle\delta \hat{\mathcal{M}}_{\theta}(\mathbf{r})\right\rangle \delta\left(\mathbf{r}-\mathbf{r}_{1}\right)-\omega\int d^2r''\delta \hat{\mathpzc{M}}(\mathbf{r},\mathbf{r}'') \hat{G}_{0}\left(\mathbf{r}'', \mathbf{r}_{1}\right)\left\langle\delta \hat{\mathcal{M}}_{\theta}\left(\mathbf{r}_{1}\right)\right\rangle
\quad+\omega\left\langle\delta \hat{\mathcal{M}}_{\theta}(\mathbf{r}) \hat{G}_{0}\left(\mathbf{r}, \mathbf{r}_{1}\right) \delta \hat{\mathcal{M}}_{\theta}\left(\mathbf{r}_{1}\right)\right\rangle+\mathcal{O}(\delta M^3)\\
&=\underbrace{\left\langle\delta \hat{\mathcal{M}}_{\theta}(\mathbf{r})\right\rangle \delta\left(\mathbf{r}-\mathbf{r}_{1}\right)}_{\displaystyle\text{1st-order local average }}-\omega\left\langle\delta \hat{\mathcal{M}}_{\theta}(\mathbf{r})\right\rangle \hat{G}_{0}\left(\mathbf{r}, \mathbf{r}_{1}\right)\left\langle\delta \hat{\mathcal{M}}_{\theta}\left(\mathbf{r}_{1}\right)\right\rangle+\underbrace{\omega\left\langle\delta \hat{\mathcal{M}}_{\theta}(\mathbf{r}) \hat{G}_{0}\left(\mathbf{r}, \mathbf{r}_{1}\right) \delta \hat{\mathcal{M}}_{\theta}\left(\mathbf{r}_{1}\right)\right\rangle}_{\displaystyle\text {nonlocal Born approximation }}+\mathcal{O}(\delta M^3),\label{2nd order approximation}
\end{split}
\end{equation}
where the second term in the last step is obtained by replacing $\delta\mathpzc{M}$ with the first-order term. In Eq.~\eqref{2nd order approximation}, the first term is just the configuration average of the perturbed constitutive tensor which is spatially local and nondispersive; while the other two second order modification terms are both spatially nonlocal and dependent on the frequency. The last second order term precisely correspond to the \textbf{\color{Blue}Born approximation} for the Schrodinger equation. In the original TAI system with Anderson type disorders \cite{TAI_Shen_2009_Phys.Rev.Lett.,TAI_Beenakker_2009_Phys.Rev.Lett.}, the first order configuration average of the random onsite energy vanishes (corresponding to the vanishing of the first two terms in our case). Consequently, the self-energy is entirely determined by the Born approximation term. In contrast, the first order configuration average $\delta \hat{\mathcal{M}}_\theta (\mathbf{r})$ of our system is nonvanishing and dependent on the disorder strength $\theta_d$, therefore, we can expect that {\color{Blue}\textbf{the first order configuration average} dominates the wave behavior in the PC ensemble and gives rise to the topological phase transition in our case}.

\begin{figure}[t!]
\includegraphics[width=0.8\textwidth]{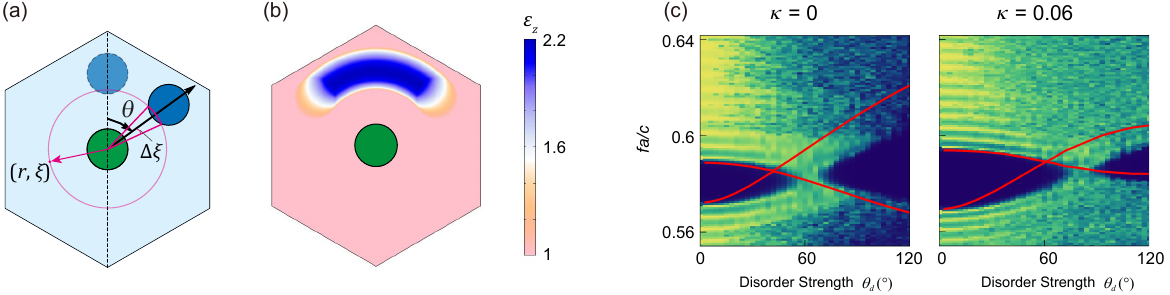}
\caption{\label{fig-effectiverandom}
Theoretical prediction for TAI phase transition. (a) Schematic for calculating the configuration average $\hat{\mathcal{M}}_\theta(\mathbf{r})_{\theta_d}$. (b) Configuration average distribution of $\varepsilon_z(\mathbf{r}) $ outside the central cylinder with disorder strength $\theta_d = 150 \degree$. (c) Comparison between the theoretical band edges (red curves) based on the first order configuration average constitutive tensor $\hat{\mathcal{M}}_\theta(\mathbf{r})_{\theta_d}$ and the numerical phase diagram of typical transmittance (background density plot) for both spin-decoupled ($\kappa =0$) and spin-coupled ($\kappa =0.06$) cases. }
\end{figure}

To demonstrate our prediction, we construct the averaged PC with the effective constitutive tensor
\begin{equation}
\hat{\mathcal{M}}_{\mathrm{eff}}\left(\mathbf{r} ; \theta_{d}\right)=\hat{\mathcal{M}}_{0}(\mathbf{r})+\left\langle\delta \hat{\mathcal{M}}_{\theta}(\mathbf{r})\right\rangle_{\theta_{d}}=\left\langle\hat{\mathcal{M}}_{\theta}(\mathbf{r})\right\rangle_{\theta_{d}}=\frac{1}{\theta_{d}} \int_{-\theta_{d} / 2}^{\theta_{d} / 2} d \theta \hat{\mathcal{M}}_{\theta}(\mathbf{r}).
\end{equation}
Suppose that an element of $\hat{\mathcal{M}}_{\theta}(\mathbf{r})$, denoted $f$, equals $f_0$ and $f_1$ in the background medium and in the off-center cylinder, respectively. As shown in Fig. \ref{fig-effectiverandom}(a), using the polar coordinates $(r, \xi)$, its distribution can be expressed as
\begin{equation}
f(r, \xi ; \theta)=f_{0}+\left\{\begin{array}{cc}
\left(f_{1}-f_{0}\right) & |\xi-\theta| \leq \Delta \xi(r) \\[3pt]
0 & \text { otherwise }
\end{array}\right.=f_{0}+\left(f_{1}-f_{0}\right) \cdot \operatorname{rect}_{2 \Delta \xi(r)}(\xi-\theta),
\end{equation}
where $\Delta \xi(r)=\arccos \left(\frac{d^{2}+r^{2}-r_{c}^{2}}{2 r_{0} r}\right)$ denotes half of the central angle of the intersection between a circle (light magenta) of radius $r$ and the off-center cylinder (see the schematic in Fig. \ref{fig-effectiverandom}(a)), and $\operatorname{rect}_{T}(t)=u(t+T / 2)-u(t-T / 2)$ denotes a rectangle function with width $T$ and $u(t)=(1+\operatorname{sign}(t)) / 2$ denotes the step function. Thus, the average of the element $f(r, \xi ; \theta)$ can be derived as
$$
\begin{aligned}
\langle f(r, \xi ; \theta)\rangle_{\theta_{d}} &=\frac{1}{\theta_{d}} \int_{-\frac{\theta_{d}}{2}}^{\frac{\theta_{d}}{2}} d \theta f(r, \xi ; \theta)\\
&=f_{0}+\frac{\left(f_{1}-f_{0}\right)}{\theta_{d}} \int_{-\frac{\theta_{d}}{2}}^{\frac{\theta_{d}}{2}} d \theta \operatorname{rect}_{\Delta \xi(r)}(\xi-\theta) \\
&=f_{0}+\frac{\left(f_{1}-f_{0}\right)}{\theta_{d}} \int_{-\infty}^{+\infty} d \theta \operatorname{rect}_{\theta_{d}}(\theta) \operatorname{rect}_{2 \Delta \xi(r)}(\xi-\theta) \\
&=f_{0}+\frac{\left(f_{1}-f_{0}\right)}{\theta_{d}} \operatorname{rect}_{\theta_{d}}(\xi) \star \operatorname{rect}_{2 \Delta \xi(r)}(\xi),
\end{aligned}
$$
where $\star$ in the last step represents the convolution on two functions. In terms of the property of the step function $u(\xi) \star u(\xi)=\xi u(\xi)$ and the translational equivalence of the convolution $(u \star u)(t-a)=u(t-a) \star u(t)=u(t) \star u(t-a)$, we obtain
\begin{equation}
\begin{aligned}
\langle f(r, \xi ; \theta)\rangle_{\theta_{d}}=&\ f_0+\frac{\left(f_{1}-f_{0}\right)}{\theta_{d}}\left[\left(\xi+\frac{\theta_{d}}{2}+\Delta \xi(r)\right) u\left(\xi+\frac{\theta_{d}}{2}+\Delta \xi(r)\right)-\left(\xi+\frac{\theta_{d}}{2}-\Delta \xi(r)\right) u\left(\xi+\frac{\theta_{d}}{2}-\Delta \xi(r)\right)\right. \\
&\left.-\left(\xi-\frac{\theta_{d}}{2}+\Delta \xi(r)\right) u\left(\xi-\frac{\theta_{d}}{2}+\Delta \xi(r)\right) +\left(\xi-\frac{\theta_{d}}{2}-\Delta \xi(r)\right) u\left(\xi-\frac{\theta_{d}}{2}-\Delta \xi(r)\right)\right] .
\end{aligned}
\end{equation}
As an example, the distribution of the averaged $\varepsilon_{z}$ component with $\theta_{d}=105^{\circ}$ is shown in Fig. \ref{fig-effectiverandom}(b). Using these averaged PCs without and with spin coupling, we numerically calculated the band edges of the two concerned bands, as shown by the red curves in Fig. \ref{fig-effectiverandom}(c), which correspond to the first order approximation of the mobility band edges of the typical transmittance spectra. Therefore, the crossing point of the two band edges offers a theoretical estimation of the TAI phase transition threshold. Although deviating a little bit from the full-wave result, the qualitatively estimation demonstrates that the TAI phase transition is indeed a first-order configuration average effect in our systems.

\section{Full-wave calculation of the local QSH-Chern numbers}

\begin{figure}[h!]
\includegraphics[width=0.9\textwidth]{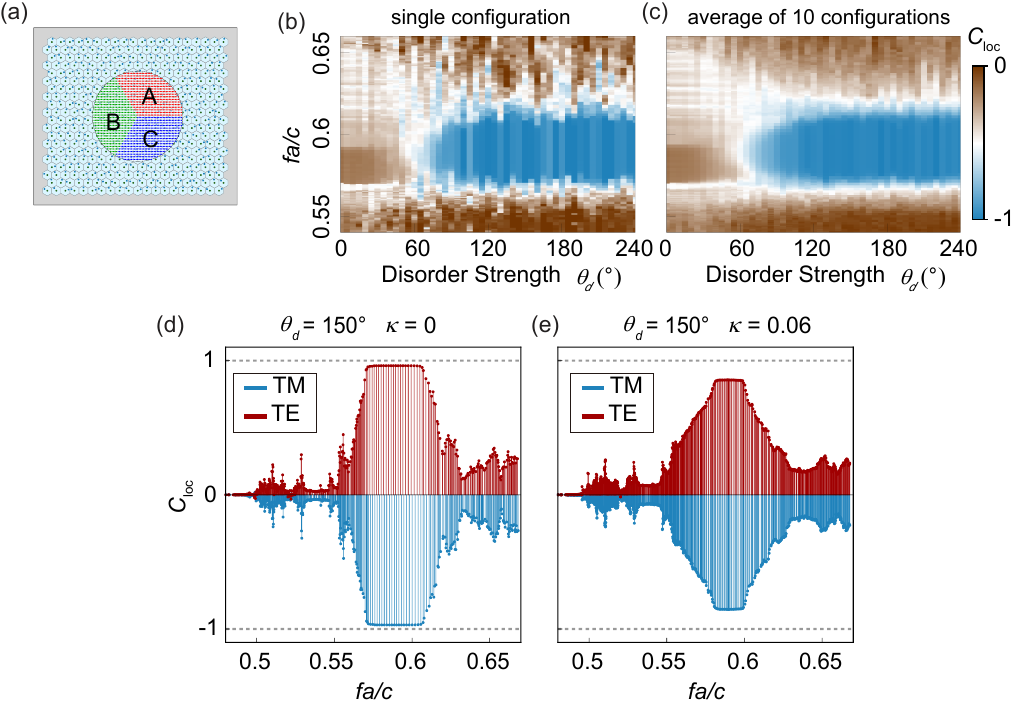}
\caption{\label{fig-s3}
(a) The schematic for calculating the local QSH-Chern number of a disordered PC (the parameters are the same as Fig.~1 of the main text). A disk with radius $5a$ is chosen as the summation region, which are discritized into grids with step size $dx=a/6, dy=0.5l$. The disk is divided into three regions (red, green and blue), and each region has $542$ grid points. The disordered PC (sample size $L_x\times L_y = 20a \times 30 l$) is surrounded by TSIC layers (gray). %with $\tensor{\varepsilon}/\varepsilon_0=\tensor{\mu}/\mu_0=\mathrm{diag}(1,1,-1)$, and the outtest boundaries are set as scattering boundary. 
(b,c) The phase diagrams of the local QSH-Chern number of a $M_z$-symmetric PC ($\kappa=0$) for (b) a single configuration (the angle $\theta$ is randomly chosen at every point in the diagram) and (c) averaged over $10$ configurations. %The local QSH-Chern number $C_{\text{loc}}$ for the TM modes of the spin-conservation case ($\kappa=0$) is converged to $-1$ (blue) in the topologically nontrivial gap. 
The profiles of the local QSH-Chern number at the disorder strength $\theta_d=150\degree$ of two PCs (d) without ($\kappa=0$) and (e) with ($\kappa=0.06$) spin-coupling. %When $\kappa=0$, the local QSH-Chern number is $-1$($1$) in the bandgap for the TM (TE) polarized projection operator. When $\kappa=0.06$, the local QSH-Chern numbers calculated using the TM (TE) polarized projection operators are no longer quantized.
}
\end{figure}

For disordered systems without periodicity, the topological invariants cannot be calculated using the Bloch states in the momentum space. A method~\cite{LocalChernNumber_Kitaev_2006_AnnalsofPhysics,TAI-experiment_Mitchell_2018_Nat.Phys.} for computing the Chern number in real space has been proposed to characterize the topology of disordered systems. However, most of the calculations of the local Chern number are applied in tight-binding models. Here, we extend this method to  full-wave PC systems.
% Since the disordered PC is a continuous model, we need to transform it into a discrete model.
As shown in Fig. \ref{fig-s3}(a), we choose a circular area with a radius of $5a$ at the center of the disordered system as the summation area. The area is divided into three sectors labeled as $A$ (red), $B$ (green), and $C$ (blue).%, which are discretized into finite square mesh points.

By replacing the lattice sum in the original definition~\cite{LocalChernNumber_Kitaev_2006_AnnalsofPhysics,TAI-experiment_Mitchell_2018_Nat.Phys.} with the integration over the corresponding regions,  the local Chern number at a frequency $\omega_0$ can be generalized to the disordered full-wave systems as% at a frequency $\omega_0$ can be calculated as
\begin{equation}\label{eq-LocChern}
    C_{\text{loc} }(\omega_0)=12 \pi i \int\limits_{\mathbf{r}_a\in A}d^2{r}_a \int\limits_{\mathbf{r}_b\in A}d^2{r}_b \int\limits_{\mathbf{r}_c\in C}d^2{r}_c\Big(P(\mathbf{r}_a,\mathbf{r}_b) P(\mathbf{r}_b,\mathbf{r}_c) P(\mathbf{r}_c,\mathbf{r}_a)-P(\mathbf{r}_a,\mathbf{r}_c) P(\mathbf{r}_c,\mathbf{r}_b) P(\mathbf{r}_b,\mathbf{r}_a)\Big),
\end{equation}
% \begin{equation}\label{eq-LocChern}
%     C_{\text{loc} }(\omega_0)=12 \pi i \sum_{j \in A} \sum_{k \in B} \sum_{l \in C}\left(P_{j k} P_{k l} P_{l j}-P_{j l} P_{l k} P_{k j}\right),
% \end{equation}
where %$j$, $k$, $l$ are the indices of the mesh points, and 
$P$ is the projection operator of real-space eigenstates below a certian frequency $\omega_0$
\begin{equation}
    P(\mathbf{r},\mathbf{r}')=\sum_{\omega<\omega_0}\langle\mathbf{r}\left|\mathcal{R}_{\omega}\right\rangle\left\langle \mathcal{L}_{\omega}\right|\mathbf{r}'\rangle .
\end{equation}
%where $|\mathcal{R}\rangle_\omega$ and $|\mathcal{L}\rangle_\omega$ denote the a pair of right and left eigenstates with eigenfrequency $\omega$. 
Physically, if magnetic fluxes are inserted through the central vertex and adiabatically change from 0 to a magnetic flux quantum, the accumulated charge at the center point is just characterized by the local Chern number of   Eq.~\eqref{eq-LocChern}~\cite{TAI-experiment_Mitchell_2018_Nat.Phys.}, which is consistent with Laughlin's charge pumping argument for quantum Hall effect~\cite{Laughlin-PRB-1981}. %The bases $|\mathcal{R}\rangle = (u\mathbf{E},u\mathbf{H})$ and $|\mathcal{L} \rangle = (u\mathbf{D},u\mathbf{B})$ can be obtained by solving the eigenvalue problem in COMSOL.
In a 2D photonic system with $M_z$ symmetry, the projection operator can be defined for the TM and TE polarizations separately. For the TM polarization projection operator, we have
\begin{equation}
    \left|\mathcal{R}^\mathrm{TM}_{\omega}\right\rangle = (uE_z, uH_x, uH_y)^\intercal, \quad
    \left|\mathcal{L}^\mathrm{TM}_{\omega}\right\rangle = (uD_z, uB_x, uB_y)^\intercal,
\end{equation}
and for the TE polarization projection operator, we have
\begin{equation}
    \left|\mathcal{R}^\mathrm{TE}_{\omega}\right\rangle = (uE_x, uE_y, uH_z)^\intercal, \quad
    \left|\mathcal{L}^\mathrm{TE}_{\omega}\right\rangle = (uD_x, uD_y, uB_z)^\intercal.
\end{equation}
The $\mathcal{T}_f$ symmetry of the system ensures the local Chern numbers of TM and TE modes take precisely opposite values $C^{\mathrm{TM}}_\mathrm{loc}=-C^{\mathrm{TE}}_\mathrm{loc}$ in any single configuration. And the local QSH-Chern number is given by $C_s=(C^{\mathrm{TM}}_\mathrm{loc}-C^{\mathrm{TE}}_\mathrm{loc})/2=C^{\mathrm{TM}}_\mathrm{loc}$.
In the numerical calculations, the continuous coordinates are discretized into finite square mesh points (see Fig.~\ref{fig-s3}(a)), and the integrations in Eq.~\eqref{eq-LocChern} are replaced by summations over the mesh points in the corresponding regions.

In disordered systems, to obtain a numerically stable local Chern number, it is necessary to average over different random configurations.
%To quantitatively characterize the QSH TAI phase shown in Fig. 2(a) of the main text,
In Figs. \ref{fig-s3}(b) and \ref{fig-s3}(c), we plot the local QSH-Chern numbers of a single configuration and averaged over $10$ configurations, respectively, and find that 10 configurations are enough to achieve a smooth phase diagram with well-quantized QSH-Chern numbers inside the mobility gaps. Figure~\ref{fig-s3}(c) exhibits the fairly same phase transition as that shown by Fig. 2(a) of the main text. It shows the local QSH-Chern numbers in the mobility bandgaps before and after phase transition (around $\theta_d=60^{\circ}$) approach to 0 and -1, respectively; thereby quantitatively confirming that the insulating phase at the large disorder side is truly a QSH TAI.
In Figs. \ref{fig-s3}(d) and \ref{fig-s3}(e), we show the profiles of the local QSH-Chern numbers at the disorder strength $\theta_d=150\degree$ for the spin-decoupled and coupled cases, respectively, where the result of spin-coupled case  ($\kappa=0.06$) is obtained by projecting the eigenstates onto the TM and TE bases. We find that for the spin-coupled case, the computed Chern numbers are no longer quantized inside the mobility gap, indicating the method fails to compute the $\mathbb{Z}_2$ index.  %but it is also restricted to tight-binding models. 

% We found that the local QSH-Chern number of the TM mode is quantized to $-1$ (blue) in the reopened gap, demonstrating that as the disorder strength increases, the disordered PC becomes a TAI with a topologically nontrivial gap. In Fig. \ref{fig-s3}(c), we plot the local QSH-Chern number of the spin-decoupled PC with disorder strength $\theta_d=150\degree$. 

From the above discussion, it can be seen that the local Chern number method based on real space information can be used to calculate the QSH-Chern number of disordered PC with $M_z$ symmetry. However, the disadvantages of this approach are also obvious. First, this method needs to consume huge computing resources and computation time, including solving the eigenstates of a large enough lattice and the matrix manipulations for the huge dimensional projection operators in full-wave systems, let alone the multiplied workload 
for configuration average. %In order to avoid the finite size effect and ensure the accuracy of the calculation, the disordered sample size cannot be small, but a larger sample size usually means that more calculation degrees of freedom need to be solved in COMSOL. For example, when calculating the local spin Chen number of the TM mode in the graph \ref{fig-s3} (d), we need to calculate $500$ eigenstates in COMSOL at a time, which requires a computer with large memory and CPU. 
Second, the choice of the circular summation region is subtle. %The projector in Eq.\eqref{eq-LocChern} is a local operator~\cite{TAI-experiment_Mitchell_2018_Nat.Phys.}. 
%Strictly speaking, the setting in Fig. \ref{fig-s3} (a) only calculates the Chern number of the site at the center of the circle, meaning that it only contains local information. And we use this local information at the center to represent the topology of the entire disorderd system. 
The radius of the region should be appropriate so as to involve all the bulk state information but to exclude the contribution of boundary states. In fact, if we choose the whole system for summation, the outcome vanishes~\cite{TAI-experiment_Mitchell_2018_Nat.Phys.}. As the influence of edge states cannot be completely eliminated, the computed local Chern numbers deviate from rigorous integers inside the gaps. And most importantly, the method is inapplicable to spin-coupled PCs. 
%Because the contribution of the edge state at the boundary and the bulk state to the Chen number cancel each other out~\cite{TAI-experiment_Mitchell_2018_Nat.Phys.}.

%Thirdly, since the continuous disordered PC needs to be discretized, the projector matrices in Eq. \eqref{eq-LocChern} are very large. Therefore, the calculation of $C_\text{loc}$ is very time-consuming and usually requires much computing resources when doing matrix multiplication. Finally, in order to obtain a quantized local number and determine the phase boundary between different insulators, it is necessary to average multiple samples with different random configurations, which inevitably increases the amount of calculation several times.
% For example, in the above calculation, each summation sector compromise $542$ summation points, and

Very recently, a generalized method for computing the local $\mathbb{Z}_2$ index in disordered TIs was developed following a similar strategy of region trisection~\cite{Z2-calculation_Mong_2019_Phys.Rev.B}. However, since a crucial step in this approach is to simultaneously diagonalize two real-space projection operators, the huge dimension of the operators makes the method nearly unfeasible for full-wave models. %In this work, we proposed a new method based on the scattering matrix to calculate the $\mathbb{Z}_2$ invariants, which will be explained in detail in the following sections.
Considering the limitations of the real space calculation methods, we will introduce a new efficient approach based on scattering matrix to calculate the $\mathbb{Z}$ and $\mathbb{Z}_2$ topological invariants for PCs in the following sections.

\section{Scattering approach for classifying the QSH phases of 2D spin-decoupled photonic crystals}

Instead of directly computing the topological indices with bulk states, scattering approaches have been proposed as an alternative way to retrieve the bulk topology in tight-binding models~\cite{ReflectionPhase_Brouwer_2010_Phys.Rev.B,ReflectionPhase_Brouwer_2011_Phys.Rev.B,Z2-calculation_Brouwer_2014_Phys.Rev.B,Z2-calculation_Akhmerov_2012_Phys.Rev.B,xiao2014surface,gao2015determination,ReflectionPhase_Chong_2014_Phys.Rev.B,ReflectionPhase_Chong_2015_Phys.Rev.X,Z2-calculation_Hafezi_2015_Phys.Rev.A,mittal2016measurement,ReflectionPhase_Chong_2016_Phys.Rev.B,ReflectionPhase_Zhang_2020_Phys.Rev.Lett.}.
% The calculations of $\mathbb{Z}$ topological invariants based on the scattering matrix method have been applied in tight-binding models. 
In this section, we will extend this method to full-wave photonic systems. 
Following the methodology of Laughlin's charge pumping~\cite{Laughlin-PRB-1981}, a twisted boundary condition with $\Psi(y=L)=\Psi(y=0)e^{i\phi_y}$  is imposed to the disordered PC's transverse boundaries ($y=0,L$)~\cite{ReflectionPhase_Brouwer_2011_Phys.Rev.B,Z2-calculation_Brouwer_2014_Phys.Rev.B}, which can be viewed as a gauge flux $\phi_y$ threading a cylinder formed by rolling up the 2D PC, as shown in Fig. \ref{fig-s4}. %Recalling Laughlin's argument about the quantum Hall effect~\cite{Laughlin-PRB-1981}, the topological pump is quantized when the flux varying periodically in time completes one cycle, which can be used to characterize the topological invariants of the insulator.
% In this section, we propose a ``topological pump" setup to measure the reflection coefficients of the disordered spin-decoupled PC and relate the winding number of reflection phases to the number of topologically protected helical edge states.
We first expound our idea via the relatively simple case, namely the PCs with both $\mathcal{T}_f$ and $M_z$ symmetries. Since the TM and TE modes are decoupled and related by $\mathcal{T}_f$ symmetry, it is sufficient to investigate only one spin sector, say TM.

\subsection{The setup for classifying the QSH phases of  spin-decoupled PCs }

\begin{figure}[bh!]
\includegraphics[width=0.8\textwidth]{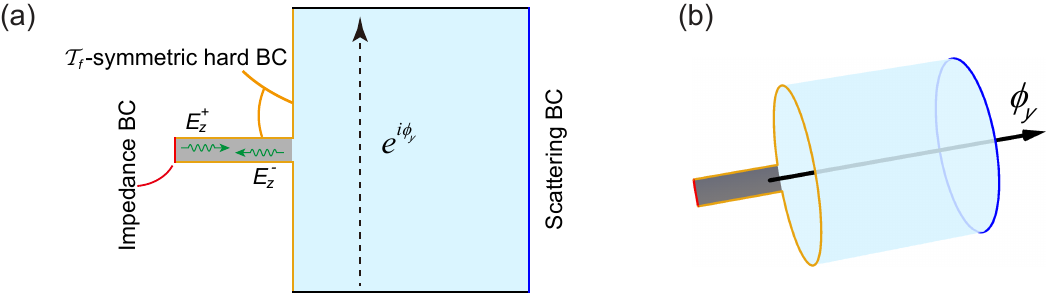}
\caption{\label{fig-s4}
(a) The schematic for calculating the topological invariants of a $\mathcal{T}_f$-symmetric disordered PC via reflection matrix. The left side of the PC (blue region) is connected with an air lead (gray) bounded by topologically trivial $\mathcal{T}_f$-symmetric hard boundaries (yellow). For purely TE (TM) reflection process of a $M_z$-symmetric PC, the hard boundaries can be simply use PEC (PMC) in numerical simulations. The upper and lower boundaries (black) are connected with a twisted periodic boundary condition with an additional phase factor $e^{i\phi_y}$, and the right (blue) boundary is set as scattering BC. %The reflection coefficients of the PC can be obtained by calculating the forward-propagating wave $E_z^+$ and the backward-propagating wave $E_z^-$.
(b) The application of twisted boundary conditions in (a) can be viewed as an adiabatically changing gauge flux (black) $\phi_y$ threading the hollow of a rolled-up PC. %formed by rolling the 2D PC into a cylinder.
}
\end{figure}

% Considering the drawbacks of the real space calculation method for computing the Chern number, we propose a new method of classifying the topological phases of a 2D photonic system via the scattering matrix. 

% \subsection{The topological classification of the spin-decoupled case}

% Here, we propose a new method for calculating the topological invariants of the disordered PCs using spin reflection. % can be calculated via a scattering matrix. 

%Now let us introduce the setup for calculating the topological invariants of the disordered PCs using spin reflection.
%For $\mathcal{T}_f$-symmetric PCs with $M_z$ symmetry, the TM and TE modes are decoupled and can be viewed as pseudo-spin up and down states. We study one spin sector, say TM, firstly. 
The Gedankenexperiment system is shown in Fig.\ref{fig-s4} (a), where a $\mathcal{T}_f$ symmetric waveguide lead (gray) is connected to the left side of the disordered PC (light blue). In the frequency range studied, we require the width of the lead to be less than half wavelength in order that the lead is a single-mode waveguide for both TM and TE polarizations.
Therefore, the wave in the lead can always be written as the superposition of  two oppositely propagating TM waveguide modes, $E_z^+ e^{ikx}$ and $E_z^-e^{-ikx}$, where $k$ is the wavenumber of the waveguide modes at the frequency $\omega$.
Imagining $E_z^+ e^{ikx}$ is the  wave incident from the far end of the lead, and $E_z^-e^{-ikx}$ denotes the  wave reflected from the PC, then the TM reflection coefficient of the PC is defined as 
\begin{equation}\label{eq-rPC}
    r(\phi_y,\omega)=E_z^{-}/E_z^+,
\end{equation}
which is a function of the frequency $\omega$ as well as the twist angle $\phi_y$ of the boundary condition.

Inside the mobility gap of the PC, the incident waves are totally reflected, and hence the reflection coefficients of the PC should be unitary: $r(\phi_y)=e^{i\varphi(\phi_y)}\in U(1)$.
Indeed, when we adiabatically vary $\phi_y$ over a cycle (from $-\pi$ to $\pi$), the first homotopy group of $U(1)$,
$   \pi_1(U(1))=\mathbb{Z}$, 
determines the topological classification of the evolutions of $r(\phi_y)$, denoted by the integer winding number of the reflection phase $\varphi(\phi_y)=\arg(r(\phi_y))$:
\begin{equation}\label{reflection phase winding}
\tilde{C}_s =\frac{-i}{2 \pi} \int_{-\pi}^{ \pi} \frac{\partial \ln r}{\partial \phi_{y}} d \phi_{y} 
=\frac{1}{2 \pi} \int_{-\pi}^{ \pi} \frac{\partial \varphi}{\partial \phi_{y}} d \phi_{y} \in \mathbb{Z}.
\end{equation}
And the $\mathcal{T}_f$-symmetry ensures that the winding number of the TE reflection phase is always opposite to the TM result.

In principle, the single-mode waveguide lead can be replaced by a multi-mode waveguide or other more complex scattering setups (\textit{e.g.}, plane wave scattering at an open boundary). Then the reflection coefficient should be replaced by a reflection matrix $R$ whose dimension is determined by the number of input and reflection channels of the system (for plane wave, the dimension is infinite). Since the unitarity of $R$, \textit{i.e.}, $R\in U(N)$, is universal inside the mobility gap of the PC, the spectral flow of $R(\phi_y)$ over a cycle of the twisted boundary always follows a $\mathbb{Z}$ classification according to the fundamental group of the unitary groups $\pi_1(U(N))\equiv\mathbb{Z}$~\cite{hatcher2005algebraic}, which can be calculated by the similar winding number integration as Eq.~\eqref{reflection phase winding} via replacing $r$ with $\det(R)$.  However, the essential topology of the reflection spectrum should be purely determined by the disordered PC but irrelevant to the number of channels. Therefore, a single mode waveguide is the most simple and efficient setup for numerical examinations.

\subsection{Correspondence between reflection phase and gapless edge states}

\begin{figure}[h!]
\includegraphics[width=0.6\textwidth]{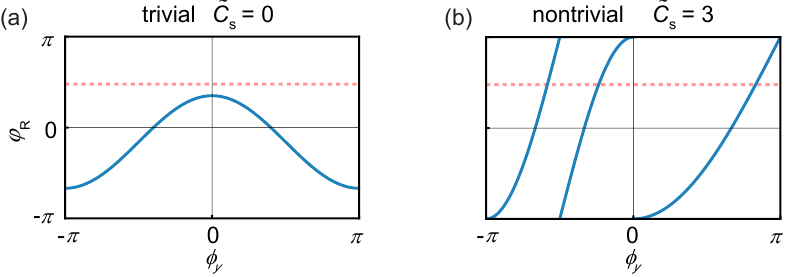}
\caption{\label{fig-s5}
The examples of the (a) trivial and (b) nontrivial evolutions of the TM reflection phases $\varphi(\phi_y)$ over a cycle of the twist angle $\phi_y$ for a $M_z$-symmetric PC.  The red dashed line denotes a constant reflection phase of the impedance boundary encased in waveguide lead in Fig.~\ref{fig-s4}(a), whose crossing points with the reflection phases of the PC correspond to the edge states localized inside the waveguide.
}
\end{figure}

Though we have shown that the flow of the reflection phase follows the same topological classification as the PCs' bulk, it is still necessary to find a one-to-one correspondence between the two classifications. Now, we will demonstrate that the winding number of the reflection phases can exactly predict the number of gapless edge states of the disordered PC with twisted boundary conditions so as to connect the two classifications. 

Imagine the left entry of the lead is closed by an impedance boundary, whose reflection coefficient is given by 
%In analogy to the reflection coefficient of the PC defined in Eq. \eqref{eq-rPC}, 
%we can define another reflection coefficients corresponding to the lead as
\begin{equation}\label{eq-GammaDef}
{r _L}=e^{i\varphi_L} = \frac{{E_z^ + }}{{E_z^ - }}.
\end{equation}
where $E_z^ -$ (${E_z^ + }$) denotes the amplitude of the leftward incident (rightward reflected) waveguide mode, and we require that the reference plane for defining the reflection phases of $r_L$ and $r$ are identical.
The impedance boundary can be achieved by any optical insulating materials respecting TRS, say a PMC with $r_L=1$. More generally, it can be formed by another gapped PC subject to a twisted boundary condition such that $r_L=r_L(\phi_y)$ also depends on the twist angle $\phi_y$ but obeys $r_L(\phi_y)=r_L(-\phi_y)$ due to TRS. Hence, TRS ensures that the impedance boundary is topologically trivial. 

Since the closed waveguide lead can be regarded as a special TRS hard seal of the disordered PC,    %If a harmonic TM field can subsist in the waveguide closed by the impedance boundary and the PC from two sides, the following consistency relation between the two reflection coefficients should be satisfied, 
the existence of edge states of the PC is equivalent to that a harmonic TM field can subsist in the closed waveguide.
From Eqs.\eqref{eq-GammaDef} and \eqref{eq-rPC}, we obtain the impedance match condition for the subsistence of waves inside the waveguide%the existence of an edge state requires $rr_L=1$, which can be written as 
\begin{equation}\label{impedance match}
    r(\phi_y)\,r_L(\phi_y)=1\quad \Leftrightarrow\quad \varphi(\phi_y)=-\varphi_L(\phi_y)  \bmod  2\pi .
\end{equation}

For a fixed frequency in the mobility gap, the schematics of reflection phases of the PC with trivial and nontrivial winding numbers are shown by the blue curves in Figs. \ref{fig-s5}(a) and (b), respectively. % to explain the relation between the winding number of the reflection phase,  $\tilde C_s$, and the number of topologically protected gapless edge states, 
And the red dashed lines exemplify
%We assume that the lead is closed by any time-reversal ($\mathcal{T}$) invariant hard boundary (to ensure that $\tilde{C}_{\rm{s}}=0$ for the lead), meaning that $\varphi_L$ can be any function satisfying $\varphi_L(\phi_y)=\varphi_L(-\phi_y)$. In Fig. \ref{fig-s5}, we plot 
 a constant reflection phase  $-\varphi_L(\phi_y)\equiv0.5\pi$ for the TRS impedance boundary. % as an example. 
%Figures \ref{fig-s5}(a) and (b) illustrate the trivial and nontrivial reflection phase windings of the PC (blue curves) with $ \tilde{C}_{\rm{s}}=0$ and $ \tilde{C}_{\rm{s}}=3$, respectively. 
The intersections between the blue and red lines represent the solutions of the impedance match condition ~\eqref{impedance match} and hence determine the number and loci of edge states at that frequency. For the trivial case, the absence of intersection indicates that no edge states are guaranteed to exist inside the bandgap. 
% However, we know that slight deformations of the PC can change the diagrams of $\varphi (\phi_y)$, indicating that the two solutions can be removed adiabatically. 
% Therefore, inside the bandgap of PCs with $ \tilde{C}_{\rm{s}}=0 $, there exist $0$ topologically protected edge states.
In contrast, the intersections of the curves in Fig. \ref{fig-s5} (b) demonstrate that the number of TM (TE) edge states is at least as many as the winding number $\tilde{C}_{\rm{s}}$ of the PC's reflection phase at any frequency inside the bandgap. % indicating the number of topologically protected gapless edge states for the PC is just determined by the winding number of the reflection phase $\tilde{C}_s$. 
% In general, the lead can be closed by any time-reversal ($\mathcal{T}$) invariant boundary condition (to ensure that $\tilde{C}_{\rm{s}}=0$ for the boundary), meaning that $\varphi_L$ can be any function (as shown by the red dashed line) satisfying $\varphi_L(\phi_y)=\varphi_L(-\phi_y)$. 

In summary, we can conclude that {\color{Blue} the nontrivial winding of the reflection phase ($\tilde{C}_s\neq0$) directly indicates the PC can support $|\tilde{C}_s|$ pairs of gapless helical edge states}. Hence, according to the bulk-edge correspondence between edge states and bulk QSH-Chern numbers, the topological classes of the reflection coefficient $r(\phi_y)$ can one-to-one map to the bulk QSH phases, \textit{i.e.}, $\tilde{C}_s\equiv C_s$.%, according to the bulk-edge correspondence.

% there must exist at least $|\tilde{C}_{\rm{s}}|$ solutions satisfying $\varphi (\omega, \phi_y)=-\varphi_L \bmod 2\pi$  for $\phi_y \in (-\pi,\pi]$ at any frequency inside the bandgap of a PC with $\tilde{C}_{\rm{s}}$.
% And according to the bulk-edge correspondence of the Chern insulator, we know that $\tilde{C}_{\rm{s}}$ is exactly the QSH-Chern number of the PC. 

\subsection{Numerical results of classifying QSH phases in disordered PCs}
% \textit{Scattering approach for classifying QSH phases in disordered PCs.-} 

\begin{figure}[t!]
\includegraphics[width=0.75\textwidth]{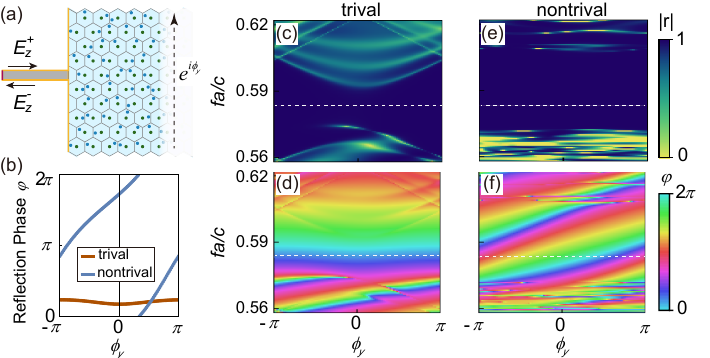}
\caption{\label{fig-3}
 (a) The schematic for calculating reflection coefficient of the TM wave. The upper and lower boundaries of the PC (size: $18a \times 15l$) are set as twisted periodic boundary condition with an additional phase factor $e^{i\phi_y}$, and the right boundary (not shown) is set as scattering boundary condition. The orange boundaries are PMC for TM waves, and the left boundary (red) of the lead (gray) is set as a plane-wave port with $E_z$-polarization. 
(b) Reflection phases along the white dashed lines in (d) and (f) at $fa/c=0.583$ in the trivial ($\theta_d=20\degree$) and nontrivial ($\theta_d=220\degree$) gaps.
 (c-f) Reflection magnitude and phase of $r(\phi_y,f)$ for (c-d) $\theta_d=20\degree$ and for (e-f) $\theta_d=220\degree$.
 }
\end{figure}

%Now let us turn to the spin-decoupled disordered PCs in Fig. 2(a) of the main text. Fig.~\ref{fig-3} (a) shows the calculation setup in COMSOL.
We have numerically simulated the reflection process inside the composite structure of the disordered PC with $\mathcal{T}_f$ and $M_z$ symmetries and the waveguide lead as shown in Fig.~\ref{fig-3}(a).
In Fig.~\ref{fig-3}(c-f), we plot the magnitude $|r(\phi_y,\omega)|$ and phase $\varphi(\phi_y,\omega)$ of the reflection coefficient for two PCs with weak ($\theta_d=20^\circ$) and strong ($\theta_d=220^\circ$) disorders, respectively. %norm $|r(\phi_y)|=1$ in the frequency range $fa/c\in (5.76,5.9)$ is near $1$, 
The frequency ranges of $|r(\phi_y)|=1$ in Fig.~\ref{fig-3}(c,e) denote the mobility gaps of the PCs.

In Fig.~\ref{fig-3}(b), the trivial and nontrivial windings of the reflection phases are exemplified at a given frequency inside two different gapped phases (see the dashed lines in Fig.~\ref{fig-3}(c,d) and (e,f)). 
According to the color map of the reflection phase in Fig.~\ref{fig-3}(d), the winding number of $\varphi(\phi_y)$  vanishes for all frequencies inside the bandgap, which verifies the weakly-disordered PC is a trivial insulator.
In comparison,  Fig.~\ref{fig-3}(f) shows that 
the winding number of the reflection phase is quantized to $\tilde{C}_s=1$ in the whole mobility gap of the strongly disordered PC, confirming that the TAI phase after disorder-induced transition in Fig. 2(a) of the main text carries a nontrivial QSH-Chern index.

% If the incident wave contains a mixture of TM and TE modes, the spins of reflected and incident waves could be different, except that their $s_z$ spin components keep identical due to the $U_s$ symmetry. Thus, after the twisted boundary evolving over a period, the reflected spin will rotate about the fixed $s_z$ axis an even number of times $2\tilde{C}_s$, offering an equivalent route to sorting QSH phases.

\section{Spin reflection approach for extracting the $\mathbb{Z}_2$ topological invariant of spin-coupled photonic crystals}

\subsection{Construction of the reflection matrix of the spin-coupled PCs}

\begin{figure}[h!]
\includegraphics[width=0.8\textwidth]{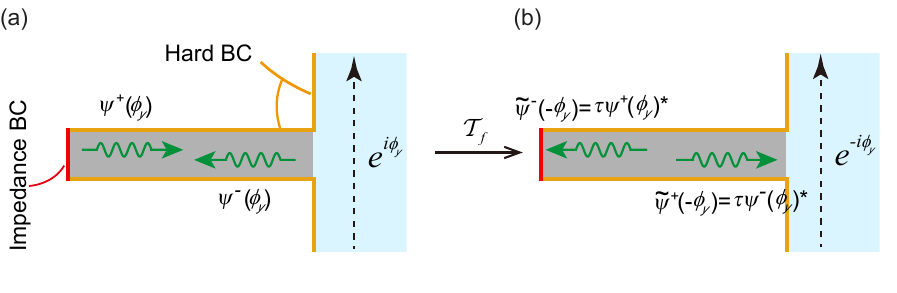}
\caption{\label{fig-s6}(a) The schematic of the reflection process inside the waveguide lead. The system (b) is obtained by operating the $\mathcal{T}_f$ reversal on the system (a). If the left open end of the waveguide is sealed by an impedance boundary (red), the blocked waveguide forms a closed trivial boundary condition for the PC.
}
\end{figure}

For a $\mathcal{T}_f$-symmetric PC without $M_z$ symmetry, we use the same waveguide lead to detect the wave reflection from the disordered PC as the $M_z$-symmetric case. We still require that the waveguide is $\mathcal{T}_f$-symmetric and supports only one TE mode and one TM mode simultaneously per propagation direction in the frequency range concerned, which can be achieved by encasing the waveguide by two thin layers (yellow in Fig.~\ref{fig-s6}) with $\tensor{\varepsilon}/\varepsilon_0=\tensor{\mu}/\mu_0=\mathrm{diag}(-s,-s,1)$ where $s\gg1$ and in COMSOL we set $s=100$. 
%the TE and TM modes are coupled, and the topological property is contained in a $2\times 2$ reflection matrix instead of the reflection coefficient. The setup for building the reflection matrix is shown in Fig. \ref{fig-s6} (a), which is the same as Fig. \ref{fig-s4} (a) except that we replace the PMC boundary condition with a $\mathcal T_f$-symmetric hard boundary (orange). 
For the TE (TM) waveguide mode, this encased boundary serves as a PEC (PMC). %, implying that the lead can support TM and TE plane wave modes simultaneously. 
Therefore, in this $\mathcal{T}_f$-symmetric dual-mode waveguide, the waves can be written as the superposition of $\psi^+ e^{ik x}$ and $\psi^- e^{-ik x}$ with $\psi^\pm = (E_z^\pm,\eta_0H_z^\pm)^\intercal$ denoting the amplitude of the TM and TE modes in a two-component spinor form. 
By virtue of the absence of $M_z$ symmetry, the TE and TM modes will be coupled by the PC after reflection. 
As a result, we have to use a $2\times 2$ reflection matrix $R(\phi_y,\omega)$ to describe the wave reflection by the PC inside the waveguide at a given frequency $\omega$ and a phase angle $\phi_y$ of the twisted boundary condition,% the PC refering to $\psi^\pm$ as
\begin{equation}\label{eq-reflectionmatrix}
    \psi^{-}\left(\phi_{y},\omega\right)=R\left( \phi_{y},\omega\right) \psi^{+}\left( \phi_{y},\omega\right),
\end{equation}
where $\psi^{+}$ and $\psi^{-}$ represent the amplitudes of the incident and reflected waves, respectively. %Hereinafter, $R(\phi_y)$ also indicates the reflectoin matrix at a certain frequency, without otherwise stated. %and $\phi_y$ is still the phase angle of the twist boundary condition.

The reflection matrix of the PC can be retrieved from the full-wave simulations using COMSOL, %this hard boundary condition can be approximated by a thin layer with $\tensor{\varepsilon}/\varepsilon_0=\tensor{\mu}/\mu_0=\mathrm{diag}(-100,-100,1)$. 
where %we replace the impedance BC with plane-wave ports as line source to excite the edge modes in the bandgap of the PC. To retrieve the $2 \times 2$ reflection matrix, 
we separately excited two incident waves with different polarizations, $\psi_1^+,\psi_2^+$, and probed the corresponding reflected waves, $\psi_1^-,\psi_2^-$. Then the reflection matrix can be extracted from
\begin{equation}
    \left[ {{\psi _1}^ -{\psi _2}^ - } \right] = R\cdot\left[ {{\psi _1}^ + {\psi _2}^ + } \right] \quad\Rightarrow\quad R = \left[ {{\psi _1}^ - {\psi _2}^ - } \right] \cdot {\left[ {{\psi _1}^ + {\psi _2}^ + } \right]^{ - 1}},
\end{equation}
where $\left[ {{\psi _1}^+{\psi _2}^+} \right]$ and $\left[ {{\psi _1}^ -{\psi _2}^ - } \right]$ represent the matrices spanned by the two linearly independent incident spinors and by the two reflected spinors, respectively. The numerical results of the reflection matrix of the spin-coupled disordered PCs are shown in Fig. 3 of the main text.

\subsection{$\mathcal{T}_f$ symmetry restriction to the reflection matrix $R$}

Now let us examine the constraints imposed on the reflection matrix by $\mathcal{T}_f$ symmetry in the bandgap of the PC. As shown in Fig. \ref{fig-s6}, after operating $\mathcal{T}_f$ on the system in Fig. \ref{fig-s6} (a), we obtain the system in Fig. \ref{fig-s6} (b), where $\mathcal{T}_f$ reverses the propagation directions of the forward-propagating and backward-propagating spinors inside the waveguide and also flips the angle of the twisted boundary condition. Given a pair of incident and reflected spinors, $\psi^{\pm}(\phi_y)$, of the original system with the twist angle $\phi_y$, the new incident and reflected spinors at $-\phi_y$ after $\mathcal{T}_f$ reversal, denoted by  $\tilde \psi^{\pm}(-\phi_y)$, can be expressed as 
\begin{equation}\label{relation of spinors by Tf}
    \tilde \psi^{\mp}(-\phi_y) = \mathcal{T}_f\Big(\psi^{\pm}\left(\phi_{y}\right)\Big)=\tau \psi^{\pm}\left(\phi_{y}\right)^*,
\end{equation}
where $\tau =\tau^*= -i\sigma_y$ denotes the unitary part of $\mathcal{T}_f$ acting on the two-component spinors.
The $\mathcal{T}_f$ symmetry of the system ensures that the reflection matrix remains unchanged after $\mathcal{T}_f$ reversal, \textit{i.e.},
\begin{align}
    \psi^{-}(\phi_y)&=R(\phi_y)\psi^{+}(\phi_y),\label{before Tf}\\
    \tilde{\psi}^{-}(\phi_y)&=R(\phi_y)\tilde{\psi}^{+}(\phi_y).\label{after Tf}
\end{align}
Note that the validity of Eq.~\eqref{after Tf} relies on the vanishing of the transmitted wave in the original system, and hence it is only true in the mobility gap of the PC. 
Plugging the relation~\eqref{relation of spinors by Tf} into Eq.~\eqref{after Tf}, we have
%Operating $\mathcal{T}_f$ on both sides of Eq.~\eqref{eq-reflectionmatrix} in the real space yields 
\begin{equation*}
\begin{aligned}
% {{\cal T}_f}{\psi ^ - }\left( {{\phi _y}} \right) &= {{\cal T}_f}R\left( {{\phi _y}} \right){{\cal T}_f}^{ - 1}{{\cal T}_f}{\psi ^ + }\left( {{\phi _y}} \right)\\
%  \Rightarrow \tau {\psi ^ - }{\left( {{\phi _y}} \right)^*} &= \tau R{\left( {{\phi _y}} \right)^*}{\tau ^{ - 1}}\tau {\psi ^ + }{\left( {{\phi _y}} \right)^*}\\
\tau{{ \psi }^+}\left( { - {\phi _y}} \right)^* &= R{\left( {{\phi _y}} \right)}{\tau}{{ \psi }^ - }\left( { - {\phi _y}} \right)^*\\
  \Rightarrow\quad {{ \psi }^+ }\left( { - {\phi _y}} \right) &= \tau^\dagger R{\left( {{\phi _y}} \right)^*}{\tau}{{ \psi }^ - }\left( { - {\phi _y}} \right).%\\
%  \Rightarrow R{\left( { - {\phi _y}} \right)^{ - 1}} &= \tau R{\left( {{\phi _y}} \right)^*}{\tau ^{ - 1}}\\
%  \Rightarrow R{\left( { - {\phi _y}} \right)^{ \dagger}}  &= {\sigma _y}R{\left( {{\phi _y}} \right)^*}{\sigma _y}.
\end{aligned}
\end{equation*}
Comparing with Eq.~\eqref{before Tf}, we obtain the requirement of $R$ in the mobility gap of a $\mathcal{T}_f$-symmetric PC:
\begin{equation}\label{TRS-dagger}
     \tau^\dagger R(\phi_y)^*\tau=R(-\phi_y)^{-1}\quad\Rightarrow\quad
     \sigma_yR(\phi_y)^*\sigma_y=R(-\phi_y)^\dagger.
\end{equation}
In the last step, we have used the unitarity ($R^\dagger R=1$) of the reflection matrix in the mobility gap. Note that the $\mathcal{T}_f$-symmetric Wilson loop operator respects the same relation as Eq.~\eqref{TRS-dagger}.

%And the reflection matrix has $\mathcal{T}_f^\dagger$ symmetry~\cite{Kawabata-PRX-2019}, 
The symmetry relation in Eq.~\eqref{TRS-dagger}, known as TRS$^\dagger$ symmetry, was recently studied as a non-Hermitian generalization of TRS~\cite{Kawabata-PRX-2019}. It enforces the spectrum of $R$ is symmetric about the $\mathcal{T}_f$-invariant points at which the eigen reflection coefficients form the Kramers' doublets. These properties are inherited by the eigen reflection phases, \textit{i.e.}, the arguments of the eigenvalues of $R$. 
% Then we define a reflection phase matrix as 
% \begin{equation}
%     A=-i\ln R,
% \end{equation}
% which is Hermitian.
% In the momentum space, $\mathcal T_f=- i\sigma_y K (\mathbf k \rightarrow - \mathbf k)$, and we have
% \begin{equation}
% \begin{aligned}
% {{\cal T}_f}A\left( {{\phi _y}} \right){{\cal T}_f}^{ - 1}=\sigma_y A(-\phi_y)^*\sigma_y =i{\sigma _y}\ln R{\left( { - {\phi _y}} \right)^*}{\sigma _y} = i\ln R{\left( {{\phi _y}} \right)^\dag } =  - i\ln R\left( {{\phi _y}} \right)=A(\phi_y)
% \end{aligned}
% \end{equation}
% implying $A(\phi_y)$ is $\mathcal{T}_f$-symmetric. 
% Supposing that $A$ has a eigenvalue $\varphi$, we have
% \begin{equation}
% \begin{aligned}
% A({\phi _y}){\psi _i}({\phi _y}) &= {\varphi _i}({\phi _y}){\psi _i}({\phi _y})\\
%  \to {{\cal T}_f}A({\phi _y}){\psi _i}({\phi _y}) &= {\varphi _i}( - {\phi _y}){{\cal T}_f}{\psi _i}({\phi _y})\\
%  \to A({\phi _y}){{\cal T}_f}{\psi _i}({\phi _y}) &= {\varphi _i}( - {\phi _y}){{\cal T}_f}{\psi _i}({\phi _y})
% \end{aligned}
% \end{equation}
% implying that $\varphi_i(-\phi_y)$ is an eigenvalue of $A(\phi_y)$ corresponding to the eigenvector $\mathcal{T}_f \psi_i(\phi_y)$. 
Therefore, the eigen reflection phases are symmetric about the $\mathcal{T}_f$-invariant points ($\phi_0=0, \pi$) and have Kramers' degeneracies, which are demonstrated by the numerical results shown in Fig. 3(b) of the main text.

\subsection{$\mathbb{Z}_2$ classification of the reflection matrix}

In the mobility gap of the PC, the reflection matrix is a $U(2)$ matrix and can be expressed with Pauli matrices as
\begin{equation}
R = e^{i q} \exp \left(i \alpha \vec n \cdot \vec \sigma\right) = e^{i q}\left(\cos \alpha \sigma_{0}+i \sin \alpha \vec{n} \cdot \vec{\sigma} \right)
\end{equation}
with the eigenvalues $r_{1,2}=\exp(i(q\pm\alpha))$, where $\vec n$ is a unit vector. Therefore, the eigen reflection phases are $\varphi_{1,2}=q\pm \alpha$. 
And because of the Kramers' degeneracy at the $\mathcal{T}_f$-invariant point $\phi_0$, we have 
\begin{equation}\label{kramers degeneracy}
    \alpha(\phi_0) = 0\ \text{or}\ \pi\quad\Rightarrow\quad R (\phi_0) =e^{iq} \sigma_0 \in U(1).
\end{equation}
Considering the evolution of $R$ in a cycle of $\phi_y\in[-\pi,\pi)$, since $R(\phi_y)$ and $R(-\phi_y)$ are related by Eq.~\eqref{TRS-dagger} due to $\mathcal{T}_f$ symmetry, a half-cycle of $\phi_y\in D^1=[0,\pi]$ ($D^1$ refers to a 1D disk, \textit{i.e.}, a line segment) is enough to determine the topology of the evolution of $R(\phi_y)$ in a whole period. Here, we adopt the concept of \textbf{\color{Blue}relative homotopy}~\cite{hatcher2005algebraic,sun2018Conversion} to derive the topological classification of the reflection matrix. The evolution of $R$ in a half-cycle under the constraint of $\mathcal{T}_f$ symmetry can be characterized by the following map
\begin{equation}
    R:\left\{\begin{aligned}
    D^1\ &\rightarrow\ U(2)\\
    \partial D^1\ &\rightarrow\ U(1)
    %\partial D^1\ni \phi_0\ &\rightarrow\ e^{iq_0}\sigma_0\in U(1)
    \end{aligned}\ ,\right.
\end{equation}
where the first line of the map indicates the reflection matrix $R(\phi_y)$ at a general point of $D^1$ is a $U(2)$ matrix, the second line indicates $R$ at the terminals of $D^1$ (\textit{i.e.}, the $\mathcal{T}_f$-invariant points) are restricted to the $U(1)$ subgroup. Two evolutions of the reflection matrix in a half-cycle, labeled as $R_1$ and $R_2$, are topologically equivalent, provided that $R_1$ can be continuously deformed into $R_2$ without breaking the $\mathcal{T}_f$ symmetry, which is mathematically described by the continuous map, so-called homotopy, relative to the subspace $U(1)$,  $\mathcal{R}:\ D^1\times[0,1]\rightarrow U(2)$,  such that $\forall\ \phi_y\in D^1$ and $t\in[0,1]$:%~\cite{hatcher2005algebraic,sun2018Conversion}
\begin{equation}
    \mathcal{R}(\phi_y,0)=R_1(\phi_y),\quad \mathcal{R}(\phi_y,1)=R_2(\phi_y),\quad
    \mathcal{R}(\partial D^1,t)\in U(1).
\end{equation}
All the trajectories of reflection matrices $R(\phi_y)$ that can be continuously deformed to each other in this way are in the same homotopy equivalence class of $U(2)$ relative to $U(1)$. {\color{Blue} And the collection of all such homotopy equivalence classes forms the relative homotopy group~\cite{hatcher2005algebraic,sun2018Conversion}, $\pi_1(U(2),U(1))$, which characterizes the topological classification of the reflection matrix' evolutions under the constraint of $\mathcal{T}_f$ symmetry.} Note that since $U(1)$ is path-connected, the relative homotopy group $\pi_1(U(2),U(1))$ is irrelevant to the choice of basepoint.

To calculate the relative homotopy group, we consider the following short exact sequence:
\begin{equation}\label{short sequence}
    U(1)\ \stackrel{f}{\longrightarrow}\ U(2)\ \stackrel{p}{\longrightarrow}\ SU(2)/\mathbb{Z}_2\cong SO(3)
\end{equation}
where $f$ and $p$ refer to two group homomorphisms defined as $f:\,U(1)\ni e^{iq}\ \rightarrow\ e^{iq}\sigma_0\in U(2)$ and 
\begin{equation}\label{projection}
    p:\ U(2)\ni e^{iq}\exp[i\alpha\vec{n}\cdot\vec{\sigma}]\ \longrightarrow\ \underset{\displaystyle\in\, SU(2)/\mathbb{Z}_2}{[\pm e^{i\alpha\vec{n}\cdot\vec{\sigma}}]}\ \cong\ \exp[-2\alpha\vec{n}\cdot\vec{\mathfrak{L}}]\in SO(3),
\end{equation}
with $\cong$ denoting the isomorphism between $SU(2)/\mathbb{Z}_2$ and $SO(3)$, and the skew-symmetric matrix
% for the pseudo-spin vector in the 3D real space $\mathbb R^3$ with the rotation angle $\Theta = -2 \alpha$ around the axis $\vec n$, and 
\begin{equation}
    \vec n \cdot \vec{\mathfrak{L}} = \left[ {\begin{array}{*{20}{c}}
0&{ - {n_z}}&{{n_y}}\\
{{n_z}}&0&{ - {n_x}}\\
{ - {n_y}}&{{n_x}}&0
\end{array}} \right]
\end{equation}
denotes the generator of the $SO(3)$ rotation.
It is obvious that the short sequence~\eqref{short sequence} is exact, which means the image of $f$ equals the kernel of $p$:  $\mathrm{im}(f)=\{e^{iq}\sigma_0| q\in [-\pi,\pi)\}=\mathrm{ker}(p)$. Furthermore, since $\forall\ G=\exp[\Theta\vec{n}\cdot\vec{\mathfrak{L}}]\in SO(3)$ the preimage $p^{-1}(G)=\{e^{iq}\exp[\frac{-1}{2}\Theta\vec{n}\cdot\vec{\sigma}]:q\in[-\pi,\pi)\}\cong U(1)$, the homomorphism $p: U(2)\, \rightarrow\, SO(3)$ forms a principal $U(1)$-bundle with $U(2)$ and $SO(3)$ serving as the bundle space and the base space, respectively, and $p^{-1}(G)\cong U(1)$ being the fiber at $G$. Hence, the sequence~\eqref{short sequence} is a Serre fibration respecting homotopy lifting property. 

\begin{figure}[t!]
\includegraphics[width=0.8\textwidth]{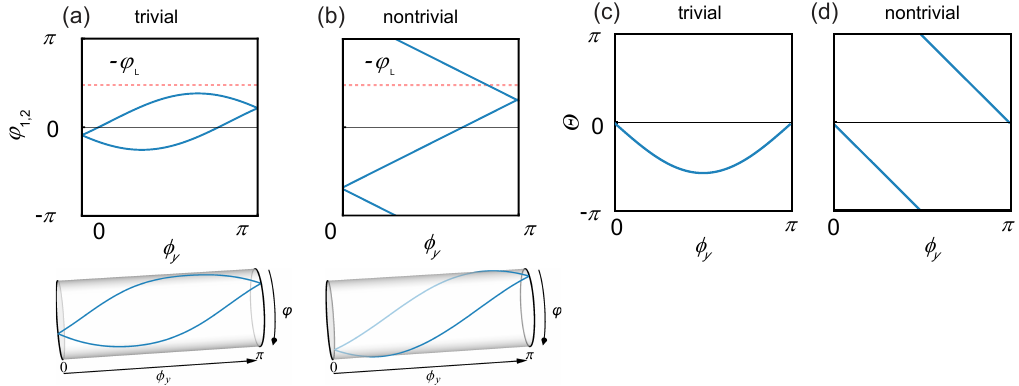}
\caption{\label{fig-s7}
(a,b) The schematics of the (a) trivial and (b) nontrival evolutions of the eigen reflection phase $\varphi_{1,2}(\phi_y)$ of the reflection matrix $R(\phi_y)$ over a half-period of $\phi_y$.
(c,d) The (c) trivial and (d) nontrivial evolutions of the relative phase $\Theta(\phi_y)=\varphi_{2}(\phi_y)-\varphi_{1}(\phi_y)$ (\textit{i.e.} the spin rotation angle) corresponding to (a) and (b).
}
\end{figure}

According to Theorem 4.41 in Ref.~\cite{hatcher2005algebraic}, for a Serre fibration $F\,\rightarrow E\,\rightarrow\,B$, the $n$-homotopy group of the bundle space $E$ relative to the fiber $F$ is always isomorphism to the $n$-homotopy group of the base space $B$, \textit{i.e.}, $\pi_n(E,F)=\pi_n(B)$. Applying the theorem to our case, we obtain %the  topological classification of $R$ bounded by $\mathcal{T}_f$ symmetry given by the relative homotopy group:
\begin{equation}\label{relative homotopy}
    \pi_1(U(2),U(1)) \stackrel{p_*}{=\joinrel=}\pi_1(SU(2)/\mathbb{Z}_2)= \pi_1(SO(3)) =\mathbb Z_2.
\end{equation}
Therefore, the evolutions of $R$ bounded by $\mathcal{T}_f$ symmetry are $\mathbb{Z}_2$ classified, labeled by the $\mathbb{Z}_2$ index $\tilde{\nu}=0\ \text{or}\ 1$. In Fig. \ref{fig-s7} (a-b), we plot the schematic spectra of the eigen reflection phases of $R(\phi_y)$ for trivial ($\tilde{\nu}=0$) and nontrivial ($\tilde{\nu}=1$) cases of the $\mathbb Z_2$ classification, which exhibits the same characteristic as the spectral flow of the $\mathcal{T}_f$-symmetric Wilson loop operator~\cite{yu2011Equivalent}. Using the same criterion for the Wilson loop, we know that the parity of the number of crossing points between the bands of the eigen reflection phases and an arbitrary reference line of $\varphi=\mathrm{const.}$ (red dashed lines in Fig.~\ref{fig-s7}) determines the $\mathbb{Z}_2$ index $\tilde{\nu}$.

% From the above result, we see that the $\mathbb Z_2$ topology is completely determined by  $\alpha$ in the $SU(2)/\mathbb{Z}_2$ part of the $R$ matrix, $[\pm\exp { \left(i \alpha \vec n \cdot \vec \sigma\right)}]$, but is irrelevant to the $U(1)$ part, $e^{iq}$. 

Since the continuous deformation of $\varphi_{1,2} = q \pm \alpha \rightarrow \varphi_{1,2}=\pm\alpha$, i.e. $q(\phi_y)\rightarrow0$, does not change the connectivity of the two bands of $\varphi_1$, $\varphi_2$, the $\mathbb{Z}_2$ index of the reflection matrix can be captured by the relative winding number of the two eigen reflection phases over a half-period of $\phi_y$
\begin{equation}
    \tilde{\nu}=\left[\frac{1}{2\pi}\int_0^\pi d\phi_y \frac{\partial\Theta (\phi_y)}{\partial\phi_y}\right]\bmod 2=\left[\frac{-1}{\pi}\int_0^\pi d\phi_y \frac{\partial\alpha (\phi_y)}{\partial\phi_y}\right]\bmod 2= 0\;\text{or}\;1,
\label{Z2-invariant}
\end{equation}
where $\Theta=\varphi_2-\varphi_1=-2\alpha$ gives the relative phase of the two eigen reflection coefficients. The quantization of the relative winding number is due to the Kramers' degeneracy at $\mathcal{T}_f$ invariant points (Eq.~\eqref{kramers degeneracy}). In Fig. \ref{fig-s7} (c-d), we plot the schematic of the relative phases for the trivial and nontrivial cases. The second equality in Eq.~\eqref{Z2-invariant} also suggests that, for a smooth gauge of $\alpha(\phi_y)$, the trivial and nontrivial $\mathbb{Z}_2$ windings can be efficiently identified by the values of $\alpha$ at $\mathcal{T}_f$-invariant points 
\begin{equation}\label{Z2-invariant 2}
\exp(i\tilde{\nu} \pi) = {e^{i\alpha (0)}}{e^{i\alpha (\pi )}} = \cos \alpha (0)\cos \alpha (\pi ) =  \pm 1.
\end{equation}
%Therefore, the $\mathbb Z_2$ index is determined by the signs of $\cos \alpha(0) $ and $\cos \alpha(\pi)$ in any smooth gauge.
Hence, we see that the $\mathbb Z_2$ topology is completely determined by  $\alpha$ in the $SU(2)/\mathbb{Z}_2$ part of the $R$ matrix, $[\pm\exp { \left(i \alpha \vec n \cdot \vec \sigma\right)}]$, but is irrelevant to the $U(1)$ part, $e^{iq}$, which is consistent with the homotopy group obtained in Eq.~\eqref{relative homotopy}. In contrast to the $\mathbb{Z}$ classified winding number (Eq.~\eqref{reflection phase winding}) for the QSH systems with $s_z$ spin conservation, the winding of relative phase in Eq.~\eqref{Z2-invariant} becomes meaningless when $\tilde{\nu}>1$ in the absence of $U(1)$ spin rotation symmetry, which can be understood from the following reason. $\tilde{\nu}>1$ indicates the bands of $\varphi_1$, $\varphi_2$ intersect at some points away from $\mathcal{T}_f$-invariant points. Without the protection of spin rotation symmetry, these degeneracies are unstable and can be easily lifted by perturbations, then the relative winding number will reduce to either 0 or 1.

% On the other hand, we can define a relative phase $ \Theta \equiv \varphi_2-\varphi_1$
% to represent the difference between the two eigen reflection phases. In Fig. \ref{fig-s7} (c-d), we plot the schematic of the relative phases for the trivial and nontrivial cases. 
% Due to the Kramers' degeneracy of spectra of $A(\phi_y)$, 
% we know that at the $\mathcal{T}_f$-invariant  point
% \begin{equation}
%   \Theta (\phi_0) =  -2\alpha (\phi_0) = (0 \;\rm{ or }\; 2\pi)\bmod 4\pi,
% \label{alpha}
% \end{equation}
% and its winding over a half-period ($\phi_y\in[0,\pi]$)
% \begin{equation}
%     \tilde{\nu}=\left[\frac{1}{2\pi}\int_0^\pi d\phi_y \frac{\partial\Theta (\phi_y)}{\partial\phi_y}\right]\bmod 2= 0\;\text{or}\;1,
% \label{Z2-invariant}
% \end{equation}
% can be defined as a $\mathbb{Z}_2$ topological invariant. 

\subsection{Physical meaning of the $\mathbb{Z}_2$ index}
%In the above section, we proved that the winding number of the relative phase $\Theta$ can be used to characterize the $\mathbb{Z}_2$ index. 
In this section, we discuss the physical meaning of the $\mathbb{Z}_2$ index of the reflection matrix from the perspective of spin reflection. Consider a pair of incident and reflected waves $\psi_\mathrm{in}$ and $\psi_\mathrm{r}$ inside the lead.  According to the Rodrigues' rotation formula, the spin of the backward wave reflected from the PC has the relationship to the spin of the incident wave as
% \begin{equation}
%     {{\vec s}_r} = \langle {\psi _\text{r}}|\vec \sigma |{\psi _{\text r}}\rangle  = \langle {\psi _\text{in}}|{R^\dag }\vec \sigma R|{\psi _\text{in}}\rangle  = G\left( {\Theta ,\vec n} \right)\langle {\psi _\text{in}}|\vec \sigma |{\psi _\text{in}}\rangle  = G\left( {\Theta ,\vec n} \right){{\vec s}_\text{in}}= \exp\big[i\Theta\, \vec{n}\cdot\vec{L}\big]\vec{s}_\mathrm{in}
% \end{equation}
\begin{equation}
\begin{aligned}\label{spin rotation}
{{\vec s}_\mathrm{r}} &= \langle {\psi _\mathrm{r}}|\vec \sigma |{\psi _\mathrm{r}}\rangle \\
 &= \langle {\psi _\mathrm{in}}|{R^\dag }\vec \sigma R|{\psi _\mathrm{in}}\rangle \\
 &= \langle {\psi _\mathrm{in}}|{e^{ - i\alpha \vec n \cdot \vec \sigma }}\vec \sigma {e^{i\alpha \vec n \cdot \vec \sigma }}|{\psi _\mathrm{in}}\rangle \\
  &= \langle {\psi _\mathrm{in}}|\vec \sigma \cos (-2\alpha ) + \vec n \times \vec \sigma \;\sin ( - 2\alpha ) + \vec n\;\vec n \cdot \vec \sigma \;(1 - \cos (-2\alpha ))|{\psi _\mathrm{in}}\rangle \\
 &=\left[ {\cos \Theta  + \sin \Theta \vec n \times \; + (1 - \cos \Theta )\vec n\;\vec n \cdot } \right] \langle {\psi _\mathrm{in}}|\vec \sigma |{\psi _\mathrm{in}}\rangle\\
 &= \exp \left[ {\Theta {\mkern 1mu} \vec n \cdot \vec{\mathfrak{L} } } \right]\langle {\psi _\mathrm{in}}|\vec \sigma |{\psi _\mathrm{in}}\rangle = G\left( {\Theta ,\vec n} \right){{\vec s}_\mathrm{in}},
\end{aligned}
\end{equation}
where $G\left( {\Theta ,\vec n} \right) = \exp \left[ {\Theta {\mkern 1mu} \vec n \cdot \vec{\mathfrak{L}} } \right]\in SO(3)$ represents the spin reflection matrix that illustrates the relative rotation between the initial and reflected spins about the axis $\vec n$ through an angle $\Theta = -2 \alpha$ counterclockwise.
Comparing with the homomorphism in Eq.~\eqref{projection}, we find that the spin reflection matrix gives the projection of the reflection matrix from $U(2)$ to $SO(3)$, $p(R)=G(\Theta,\vec{n})$, and the spin rotation angle is exactly the phase difference  $\Theta=\varphi_2-\varphi_1$ of the two eigenvalues of the reflection matrix $R$.  In this way, {\color{Blue}the $\mathbb{Z}_2$ index, $\tilde{\nu}=0$ or $1$, can be endowed with a graceful physical intuition, \textit{i.e.}, it characterizes the two possible quantized values, $0$ and $2\pi$, of the accumulated rotation angle of the spin reflection matrix in a half-cycle of $\phi_y$}, which is in accord with the result of Eq.~\eqref{relative homotopy} that $\pi_1(SO(3))=\mathbb{Z}_2$ determines the topological classification.

\begin{figure}[t!]
\includegraphics[width=0.6\textwidth]{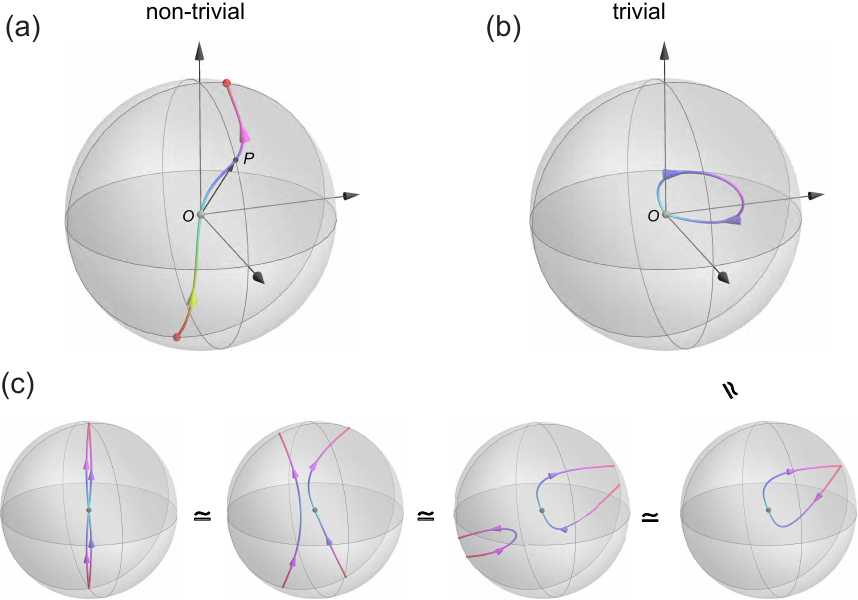}
\caption{\label{fig-s8}
The schematics of the (a) nontrivial and (b) trivial evolution loops (colored curves) of  $SO(3)$ spin rotation. %The gray ball with radius $\pi$ and identifying each pair of antipodal points on its surface as a single point denotes a 3D real projective space $\mat$
The solid ball with radius $\pi$ (identifying each pair of antipodal points on its surface as a single point) represents the manifold of the  $SO(3)$ group.
Each points on the trajectories is given by the vector $\Theta(\phi_y)\vec{n}(\phi_y)$ with its direction and length denoting the rotation axis $\vec{n}$ and the rotation angle $\Theta(\phi_y)$, respectively, hence bijectively corresponding to the spin reflection matrix $G(\Theta,  \vec{n})$. 
%(a) For the trivial case, the trajectory over a half-period $\phi \in [0,\pi]$ are inside the ball, implying the winding of $\Theta$ is $0$. (b) For the nontrivial case, the trajectory over a half-period $\phi \in [0,\pi]$ go through the ball, implying the winding of $\Theta$ is $1$.
(c) The process of continuously deforming a double-round spin rotation into a trivial nonwinding loop.
}
\end{figure}

We should note that the rotation axis $\vec{n}(\omega,\phi_y)$ is not fixed in general, but can change with frequency and twist angle. As shown in Fig. \ref{fig-s8}, we plot the schematics of the homotopically trivial and nontrivial loops of the spin rotation matrix in $SO(3)$, where the solid ball with radius $\pi$ denotes a 3D real projective space $\mathbb{R}P^3\cong SO(3)$ with identifying any pair of antipodal points on the sphere as a single point. Given a point in the ball $\vec{g}=\Theta\vec{n}$, its direction  and radius represent the rotation axis $\vec{n}$ and rotation angle $\Theta$ of $G(\Theta,\vec{n})$, respectively. Since $\Theta=0$ at $\mathcal{T}_f$-invariant points, the loops of $G(\Theta,\vec{n})$ should always start and end at the center of the ball. The non-straight trajectories imply that the rotation axis is not fixed. For the trivial case in Fig.~\ref{fig-s8}(b), after a half-cycle over $\phi_y \in [0,\pi]$, the trajectory forms a closed loop inside the ball. In comparison, for the nontrivial case in Fig.~\ref{fig-s8}(a), the trajectory passes through the solid ball from one point on the surface to its antipodal point. 
Following this graphical representation, it is easy to see why a double-round spin rotation is topologically trivial when the rotation axis can be freely tuned. Fig.~\ref{fig-s8}(c) sketches this deformation process. Initially, a double-round rotation about a fixed axis is shown by the double-line passing through the center twice. By  changing the rotation axis continuously, the double-line eventually converts to a trivial loop without touching the ball's surface.

\subsection{Correspondence between the $\mathbb{Z}_2$ index of reflection matrix and helical edge states}

Now let us relate the winding number of the relative phase $\Theta$ to the number of helical edge states.
Similar to the case with $M_z$ symmetry, we introduce an $\mathcal{T}_f$-symmetric impedance boundary to block the left open end of the waveguide lead, as depicted in Fig.~\ref{fig-s6}(a).  The reflection matrix of the impedance boundary is set as $R_L(\phi_y,\omega)$, which correlates the amplitudes of the left and right going spinors,
\begin{equation}\label{reflection matrix L}
\psi^+(\phi_y,\omega)=R_L(\phi_y,\omega)\psi^-(\phi_y,\omega).
\end{equation}
To guarantee the impedance boundary is  topologically trivial, we let $R_{L}=e^{i\varphi_L}\sigma_0$ be a constant (identical to Eq.~\eqref{eq-GammaDef} for the $M_z$ symmetric case). 
The existence of edges states confined in the waveguide requires that Eqs.~\eqref{eq-reflectionmatrix} and \eqref{reflection matrix L} are satisfied simultaneously, namely
\begin{gather}
    \psi^{+}(\phi_y)=R_{L} \psi^{-}(\phi_y)=e^{i \varphi_{L}} \psi^{-}(\phi_y) = e^{i \varphi_{L}}R(\phi_y) \psi^{+}(\phi_y),\nonumber\\
    \Rightarrow\quad R(\phi_y)\psi^{+}(\phi_y)=e^{-i\varphi_L}\psi^{+}(\phi_y).\label{criterion for helical edge states}
\end{gather}
%implying that 
% $e^{i \varphi_{\rm{I}}}R_R(\phi_y) = 1$ should have solutions. 
Therefore, {\color{Blue}an edge state appears at $\phi_y$ and $\omega$ if and only if
one of the eigenvalues of the reflection matrix $R(\phi_y,\omega)$ of the PC equals  $e^{-i\varphi_L}$, the inverse of the reflection coefficient of the impedance boundary}. %According to this criterion, a rigorous correspondence between the $\mathbb{Z}_2$ index of the reflection matrix and the existence of edge states can be established.
%As shown in Fig. \ref{fig-s6} (a), we impose a $T_f$-symmetric impedance boundary condition (red line) $R_L=e^{i\varphi_L } \sigma_0 $ to close the lead. For simplicity, we assume that $\varphi_L$ is a constant function of $\phi_y$, which are denoted by the red dashed lines in Fig. \ref{fig-s7}.
On the other hand, the $\mathbb{Z}_2$ index of the reflection matrix at a frequency $\omega$ inside the mobility gap specifies the parity of the number of eigen reflection phases at any given constant $\varphi=\varphi_L$ (the red dashed lines in Fig.~\ref{fig-s7} (a,b)) over a half-period of $\phi_y$, hence also indicating the parity of number of edge states in light of the criterion~\eqref{criterion for helical edge states}.  %$\varphi(\phi_y)=-\varphi_L$ over the half-period loop $\phi_y\in[0,\pi]$ has even number of solutions. %While for the nontrivial case, we can always find odd number of solutions $\varphi(\phi_y)=-\varphi_L$ over the half-period loop $\phi_y\in[0,\pi]$. 
Over the full-period of $\phi_y\in[-\pi,\pi]$, $\mathcal{T}_f$ symmetry forces the concurrence of the edge states at $\pm\phi_y$ forming helical edge state pairs.  Therefore, a rigorous correspondence between the $\mathbb{Z}_2$ index $\tilde{\nu}$ of the reflection matrix and the helical edge states  at any frequency inside the mobility gap can be achieved:
\begin{equation}
    \color{Blue}\tilde{\nu}= \text{parity of the number of helical edge state pairs}.
\end{equation}
According to the well-established bulk-edge correspondence between helical edge states and the Kane-Mele $\mathbb Z_2$ index of bulk states~\cite{kane2005z}, the topological invariant $\tilde v$ of the reflection matrix is identical to the Kane-Mele $\mathbb Z_2$ index.

\begin{figure}[t!]
\includegraphics[width=0.98\textwidth]{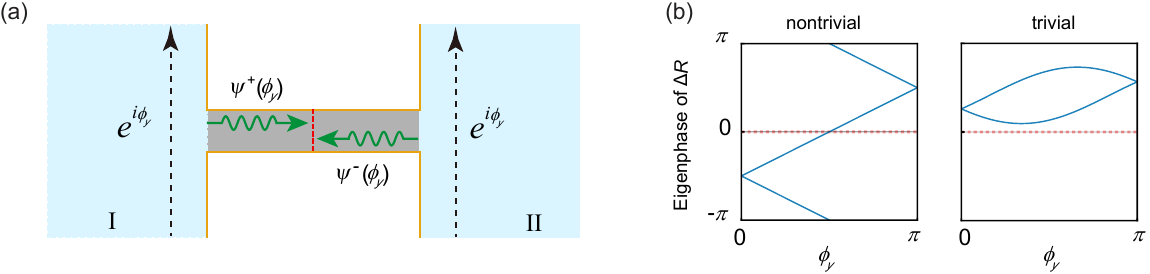}
\caption{\label{fig-interface}
(a) A specific domain wall between two different disordered PCs, I and II, formed by connecting them with a $\mathcal{T}_f$-symmetric waveguide. 
(b) The schematics of the eigenvalues' phases of the reflection matrix product $\Delta R = R_{\rm I} R_{\rm II}$, whose intersections with $\varphi=0$ (red dashed line) correspond to the interface states localized in the waveguide in (a). 
}
\end{figure}

Next, we go one step further to show the $\mathbb{Z}_2$ index $\tilde{\nu}$ of the reflection matrix can also determine the existence of helical interface states localized in-between two $\mathcal{T}_f$-symmetric PCs. As shown in Fig.~\ref{fig-interface}(a), we consider two disordered PCs, I and II, connected with a $\mathcal{T}_f$-symmetric waveguide, which serves as a specific domain wall between the two PCs. Supposing the reflection matrix of the left and right PCs are $R_{\rm I}(\phi_y,\omega)$ and $R_{\rm{II}}(\phi_y,\omega)$ with respect to a common reference plane at $x=0$ (red dashed line), the existence of an interface state $\psi=\psi^+e^{ikx}+\psi^{-}e^{-ikx}$ confined in the waveguide requires
\begin{gather}
        \psi^{+}=R_{\rm{I}}(\phi_y) \psi^{-},\quad \psi^{-}=R_{\rm{II}}(\phi_y) \psi^{+},\nonumber\\
        \Rightarrow\quad 
        R_{\rm{I}}(\phi_y)R_{\rm{II}}(\phi_y)\psi^{+}=\psi^{+},
\end{gather}
which means {\color{Blue}the existence of an interface state at $\phi_y$ is equivalent to $\Delta R(\phi_y)=R_{\rm{I}}(\phi_y)R_{\rm{II}}(\phi_y)$ has a unity eigenvalue $r_0=1$ and hence a zero eigen phase $\varphi_0=\arg(r_0)=0$}. It is easy to see that $\Delta R\in U(2)$ also respects the $\mathcal{T}_f$ symmetry restriction~\eqref{TRS-dagger}. Expressing $\Delta R=e^{i p}\exp(i\gamma\vec{m}\cdot\vec{\sigma})$, the topology of the evolution of $\Delta R$ is also $\mathbb{Z}_2$ classified and characterized by (in the same way as Eq.~\eqref{Z2-invariant 2}):
\begin{equation}\label{Delta Z2}
    \exp(i\Delta\tilde{\nu}\pi)=\cos\left(\gamma(0)\right)\cos\left(\gamma(\pi)\right)=\pm 1,
\end{equation}
under a continuous gauge of $\gamma(\phi_y)$ in the half period of $\phi_y\in[0,\pi]$.
Consequently, as shown in Fig.~\ref{fig-interface}, {\color{Blue}the $\mathbb{Z}_2$ index $\Delta\tilde{\nu}$ of $\Delta R$ determines the parity of the number of helical interface state pairs}. So we need to derive the relation between $\Delta\tilde{\nu}$ and the $\mathbb{Z}_2$ indices, $\tilde{\nu}_\mathrm{I}$, $\tilde{\nu}_\mathrm{II}$ , of the two PCs.

Plugging $R_a=e^{iq_a}\exp(i\alpha_a\vec{n}_a\cdot\vec{\sigma})$ ($a=\rm I,\rm{II}$) into $\Delta R$, we obtain
\begin{equation}
\begin{split}    
    \Delta R &=e^{ip}\exp(i\gamma\vec{m}\cdot\vec{\sigma})= R_{\rm I} R_{\rm II}=e^{i(q_{\rm I}+q_{\rm II})}\exp(i\alpha_L\vec{n}_{\rm I}\cdot\vec{\sigma})\exp(i\alpha_{\rm II}\vec{n}_{\rm II}\cdot\vec{\sigma})\\
    &= e^{i(q_{\rm I}+q_{\rm II})}\Big[\cos\alpha_{\rm I}\sigma_0+i\sin\alpha_{\rm I}\vec{n}_{\rm I}\cdot\vec{\sigma}\Big]\Big[\cos\alpha_{\rm II}\sigma_0+i\sin\alpha_{\rm II}\vec{n}_{\rm II}\cdot\vec{\sigma}\Big]\\
    &=  \underbrace{e^{i(q_{\rm I}+q_{\rm II})}}_{\displaystyle =e^{ip}}\Big[\underbrace{\left(\cos\alpha_{\rm II}\cos\alpha_{\rm I}-\sin\alpha_{\rm I}\sin\alpha_{\rm II}(\vec{n}_{\rm I}\cdot\vec{n}_{\rm II})\right)}_{\displaystyle =\cos\gamma}\sigma_0\\ 
    &\quad+i\underbrace{\left(\cos\alpha_{\rm I}\sin\alpha_{\rm II}\vec{n}_{\rm II}+\sin\alpha_{\rm I}\cos\alpha_{\rm II}\vec{n}_{\rm I}-\sin\alpha_{\rm I}\sin\alpha_{\rm II}(\vec{n}_{\rm I}\times\vec{n}_{\rm II})\right)}_{\displaystyle =\sin\gamma\ \vec{m}}\cdot\vec{\sigma}\Big].
\end{split}
\end{equation}
At $\mathcal{T}_f$-invariant points ($\phi_0=0,\pi$), $\alpha_a(\phi_0)=0$ or $\pi$ ($a=\rm I,\rm II$), which leads to
\begin{equation}\label{simple relation}
    \cos\big(\gamma(\phi_0)\big)=\cos\big(\alpha_{\rm I}(\phi_0)\big)\cos\big(\alpha_{\rm II}(\phi_0)\big).
\end{equation}
Substitution of Eq.~\eqref{simple relation} into Eq.~\eqref{Delta Z2} yields the relation between the $\mathbb{Z}_2$ index $\Delta\tilde{\nu}$ and those of the two PCs
\begin{gather}
\begin{aligned}
    \exp(i\Delta\tilde{\nu}\pi)&=\cos\big(\alpha_{\rm I}(0)\big)\cos\big(\alpha_{\rm II}(0)\big)\cos\big(\alpha_{\rm I}(\pi)\big)\cos\big(\alpha_{\rm II}(\pi)\big)\\
    &= \Big[\cos\big(\alpha_{\rm I}(0)\big)\cos\big(\alpha_{\rm I}(\pi)\big)\Big]\Big[\cos\big(\alpha_{\rm II}(0)\big)\cos\big(\alpha_{\rm II}(\pi)\big)\Big]\\
    &= \exp(i\tilde{\nu}_{\rm I})\exp(i\tilde{\nu}_{\rm II})=\exp(i(\tilde{\nu}_{\rm I}+\tilde{\nu}_{\rm II})),
\end{aligned}\\[3pt]
\Leftrightarrow\quad \Delta\tilde{\nu}= (\tilde{\nu}_{\rm I}+\tilde{\nu}_{\rm II})\bmod 2=(\tilde{\nu}_{\rm I}-\tilde{\nu}_{\rm II})\bmod 2.
\end{gather}
Therefore, we have proved that {\color{Blue}the difference between the $\mathbb{Z}_2$ indices $\Delta\tilde{\nu}=(\tilde{\nu}_{\rm I}-\tilde{\nu}_{\rm II})\bmod 2$ of the PCs at two sides determines the parity of the number of helical interface state pairs}.

\section{Experimental proposal for spin reflection with an effective twisted boundary}

\begin{figure}[b!]
\includegraphics[width=0.88\textwidth]{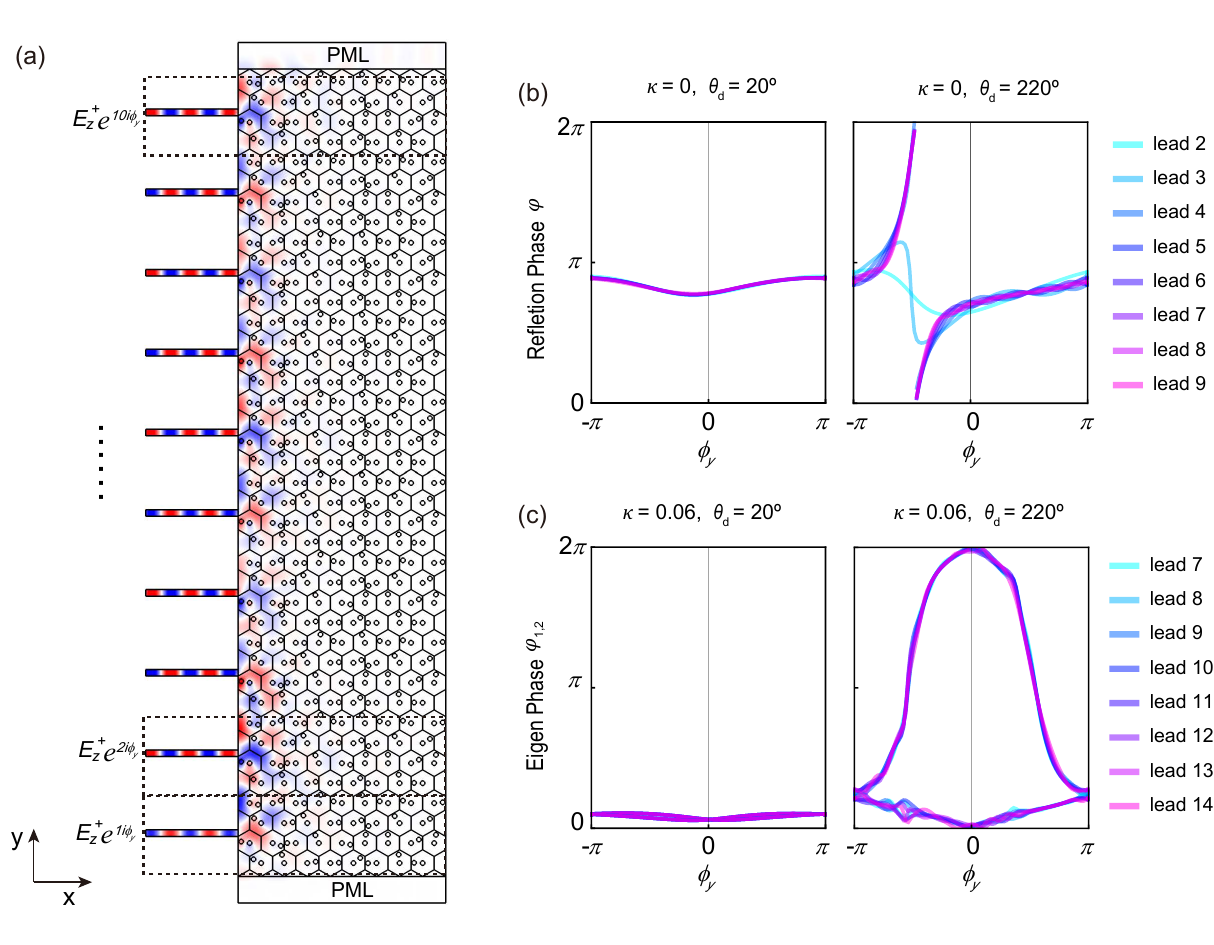}
\caption{\label{fig-twistBC}
The possible implementation of ``twisted boundary condition” in experiment. (a) A disordered PC of size $9a \times 6l$ (marked by the supercell in dashed box) is repeated $10$ times along the $y$ direction, and the phase difference of the incident waves between two adjacent leads is $\phi_y$. The upper and lower boundaries of the PC are encased by perfectly matched layers (PML) to mimic the open boundary condition in experiments. The colormap denotes the $E_z$ component for $\phi_y=\pi$, which is the numerically calculated in COMSOL. (b) Reflection phases in different leads for spin-decoupled PCs with disorder strength $\theta_d=20\degree$(trivial) and $\theta_d=220\degree$(non-trivial). (c) Eigen reflection phases in different leads for the spin-coupled PCs ($\kappa=0.06$) with $18$ supercells each of which has size $15a \times 6l$.  
}
\end{figure}

In the above two sections, we proposed an efficient way to extract the topological invariants for the full-wave disordered systems. %And the result shows that our approach not only is computationally efficient but also invests the $\mathbb{Z}$ and $\mathbb{Z}_2$ invariants with a clear physical meaning. Therefore, even if regarding our approach as a theoretical generalization of Laughlin’s charge pumping Gedanken experiment, it may advance the study of the topological disordered systems both technically and conceptually. Nevertheless, as 
Here, we show that the scheme with twisted boundary is also experimentally achievable after some minor modifications.

As shown in Fig. \ref{fig-twistBC}, we propose a possible implementation of ``twisted boundary condition” in real experiments, where the multi-lead structure is obtained by repeating the supercell (dashed box) in the $y$ direction $N$ times and each supercell is composed of a waveguide lead and a disordered PC of $M$ layers (the disorder configuration is the same in every supercell). Here, considering the feasibility of the experiments, we choose $M=4$ and $N=10$ ($N=18$) for the spin-decoupled (spin-coupled) case. To simulate the twisted boundary condition with a twist angle $\phi_y$ for each supercell, we can illuminate the waveguide array with a constant phase increment $\phi_y$ in the $y$ direction, such that the incident wave in the $n$th lead equals $E_z^+ e^{(in\phi_y )}$. Then we collect the reflected waves and compute the eigen reflection phases in each lead as what we have done in the original single-lead structure (such as Fig. \ref{fig-3}). 

For the spin-decoupled PCs with weak ($\theta_d=20\degree$) and strong ($\theta_d=220\degree$) disorders, Fig. \ref{fig-twistBC}(b) shows the TM reflection phases in lead $n$ ($n=2,3,\dots,9$) as functions of $\phi_y$, whose winding numbers reveals the topological properties of the disordered PCs. Here, we note that for the topologically non-trivial case ($\theta_d=220\degree$), the reflection phases in the upper leads ($n>4$) are nearly identical and can manifest the non-trivial topology of the PC, while the results in the lowest two leads ($n=2,3$) deviate from the idea case with infinitely many supercells. This is indeed induced by the finite size effect.  For the topologically trivial case ($\theta_d=20\degree$), there is no unidirectional edge state, and the reflection phase in all leads are nearly identical.

To justify the repeating supercell structure and understand the numerical results, let us first imagine the ideal structure of infinitely many supercells, where the total reflected wave in the $n^\mathrm{th}$ waveguide is the superposition of the components, $\tilde{E}_r^{(n,m)}$, generated by the incident waves from the leads in all supercells (labeled by the index $m$): $E_r^{(n)}=\sum_{m=-\infty}^{+\infty}\tilde{E}_r^{(n,m)}$, and should be precisely identical with the result of a single supercell with a twisted boundary condition.  Due to the energy leaking, $|\tilde{E}_r^{(n,m)}|$ generally decreases with the distance $|n-m|$ between the incident and reflected leads. Hence, the reflected wave $E_r^{(n)}$ in a lead is mainly contributed by the scattering from the several leads close to it. Therefore, for a structure with finite supercells, one may expect that the reflection phases in the leads would converge to the ideal result provided that the leads are far enough away from the upper and lower boundaries, while the result in the topmost and bottommost leads would stray from the infinite case. However, since the flow of TM edge states is clockwise (upwards on the left boundary), the reflected field in a lead to is much more affected by the incident waves from the leads below it than from those above it. Therefore, the reflected fields in the topmost leads  are nearly uninfluenced by the truncation of the upper boundary. In contrast, the reflection phases in the bottommost leads are seriously disturbed by the lower boundary. This explains why the numerical results of the reflection phases of the leads 8 and 9 are rather perfect in Fig.~\ref{fig-twistBC}(b), while the results of the leads 2 and 3 violate the prediction of the ideal case.

The results for the spin-coupled PCs are shown in Fig. \ref{fig-twistBC}(c). Due to the concurrence of the edge states in both directions, we increase the supercells to $18$, and only show the eigen reflection phases in the leads close to the structure’s midline ($n=7,8,\dots,14$). The numerical results in Fig. \ref{fig-twistBC}(c) are consistent with the Fig. 3(b) in the main text, demonstrating the effectiveness of our design to reproduce the Gedanken experiment with a twisted boundary condition.

%\bibliography{Reference}
